\documentclass[sigconf,nonacm]{acmart}
\usepackage{arydshln}
\usepackage{diagbox}
\usepackage{algorithmic}
\usepackage{algorithm}
\usepackage{multirow}
\usepackage{multicol}
\usepackage{adjustbox}
\usepackage{booktabs}
\usepackage{cleveref}
\usepackage{subfig}
\usepackage{balance}

\usepackage{enumitem}
\usepackage{xcolor}
\usepackage{booktabs}
\usepackage{tabularx}

\usepackage{xspace}

\makeatletter
\DeclareRobustCommand\onedot{\futurelet\@let@token\@onedot}
\def\@onedot{\ifx\@let@token.\else.\null\fi\xspace}
\DeclareRobustCommand\nodot{\futurelet\@let@token\@nodot}
\def\@nodot{\ifx\@let@token.\else~\null\fi\xspace}
\def\eg{\emph{e.g}\onedot}

\makeatother
\newcommand{\name}{Fast-GSVC}

\AtBeginDocument{%
  }

\newcommand{\re}[1]{\textcolor{black}{#1}}

\setcopyright{none}
\renewcommand\footnotetextcopyrightpermission[1]{}

\begin{document}

\title{Faster, Smaller, and Better-Looking: {\name} for 2D Gaussian Video Coding}

\author{Longan Wang}
\affiliation{%
  \institution{National University of Singapore}
  \city{Singapore}\country{}}
\email{wanglongan@u.nus.edu}
\orcid{0009-0001-9989-3739}

\author{Yuang Shi}
\affiliation{%
  \institution{National University of Singapore}
  \city{Singapore}\country{}}
\email{yuangshi@u.nus.edu}
\orcid{0000-0002-7893-8512}

\author{Chia Qian Tan}
\affiliation{%
  \institution{National University of Singapore}
  \city{Singapore}\country{}}
\email{tanchiaqian@gmail.com}
\orcid{0009-0003-5005-6299}

\author{Wei Tsang Ooi}
\authornote{Corresponding author}
\affiliation{%
  \institution{National University of Singapore}
  \city{Singapore}\country{}}
\email{dcsooiwt@nus.edu.sg}
\orcid{0000-0001-8994-1736}

\begin{abstract}
2D Gaussians have emerged as a promising explicit representation for videos.  However, fitting Gaussians to a video sequence requires substantial computation, resulting in long encoding times.
In this work, we present {\name}, a 2D Gaussian video codec that accelerates encoding while improving rate-distortion performance.
We identify the key bottlenecks in 2DGS training and accelerate optimization through improved initialization, smoother yet more aggressive parameter updates, coarse-to-fine progressive Gaussian optimization, and optical-flow guidance for both Gaussian initialization and temporal propagation.
We also discovered that the quantization fine-tuning process of the existing codec does not fully exploit temporal coherence between frames, leading to suboptimal compression. 
To address this, we preserve spatiotemporal correlations to maintain concentrated residuals and introduce temporal-aware quantization for further improved rate-distortion performance.
In our experiments on the UVG dataset, {\name} reduces encoding time by up to 92.5\%, corresponding to a 13.4$\times$ speedup over GSVC. At comparable PSNR, it reduces file size by up to 58\%; at comparable bit-per-pixel, it improves PSNR by up to 2.5 dB. 
Project page: \url{https://longanwang-cs.github.io/fastgsvc/}. 

\end{abstract}

\ccsdesc[500]{Computing methodologies~Image compression}

\keywords{2D Gaussian Splatting, Video Coding, System acceleration}

\maketitle

\section{Introduction \label{Sec:Introduction}}

\begin{figure}[t]
    \centering

    \subfloat[Loss convergence.]{
        \includegraphics[width=0.47\linewidth]{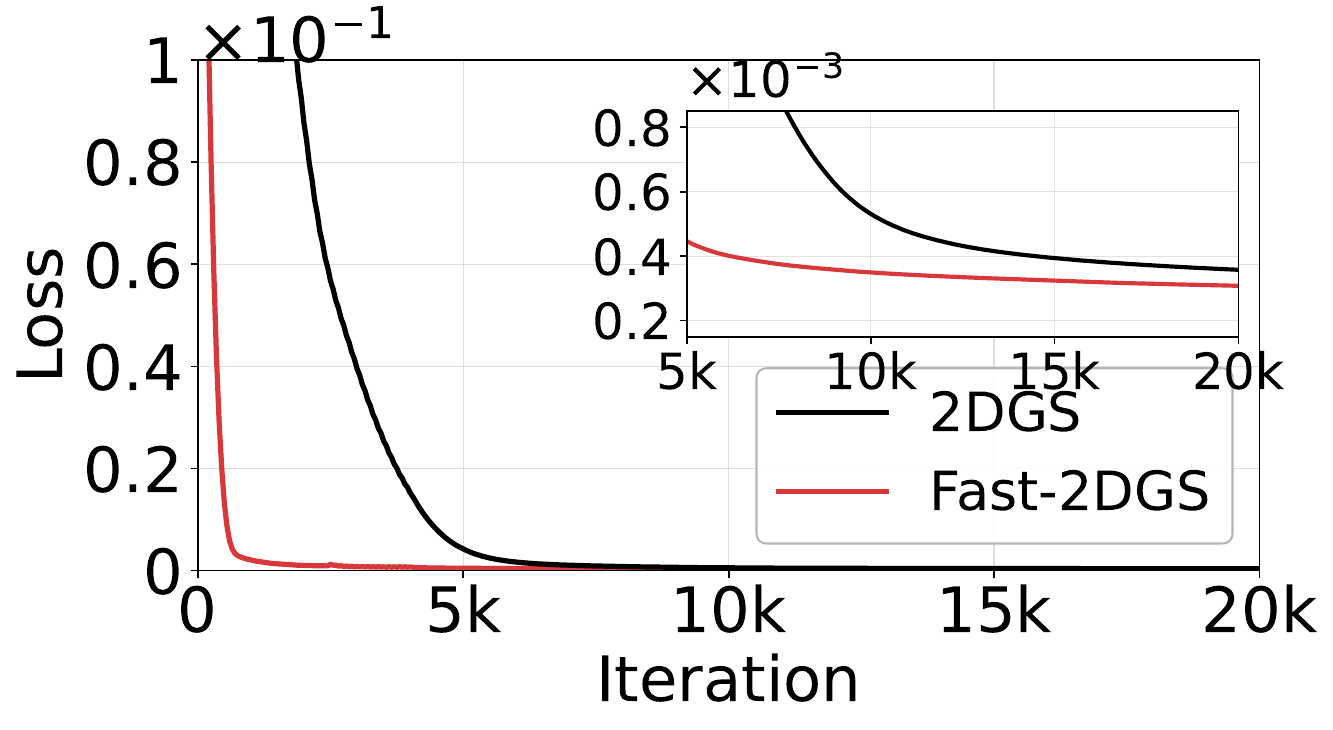}
        \label{fig:training_loss}
    }
    \subfloat[Gradient fluctuation.]{
        \includegraphics[width=0.47\linewidth]{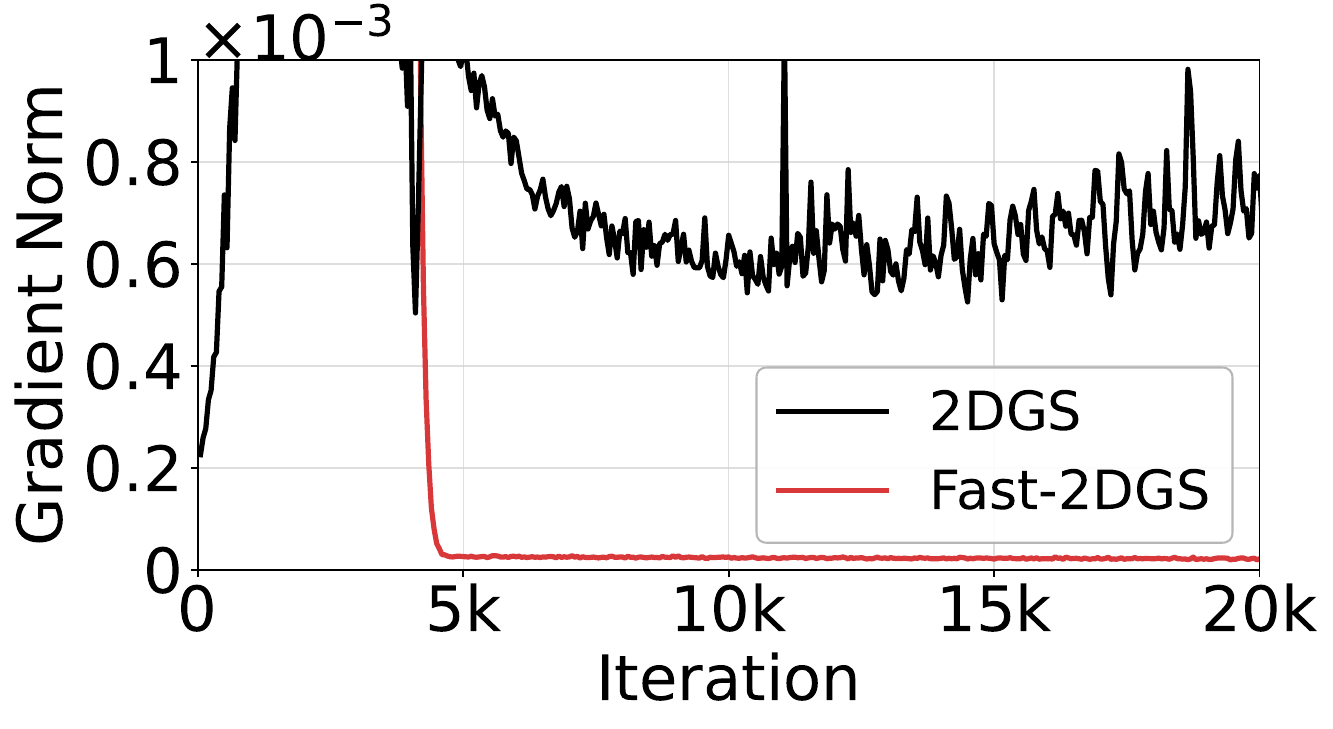}
        \label{fig:gradient_norm}
    }

    \subfloat[\re{Unstable 2DGS optimization.}]{
        \includegraphics[width=0.49\linewidth]{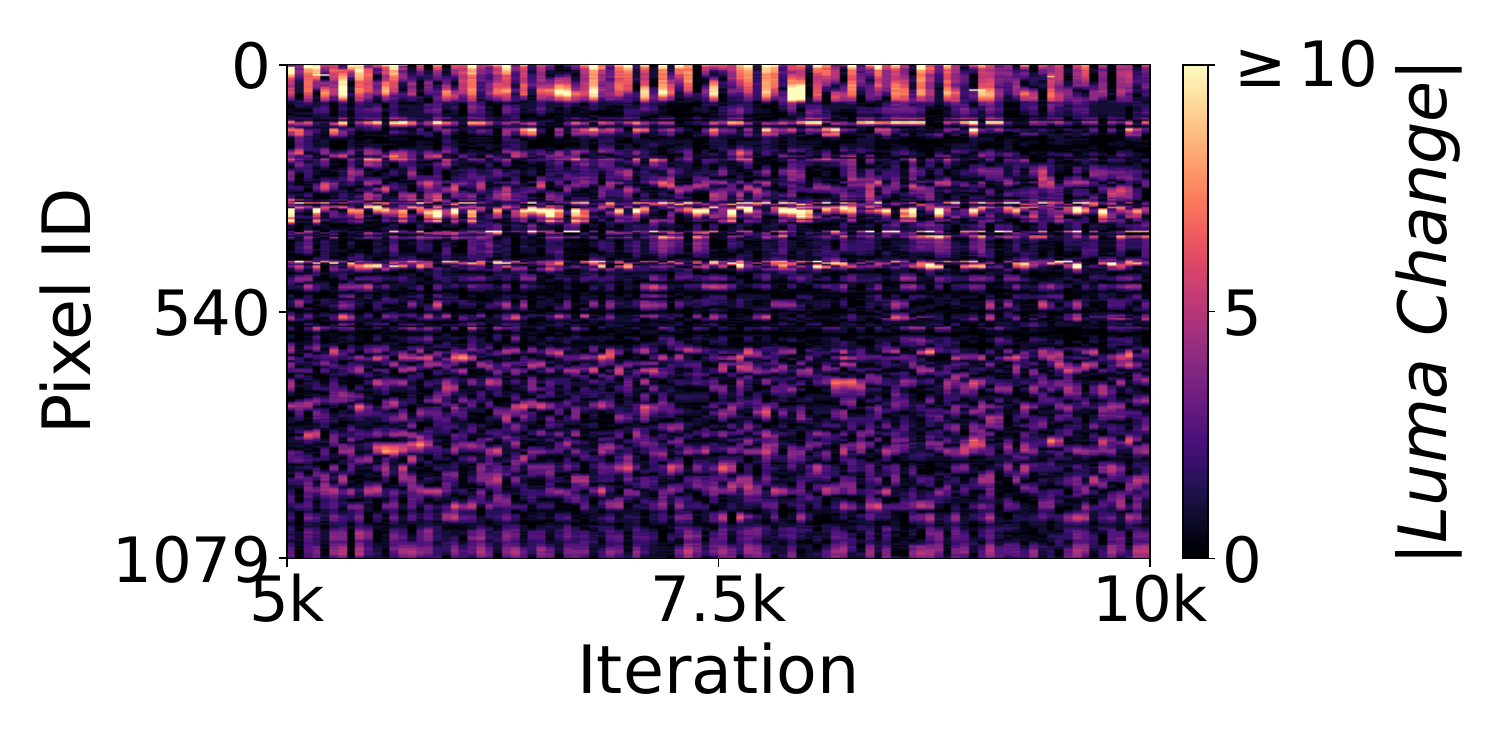}
        \label{fig:stability_gsvc}
    }
    \subfloat[\re{Stable Fast-2DGS optimization.}]{
        \includegraphics[width=0.49\linewidth]{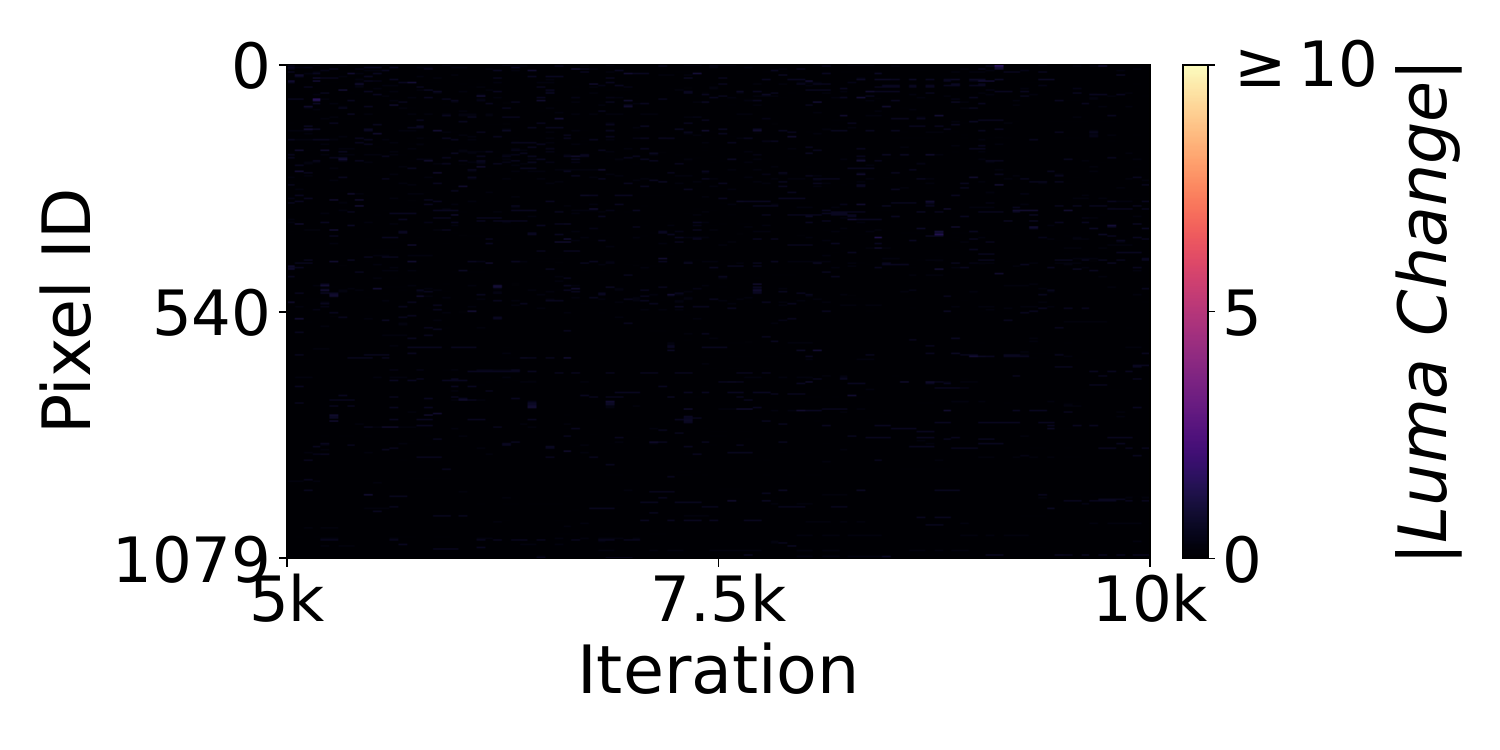}
        \label{fig:stability_fastgsvc}
    }

    \caption{
    \re{Comparison of optimization dynamics between 2DGS and Fast-2DGS.(a) Fast-2DGS achieves faster loss reduction, while the inset highlights that Fast-2DGS converges to smaller loss.
    (b) Fast-2DGS reduces gradient fluctuations.
    (c,d) Absolute pixel-wise luma changes between sampled optimization iterations along the central vertical scanline, showing substantially reduced pixel fluctuations in Fast-2DGS.
    }
    }
    \label{fig:optimization_stability}
\end{figure}

\begin{figure*}[t]
    \centering
    \setlength{\tabcolsep}{1.5pt}
    \renewcommand{\arraystretch}{0.6}

       \subfloat[GSVC (2DGS).]{
        \begin{minipage}{\textwidth}
            \centering
            \includegraphics[width=0.19\textwidth]{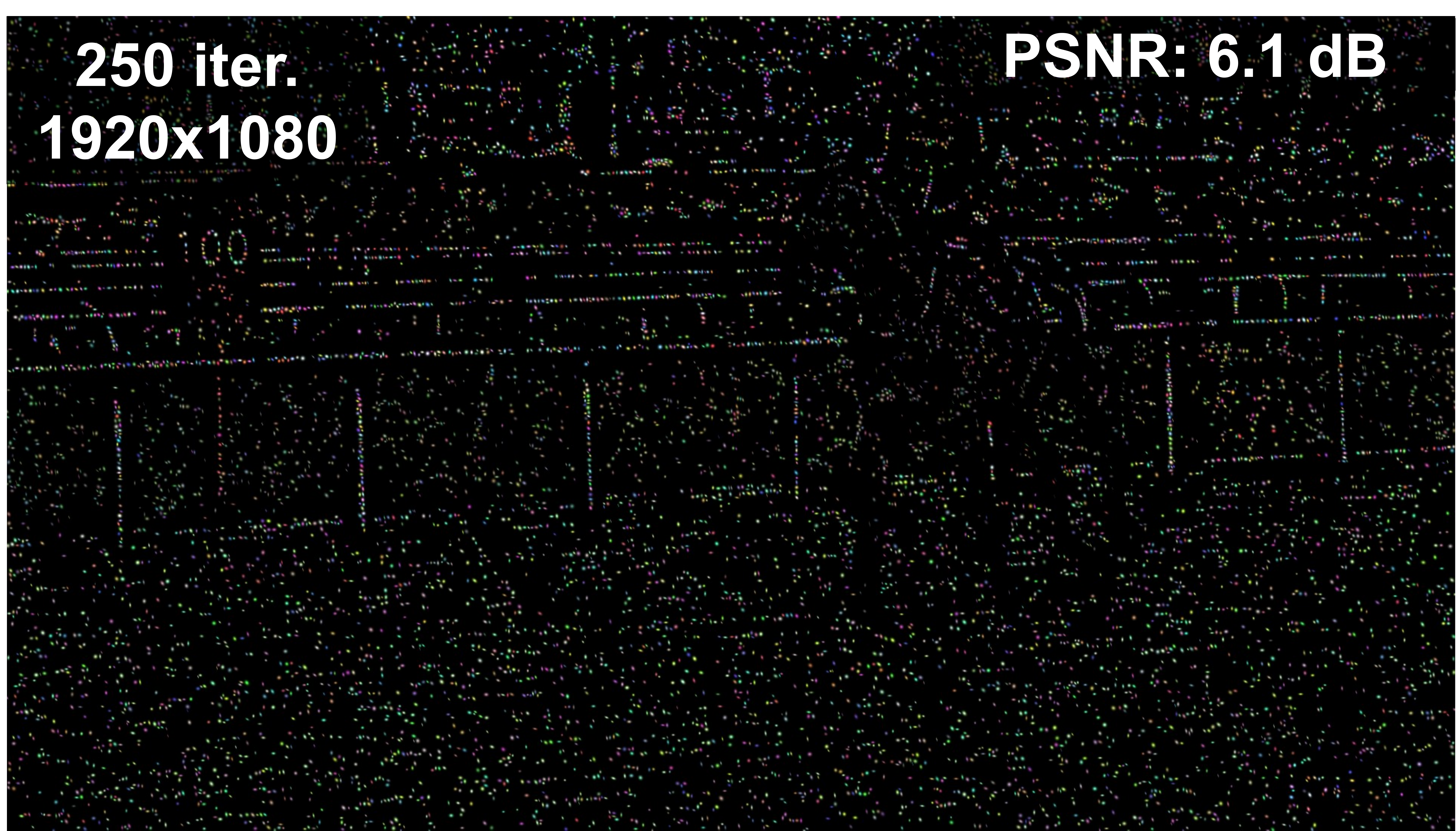}\hfill
            \includegraphics[width=0.19\textwidth]{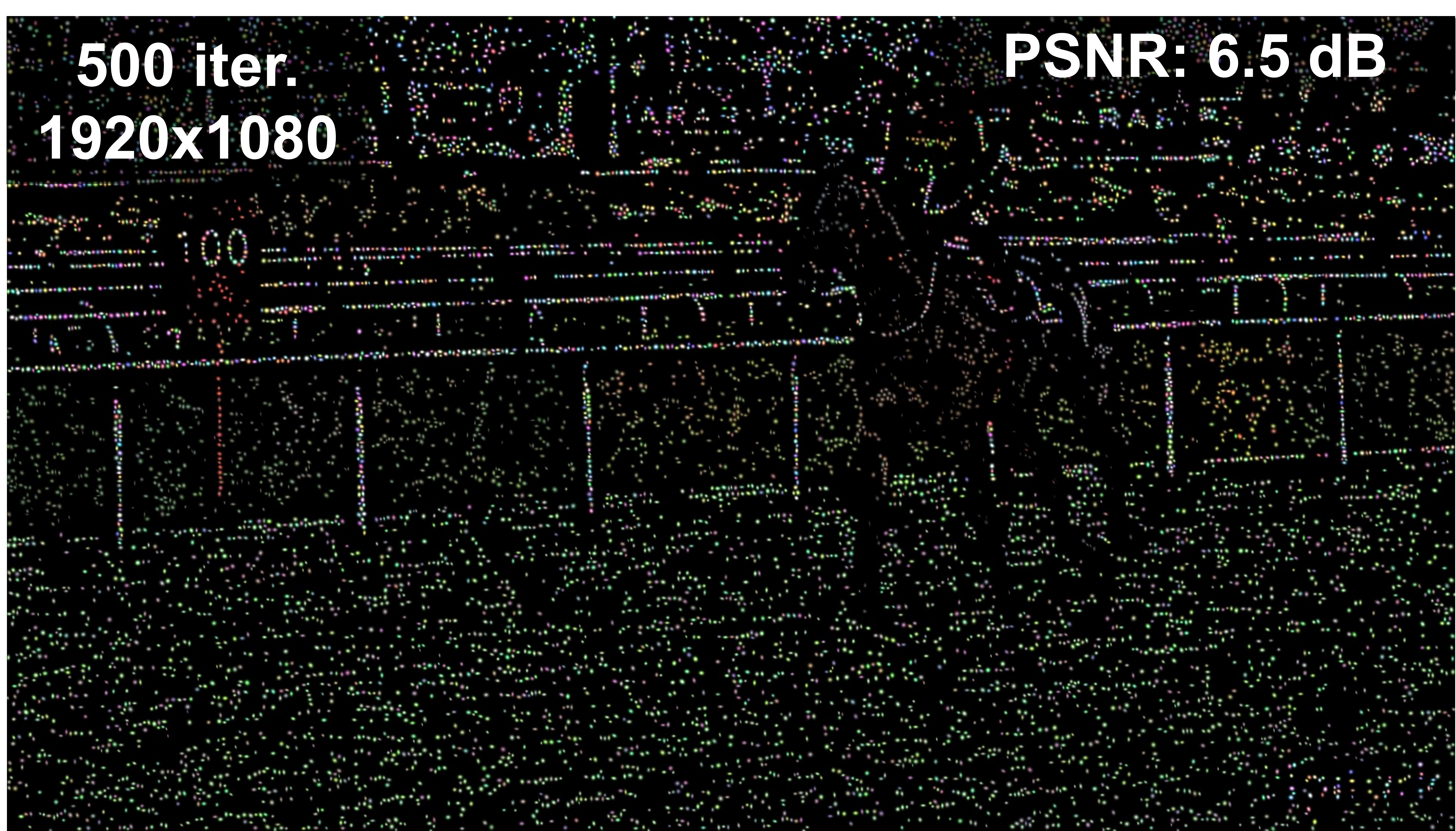}\hfill
            \includegraphics[width=0.19\textwidth]{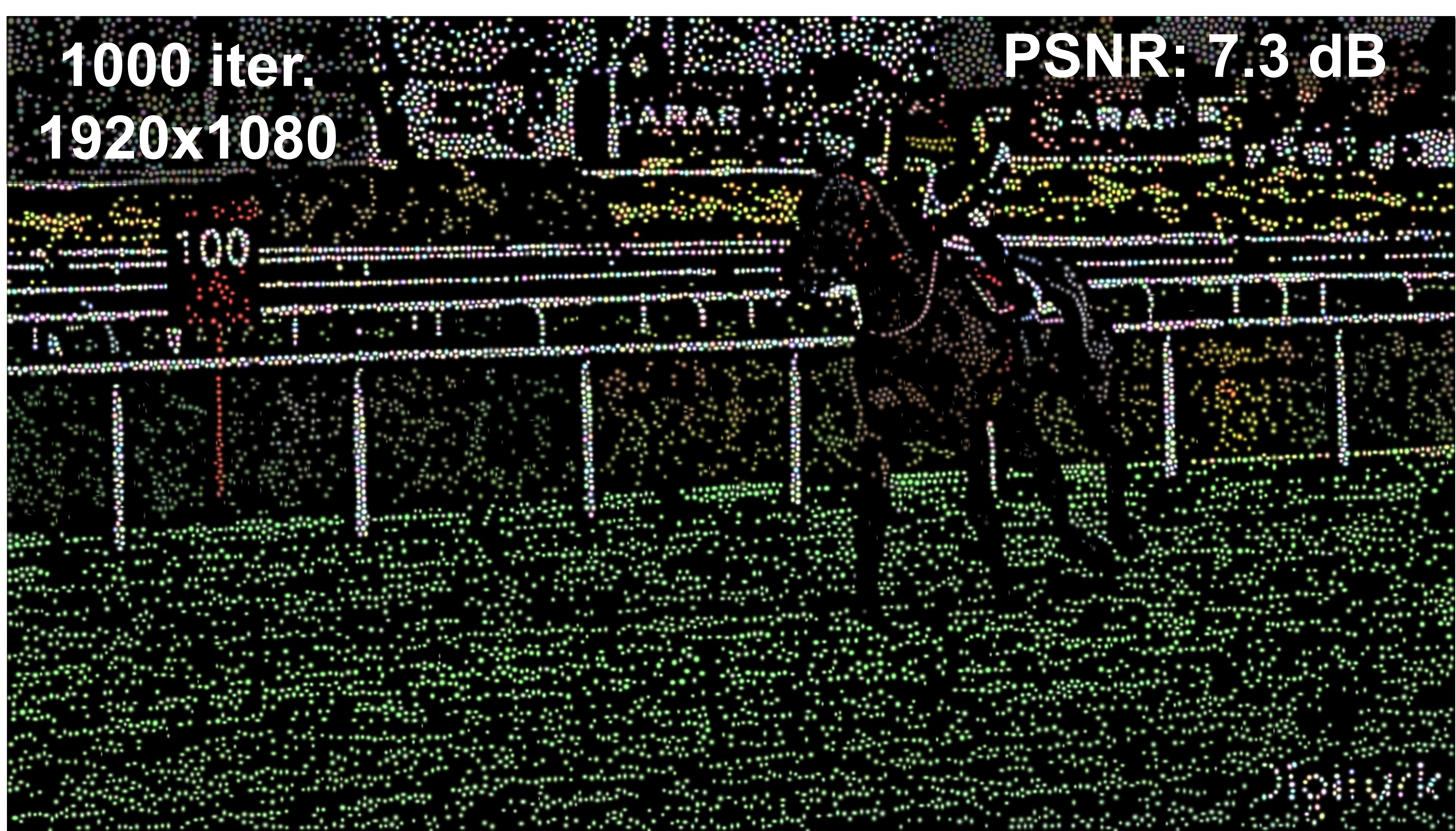}\hfill
            \includegraphics[width=0.19\textwidth]{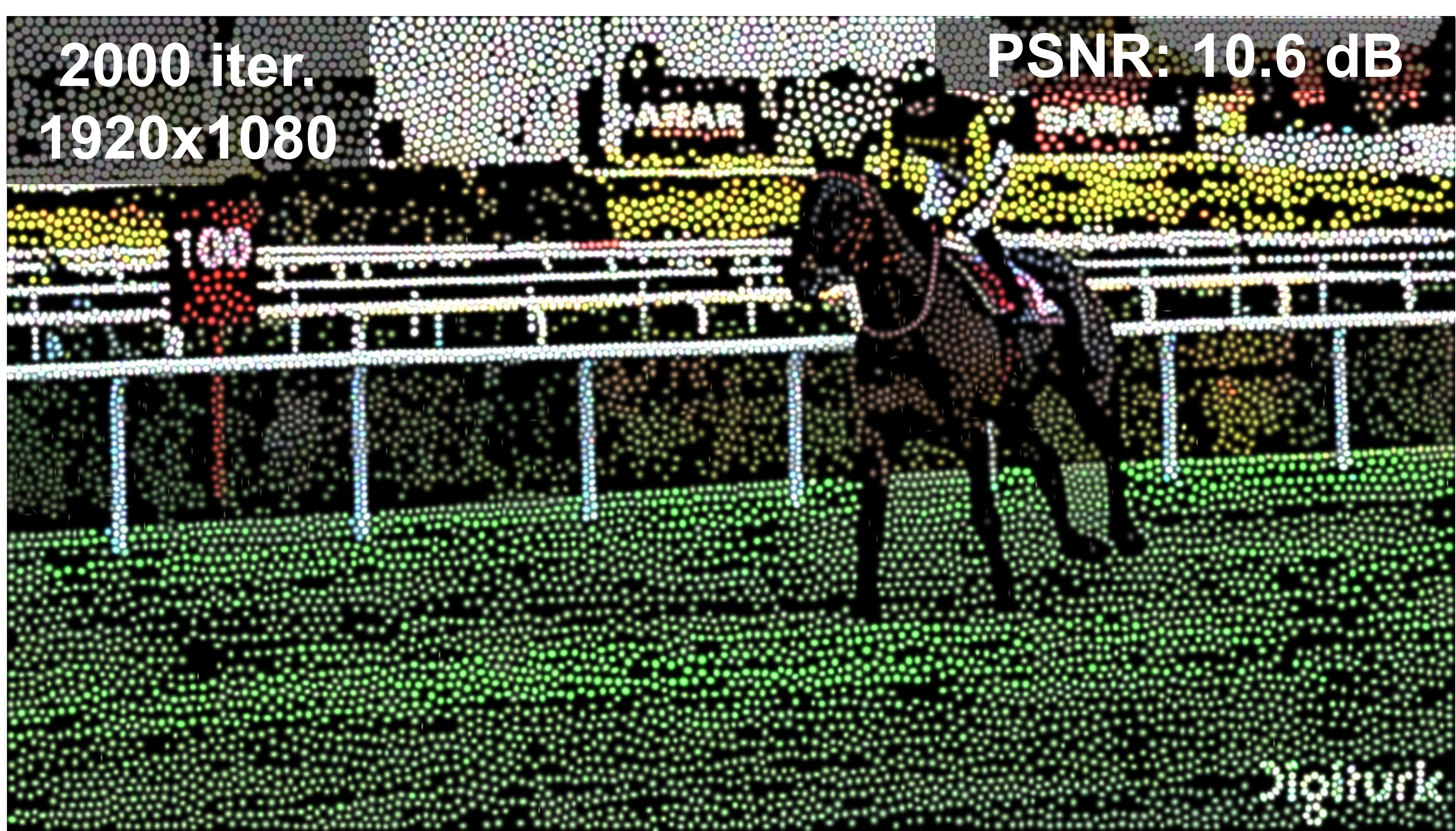}\hfill
            \includegraphics[width=0.19\textwidth]{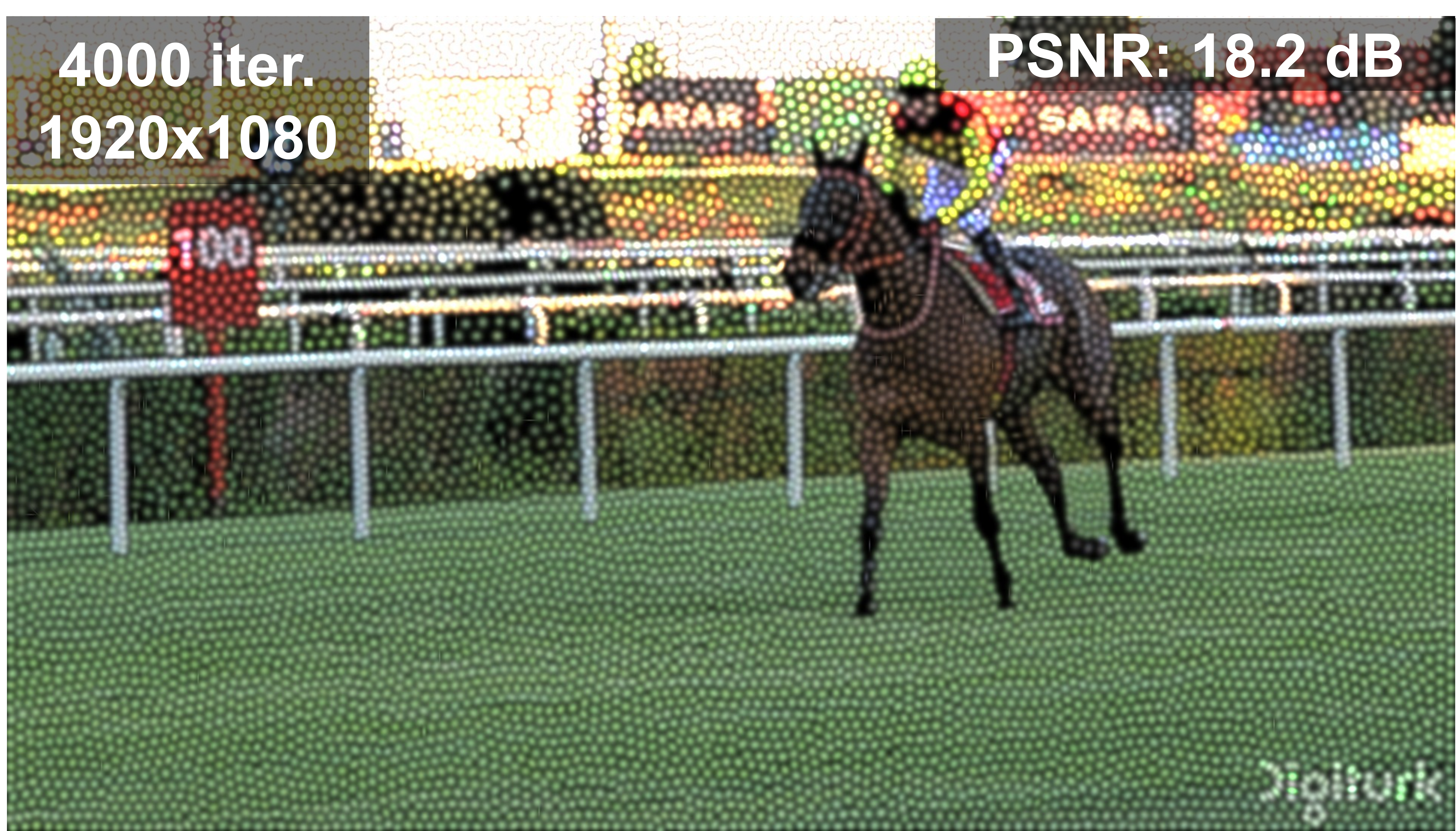}
        \end{minipage}
        \label{fig:training_dynamics_gsvc}
    }
    \hfill
     \subfloat[{\name} (Fast-2DGS).]{
        \begin{minipage}{\textwidth}
            \centering
            \includegraphics[width=0.19\textwidth]{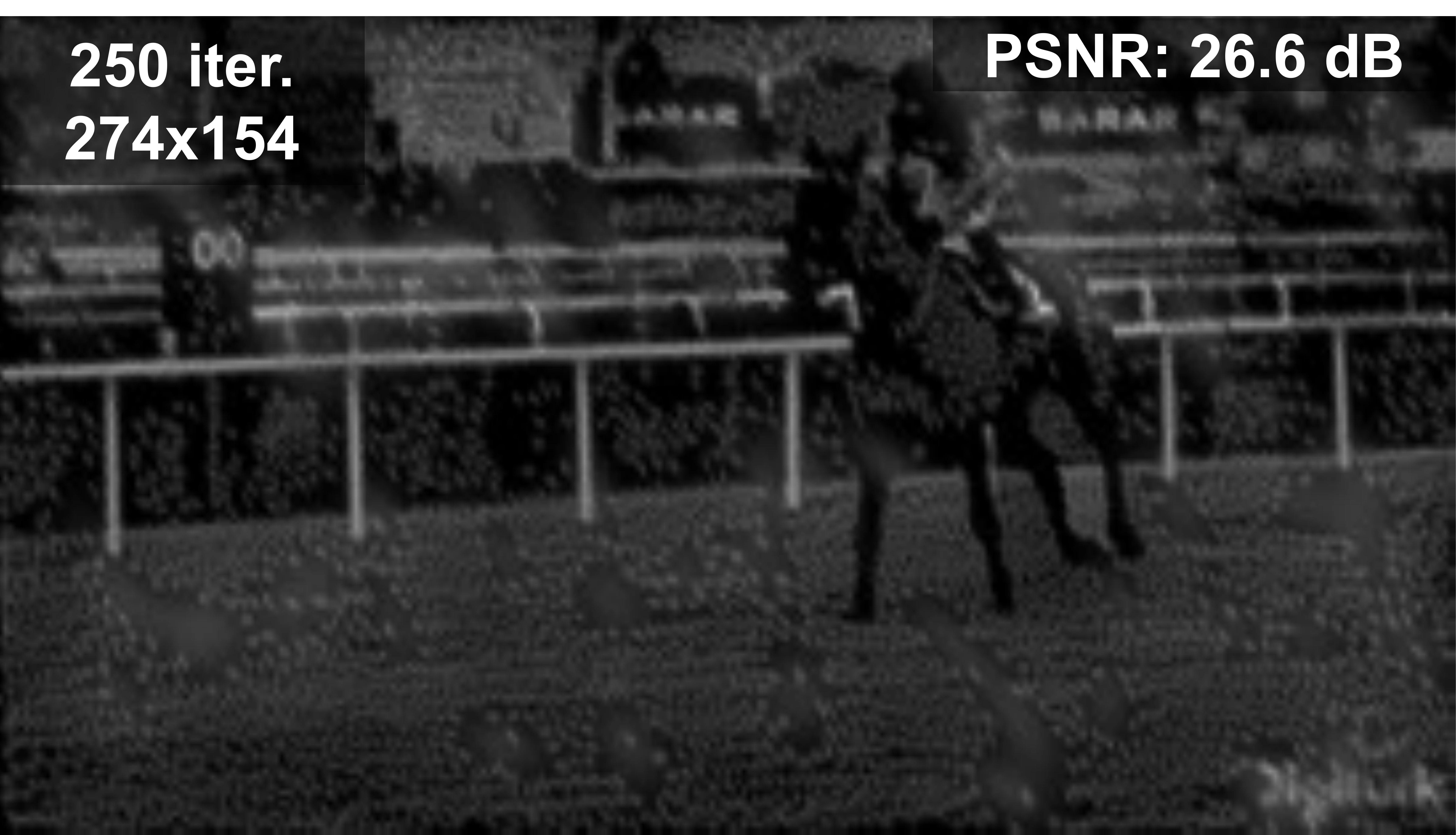}\hfill
            \includegraphics[width=0.19\textwidth]{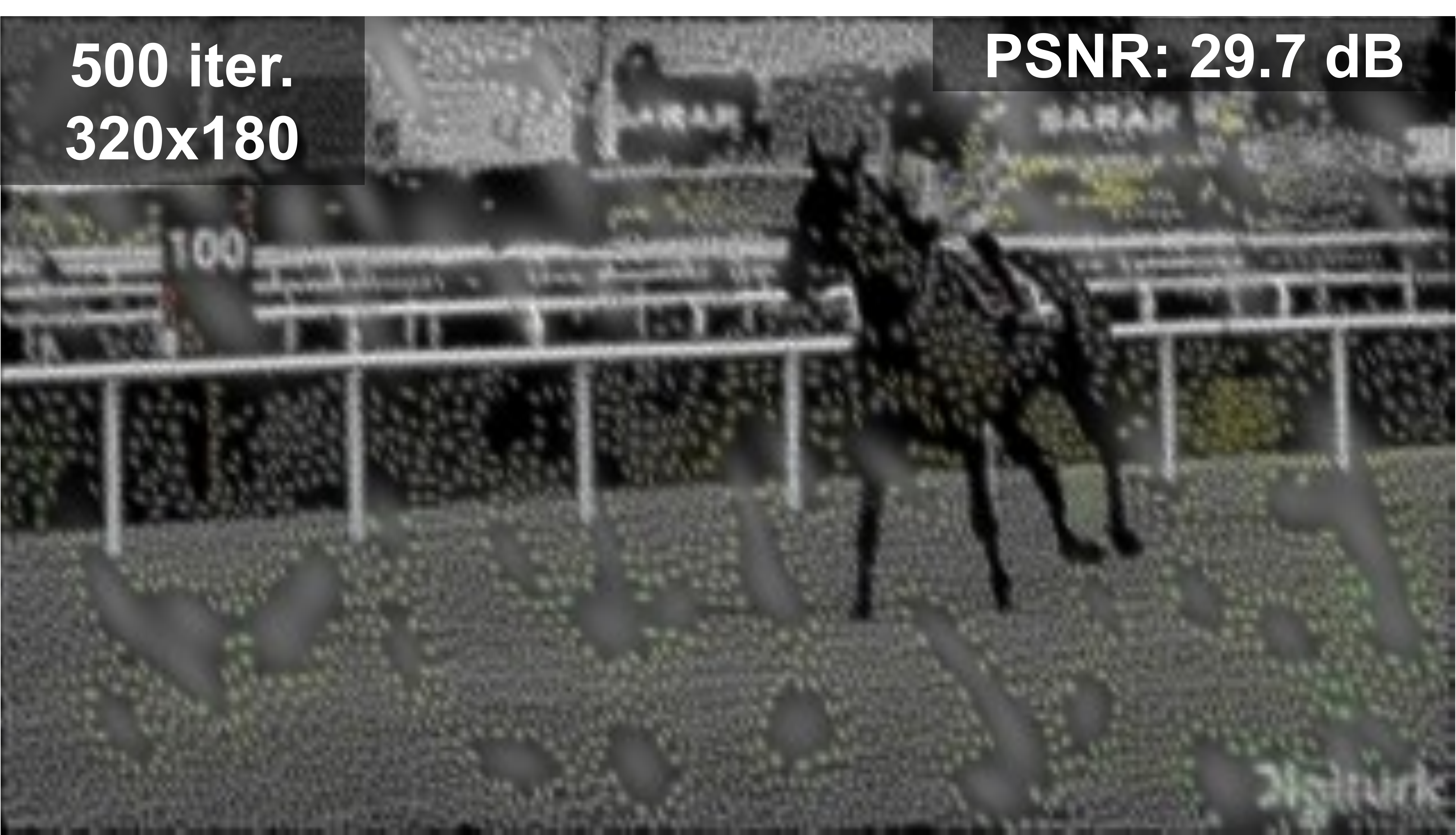}\hfill
            \includegraphics[width=0.19\textwidth]{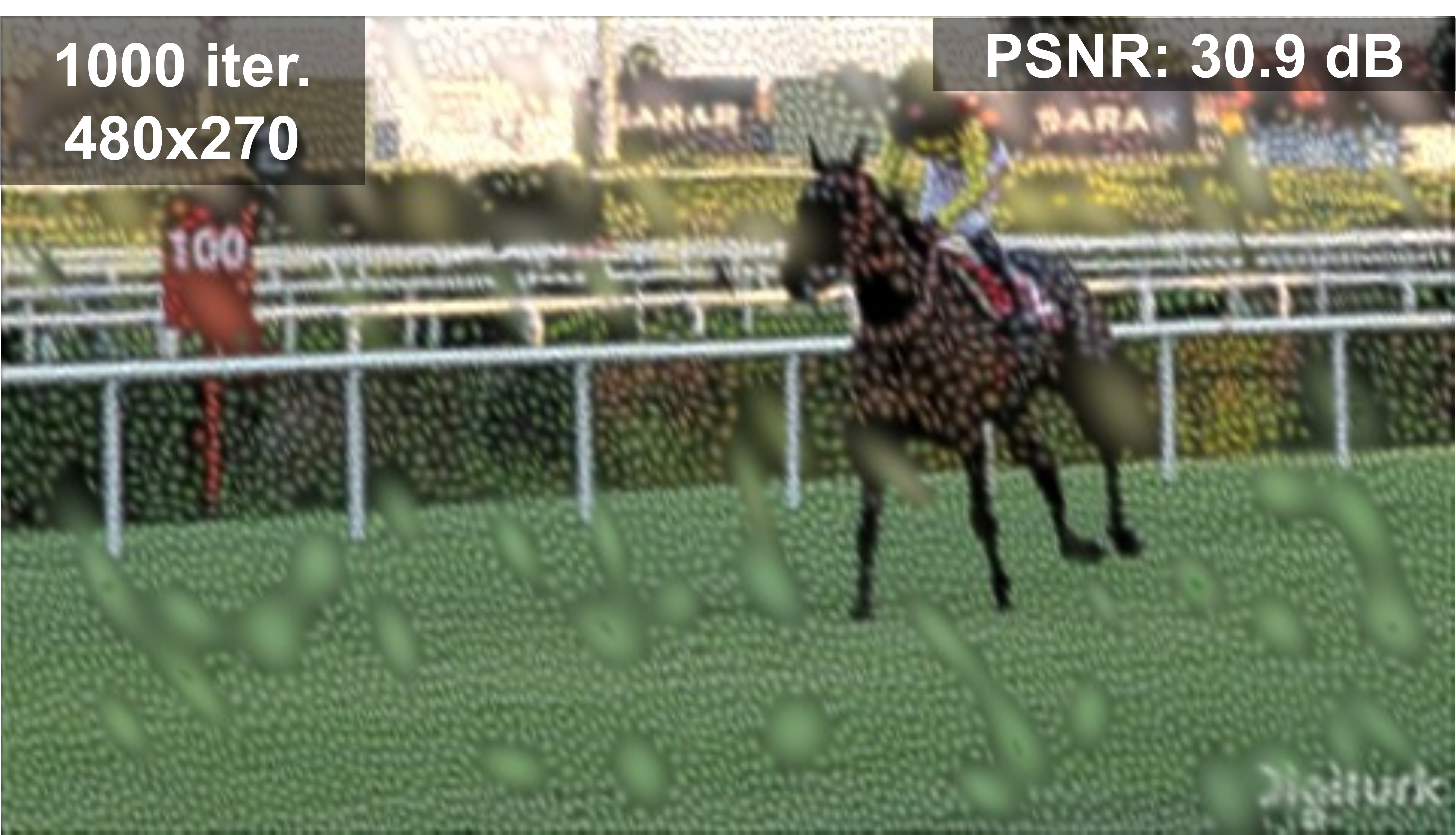}\hfill
            \includegraphics[width=0.19\textwidth]{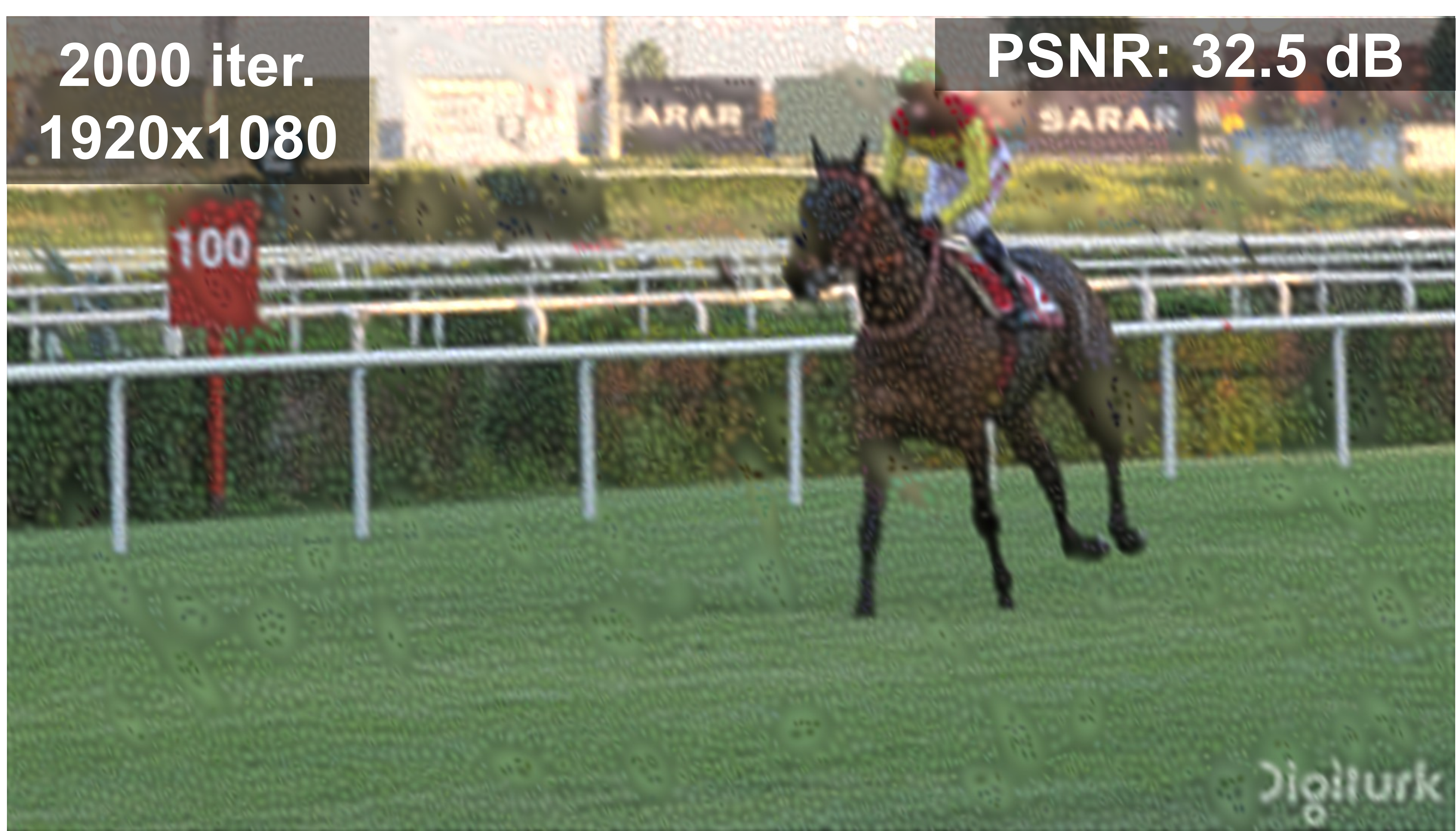}\hfill
            \includegraphics[width=0.19\textwidth]{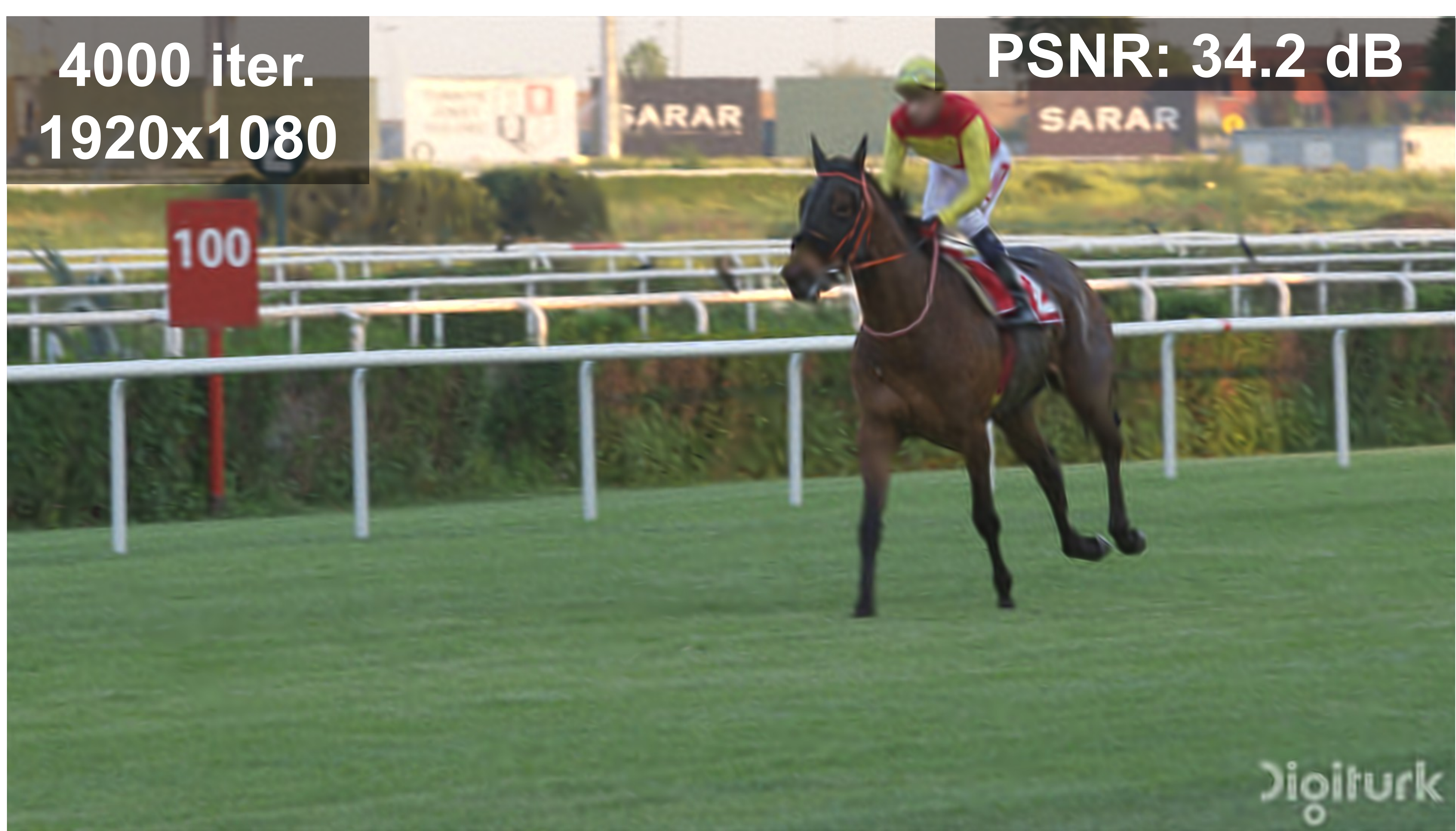}
        \end{minipage}
        \label{fig:training_dynamics_fastgsvc}
    }
    \caption{Training dynamics on Jockey with 10k Gaussians over the first 4k iterations, with PSNR measured at the corresponding training resolution. After 4k iterations, {\name} already reaches 34.2 dB and produces a well-fitted reconstruction, whereas GSVC achieves only 18.2 dB and still exhibits noticeable holes and artifacts.}
    \label{fig:training_dynamics_render}
\end{figure*}

For decades, video coding has relied on signal processing over fixed pixel grids, as in standards such as AVC~\cite{AVC}, HEVC~\cite{HEVC}, and VVC~\cite{VVC}.
Beyond this paradigm, neural methods have also been explored, encoding videos into network parameters or latent representations~\cite{DBLP:conf/mm/Li0022, DBLP:conf/cvpr/LiLL23, DBLP:conf/cvpr/LiLL24,shi2024volumetric}.
While compact, their implicit representations require neural inference for decoding, limiting fine-grained control and direct editing.
Recently, 3D Gaussian Splatting (3DGS)~\cite{kerbl20233d} has emerged as a promising explicit representation for videos~\cite{DBLP:conf/nips/Sun0MLC024, DBLP:journals/corr/abs-2411-11024}.
With directly accessible and manipulable Gaussians, 3DGS supports fast rendering~\cite{kerbl20233d}, arbitrary-resolution projection~\cite{zeng2026arbitrary}, direct editing~\cite{DBLP:conf/nips/Sun0MLC024, DBLP:journals/corr/abs-2411-11024}, progressive reconstruction for layered streaming~\cite{DBLP:conf/3dim/ShiMGO25}, and flexible size-quality control~\cite{shi2026sketch}.
However, 3DGS is computationally expensive due to its large parameter volume after densification and costly depth-sorted alpha blending.

In 2024, 2D Gaussian Splatting (2DGS)~\cite{zhang2024gaussianimage} was introduced for image representation and compression.
By operating directly on the image plane, 2DGS removes the depth and view-dependent modeling required by 3DGS, reduces the number of parameters per Gaussian from 59 to 8, and replaces depth-sorted alpha blending with cheap accumulated summation.
Building on these advantages, GSVC~\cite{wang2025gsvc} introduces the first Gaussian video codec with 2DGS.
To exploit temporal coherence, GSVC represents a video using anchor frames and propagation frames. 
Anchor frames are first independently optimized from scratch, similar to intra-coded frames, while propagation frames exploit inter-frame dependency by inheriting and refining the Gaussians from the previous frame.
Gaussians can be added and removed to adapt to dynamics and redundancy, and scene changes trigger new anchor frames.
For compression, GSVC applies quantization-aware fine-tuning with residual coding.
Together, these designs exploit temporal redundancy for compact representation, while achieving state-of-the-art rate-distortion performance.
However, even with lightweight 2DGS and an H100 GPU, GSVC requires hundreds of seconds per 1080p frame to encode, accumulating to over four days for a 600-frame video.

\re{To understand what makes 2DGS optimization slow, we examine its loss curve during iterative optimization, as shown in~\Cref{fig:training_loss}.
The process consists of two stages: an initial fitting stage, where Gaussians expand to fill uncovered regions and the loss decreases relatively quickly, followed by a refinement stage, where optimization continues to refine details while the loss decreases more slowly, as highlighted in the inset.
Examining these stages further, we identify two key bottlenecks in 2DGS optimization:}
(i) the initial fitting stage requires many iterations to establish coverage, as shown in~\Cref{fig:training_dynamics_gsvc}; and (ii) later refinement remains unstable, with large gradient fluctuations in~\Cref{fig:gradient_norm} and visible flickering in~\Cref{fig:stability_gsvc}, slowing convergence.
These observations suggest that, beyond low-level GPU acceleration, 2DGS-based Gaussian video encoding can be further accelerated by improving optimization effectiveness, achieving higher or comparable quality with fewer iterations.

\re{We first address these bottlenecks through coverage-aware initialization and aggressive parameter updates to accelerate initial fitting, reducing uncovered regions as shown in~\Cref{fig:training_dynamics_fastgsvc}.
For later refinement, historical parameter averaging suppresses short-term fluctuations while maintaining aggressive updates, as shown in~\Cref{fig:gradient_norm,fig:stability_fastgsvc}.
Together, these techniques enable faster convergence, as shown in~\Cref{fig:training_loss}, and achieve a lower reconstruction loss, as highlighted in its inset.}

Next, we extend our solution to video coding with tailored designs for training from scratch on anchor frames, consecutive optimization on propagation frames, and quantization-aware fine-tuning for compression.
For anchor frames, {\name} simplifies optimization from scratch through coarse-to-fine reconstruction, progressively increasing both target complexity and computational cost.
For propagation frames, {\name} further exploits temporal information through optical flow in two ways: additional Gaussians are introduced in dynamic regions that cannot be reliably explained by flow-based propagation, providing a better initialization; during optimization, optical flow serves as a motion prior to guide Gaussians propagated from the previous frame along the estimated motion, accelerating convergence.
For compression, {\name} preserves spatiotemporal correlations to maintain concentrated residuals and introduce temporal-aware quantization for further improved rate-distortion performance.
Finally, {\name} streamlines GPU execution across the pipeline with CUDA Graphs, reducing repeated kernel-launch overhead during optimization.
Our contributions are summarized as follows:
\begin{itemize}[leftmargin=*]
    \item We accelerate 2DGS encoding through improved initialization, smoother yet more aggressive parameter updates, coarse-to-fine progressive optimization, and optical-flow-guided initialization and propagation.
    \item We improve rate-distortion performance while preserving temporal coherence through temporal-aware quantization and spatiotemporal constraints. 
\end{itemize}
Extensive experiments on UVG show that {\name} reduces encoding time by 92.5\%, achieving a 13.4$\times$ speedup over GSVC. At comparable quality, it reduces bpp by approximately 58\%; at comparable bpp, it improves PSNR by approximately 2.5~dB.

\section{Related Work} \label{sec:related}
\subsection{Iterative Gaussian Splatting Acceleration}
Gaussian Splatting~\cite{kerbl20233d}, initially introduced as 3DGS, reconstructs static 3D scenes through iterative optimization.
Existing 3DGS acceleration methods mainly reduce per-iteration computation or improve convergence.
Taming 3DGS~\cite{mallick2024taming}, FastGS~\cite{ren2026fastgs}, and Group Training~\cite{wang2025faster} reduce computation by controlling the number of active Gaussians or selectively updating primitives. 
DashGaussian~\cite{chen2025dashgaussian} further adopts progressive resolution and Gaussian budgets to avoid unnecessary computation, while QuickSplat~\cite{liu2025quicksplat} improves initialization and densification to accelerate convergence.
These methods primarily target 3D scene and surface reconstruction, where millions of Gaussians require projection, depth sorting, and alpha blending during rendering, making each optimization step computationally expensive. 
In contrast, lightweight 2D Gaussian Splatting~\cite{zhang2024gaussianimage} (2DGS), developed for image~\cite{zhang2024gaussianimage,li2026gaussianimage++} and video~\cite{wang2025gsvc,lee2025gaussianvideo} reconstruction, operates directly in the image plane with more compact 2D Gaussians, avoiding costly 3D projection and depth sorting.
Although the per-iteration cost is low, 2DGS optimization remains inefficient, with slow fitting and unstable refinement.
In this work, beyond low-level execution acceleration, we accelerate 2DGS training by improving optimization effectiveness to reach comparable or better solutions in fewer steps.

\subsection{Video Coding}
Signal processing has long underpinned video coding. It exploits spatial and temporal redundancy through hand-crafted modules such as intra/inter prediction and motion compensation, with prediction residuals further compressed by transforms such as the Discrete Cosine Transform (DCT). Over decades of development, these techniques have evolved into widely adopted standards such as H.264/AVC~\cite{AVC}, H.265/HEVC~\cite{HEVC}, and H.266/VVC~\cite{VVC}.
Meanwhile, the community has also explored alternative approaches.
Driven by advances in deep learning, neural approaches
(\eg,~\cite{DBLP:conf/mm/Li0022,DBLP:conf/cvpr/LiLL23,DBLP:conf/cvpr/LiLL24,shi2024volumetric})
represent videos implicitly, using learned model parameters or latent representations, and achieve promising rate-distortion performance.
In this work, we explore another alternative based on 2D Gaussian Splatting, which represents video content with explicit and directly accessible Gaussian primitives.

\subsection{Gaussian Video}
Gaussian Splatting has emerged as an explicit representation for video reconstruction and processing, as demonstrated by VGR~\cite{DBLP:conf/nips/Sun0MLC024} and VeGaS~\cite{DBLP:journals/corr/abs-2411-11024} based on 3DGS.
More recently, lightweight 2DGS has been extended to video along two main directions.
One direction focuses on generalizable video reconstruction.
GaussianVideo~\cite{lee2025gaussianvideo} models temporal variation by deforming 2D Gaussians over time, GVT~\cite{chen2025versatile} generates Gaussian tokens with learned spatiotemporal features, and HyperGS~\cite{zohra2026hypergs} predicts frame-wise Gaussian representations using a feed-forward model.
The other direction adopts per-video iterative optimization for video coding, naturally supporting flexible rate-quality control: videos can be optimized with arbitrary Gaussian budgets for rate control, while progressive pruning can meet a target quality for constant-quality encoding~\cite{wang2025gsvc}.
Following this paradigm, GSVC~\cite{wang2025gsvc} introduced the first 2D Gaussian video codec, integrating temporal propagation, consecutive learning, quantization, entropy coding, scene-change detection, and Gaussian addition/removal to adapt to video dynamics.
Leveraging this fine-grained control over Gaussian budgets, P-GSVC~\cite{wang2026p} further extends GSVC with progressive 2DGS for scalable and layered coding, with less than 3\% layering overhead.
However, such rate-quality flexibility requires extensive iterative optimization, resulting in long encoding times.
In {\name}, we accelerate end-to-end Gaussian video encoding while preserving fine-grained rate-quality control and further improving compression performance. We also investigate feed-forward prediction as a potential acceleration route through initialization, as analyzed in~\Cref{sec:experiment}.
\section{\name} \label{sec:methodology}

\subsection{Problem Formulation \label{sec:formulation}}
Given a 2D video $\mathcal{F}=\{\boldsymbol{F}_t\}_{t=1}^{T}$, a Gaussian video codec encodes each frame into a compact representation $\tilde{\mathcal{G}}_t$, from which the reconstructed frame is obtained as
\begin{equation}
    \hat{\mathcal{F}}
    = \left\{\hat{\boldsymbol{F}}_t\right\}_{t=1}^{T}
    = \left\{\mathcal{D}(\tilde{\mathcal{G}}_t)\right\}_{t=1}^{T},
\end{equation}
where $\mathcal{D}(\cdot)$ denotes the decoding process.
For $\boldsymbol{F}_t$, our goal is to improve optimization effectiveness such that a better solution can be reached with fewer iterations.
However, the actual computation per iteration varies with the number of Gaussians and their overlaps.
We therefore use the wall-clock time as a practical measure of optimization effectiveness:
$T_{\mathrm{opt}}^{t}
=
\sum_{k=1}^{K_t} \tau_{t,k}$,
where $K_t$ is the number of optimization iterations for frame $\boldsymbol{F}_t$, and $\tau_{t,k}$ is the runtime of its $k$-th iteration.
Accordingly, our objective is to minimize the per-frame optimization time while maintaining the target rate-distortion performance:
\begin{equation}
\min \; T_{\mathrm{opt}}^{t}
\quad
\text{s.t.}
\quad
\mathcal{L}(\hat{\boldsymbol{F}}_t,\boldsymbol{F}_t)
\leq
\mathcal{L}_{\mathrm{target}},
\qquad
B_t \leq B_{\mathrm{target}},
\label{eq:objective}
\end{equation}
where $\mathcal{L}(\cdot,\cdot)$ denotes the reconstruction loss between two frames, $B_t$ is the encoded frame size, and $\mathcal{L}_{\mathrm{target}}$ and $B_{\mathrm{target}}$ define the target rate-distortion operating point based on the converged performance of GSVC after extensive iterative optimization.

\begin{figure}[t]
    \centering
    \subfloat[\re{w/o representation-stage CI}]{
        \includegraphics[width=0.23\textwidth]{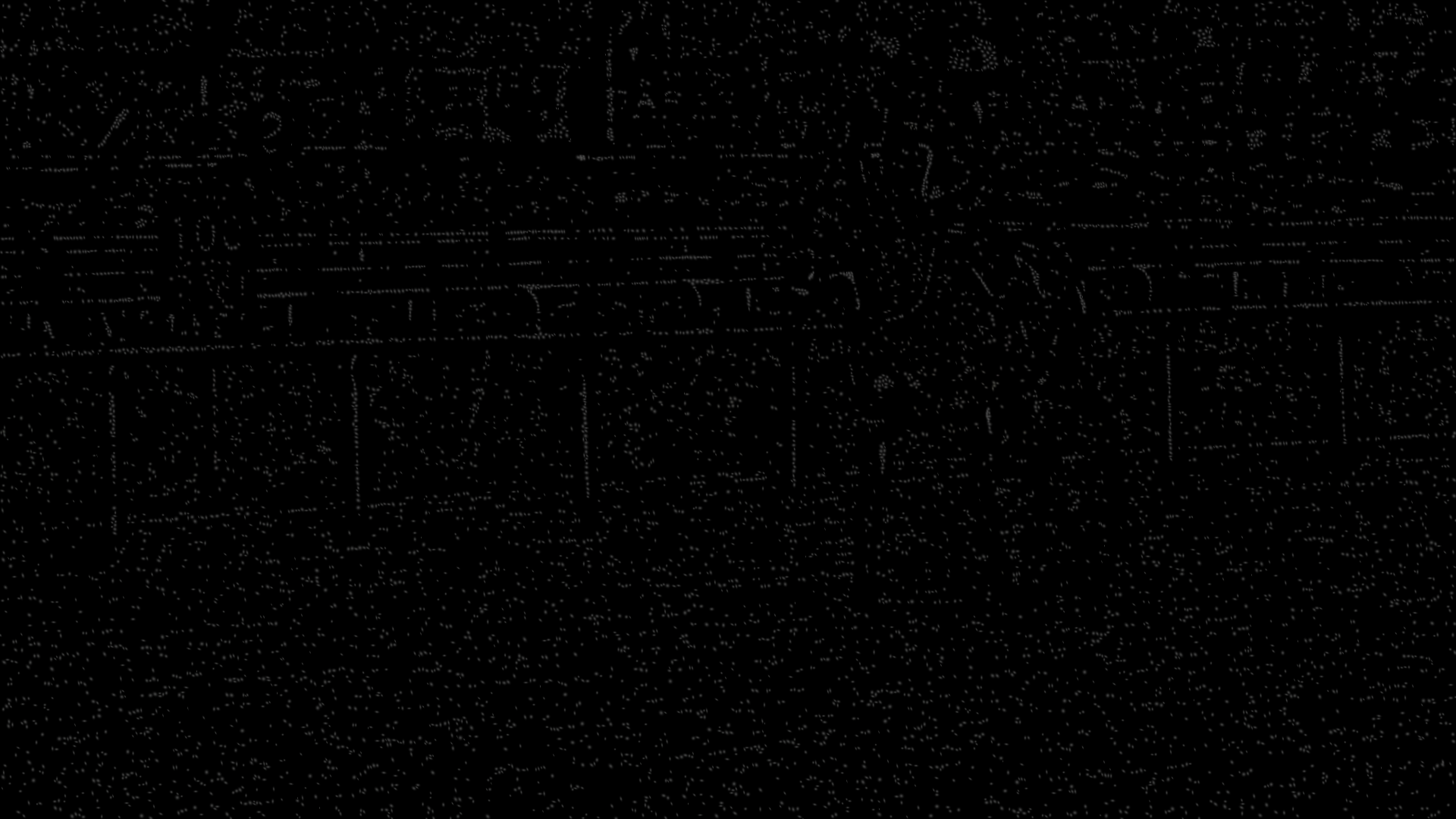}
        \label{fig:cov_init_render_wo}
    }
    \subfloat[\re{w/ representation-stage CI}]{
        \includegraphics[width=0.23\textwidth]{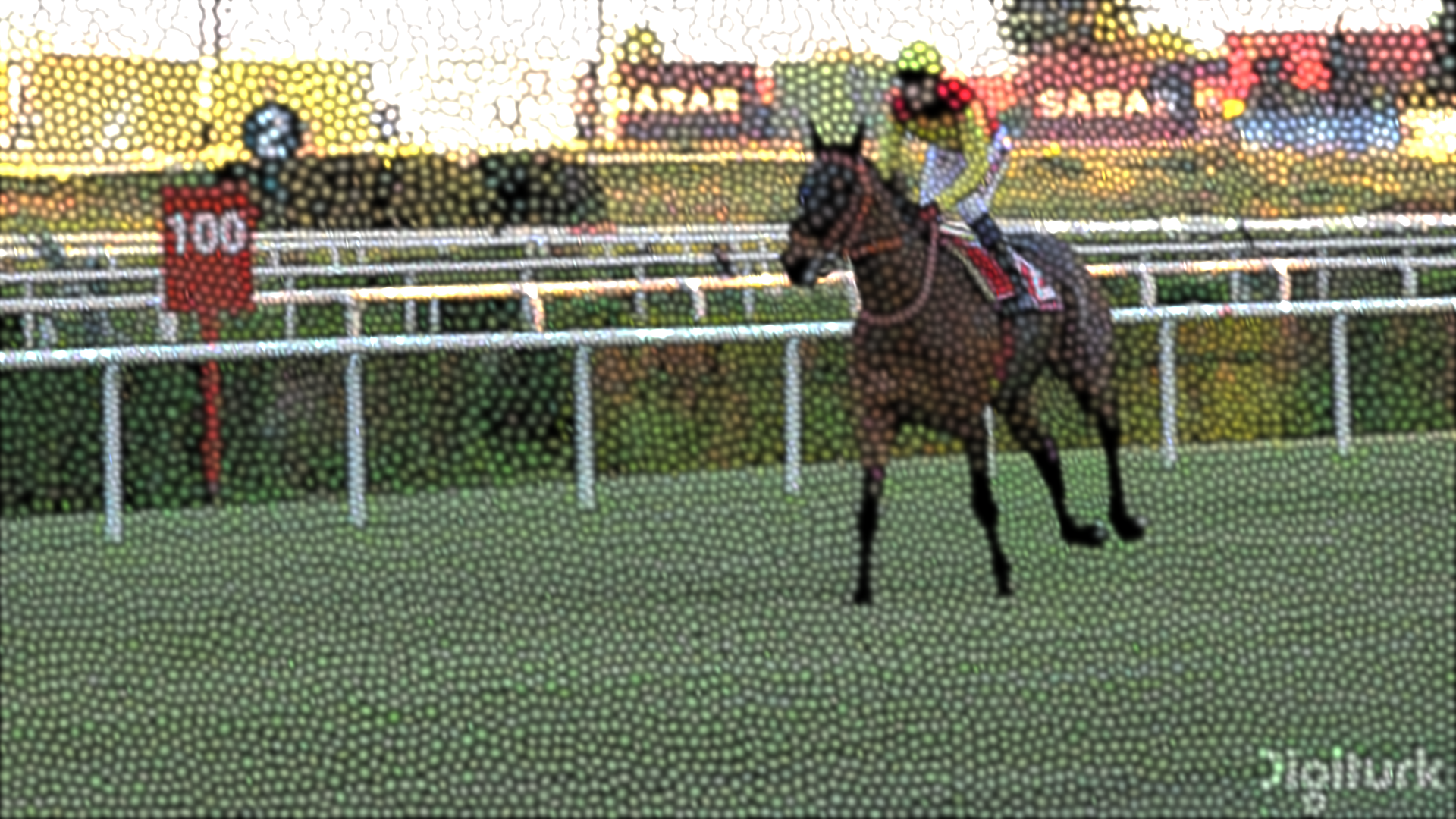}
        \label{fig:cov_init_render_w}
    }
    \caption{\re{Reconstruction results after 250 iterations with and without representation-stage CI.}}
    \label{fig:cov_init_rendering}
\end{figure}

\subsection{Fast-2DGS \label{sec:2d_gs}}

\textbf{2DGS representation.}
We adopt lightweight 2DGS as the representation primitive.
Taking frame $\boldsymbol{F}_t$ as an example, we omit the frame and Gaussian indices for simplicity and write
$\boldsymbol{G}_i^t=(\boldsymbol{\mu}_i^t,\boldsymbol{\Sigma}_i^t,\boldsymbol{c}_i^{\prime t})$
as
$\boldsymbol{G}=(\boldsymbol{\mu},\boldsymbol{\Sigma},\boldsymbol{c}')$,
where $\boldsymbol{\mu}\in\mathbb{R}^2$ denotes the Gaussian center on the image plane,
$\boldsymbol{\Sigma}\in\mathbb{R}^{2\times2}$ controls its shape and orientation, and \re{$\boldsymbol{c}'\in\mathbb{R}^3$ denotes the opacity-weighted RGB color, where opacity is absorbed into the color coefficients. Following GaussianImage~\cite{zhang2024gaussianimage}, these coefficients are not restricted to $[0,1]$, allowing Gaussians to both add and remove content.}
To guarantee a positive semi-definite covariance matrix, we parameterize $\boldsymbol{\Sigma}=\boldsymbol{L}\boldsymbol{L}^{\top}$ using Cholesky decomposition, where $\boldsymbol{L}$ is a lower-triangular matrix parameterized by the Cholesky vector $\boldsymbol{l}=[\ell_0,\ell_1,\ell_2]^{\top}$.
Thus, each Gaussian contains eight parameters: two for position, three for shape, and three for color.
Without depth sorting, 2DGS reconstructs each pixel as a weighted sum of the overlapping Gaussians.
For pixel $j$, its reconstructed color is
\begin{equation}
\boldsymbol{C}_j
=
\sum_{n\in\mathcal{N}_j}
\boldsymbol{c}'_n
\exp\left(
-\frac{1}{2}
\boldsymbol{d}_n^{\top}
\boldsymbol{\Sigma}_n^{-1}
\boldsymbol{d}_n
\right),
\label{eq:2dgs_rendering}
\end{equation}
where $\mathcal{N}_j$ denotes the set of Gaussians contributing to pixel $j$, and
$\boldsymbol{d}_n$ is the displacement between pixel $j$ and the center of Gaussian $n$.
The Gaussian representation is obtained through differentiable iterative optimization,
\begin{equation}
\mathcal{G}_t
=
\arg\min_{\mathcal{G}_t}
\mathcal{L}\big(
\mathcal{R}(\mathcal{G}_t),\boldsymbol{F}_t
\big),
\label{eq:2dgs_optimization}
\end{equation}
where $\mathcal{R}(\cdot)$ denotes 2D Gaussian rendering.
As discussed in Section~\ref{Sec:Introduction}, better initialization and aggressive yet stable parameter updates improve optimization effectiveness. We next describe how this acceleration principle is applied to 2DGS.

\textbf{Coverage-aware initialization.}
We first improve the initial Gaussian distribution through coverage-aware initialization (CI).
\re{Inspired by Image-GS~\cite{DBLP:conf/siggraph/ZhangLKJDCSKS25}, we sample Gaussian centers $\boldsymbol{\mu}$ from a distribution derived from a blurred image-gradient map, with a uniform sampling component. This places more Gaussians in detailed regions while retaining coverage in textureless areas. We initialize the color $\mathbf{c}'$ to zero.}
For shape, we sample $\ell_1$ from $[0,1]$ and $\ell_0,\ell_2$ from $[0.5,1]$ to avoid overly small or elongated Gaussians and improve initial coverage.
To adapt the initial spatial extent of Gaussians to different resolutions and Gaussian budgets, we define the coverage scale as $s_{\mathrm{cov}}=\min(H,W)^2/(\alpha N_t)$, where $H$ and $W$ denote the frame height and width, $N_t$ is the Gaussian budget, and $\alpha$ is a balancing coefficient.
We scale the initial Cholesky vector by $s_{\mathrm{cov}}$ to adjust the spatial extent of Gaussians.
\re{We use the shorter image dimension to prevent excessively
large Gaussian footprints and overlaps in images with
high aspect ratios.}
For a given resolution, a smaller Gaussian budget uses larger footprints to improve initial coverage and reduce the iterations needed for Gaussian growth, while a larger budget uses smaller footprints to avoid excessive overlap and the resulting increase in computational cost. 
\re{As shown in~\Cref{fig:cov_init_rendering}, CI improves spatial coverage during initial fitting, as illustrated by the reconstructions after 250 iterations.}

\begin{figure}[t]
    \centering
    \subfloat[\re{w/o quantization-stage CI}]{
        \includegraphics[width=0.23\textwidth]{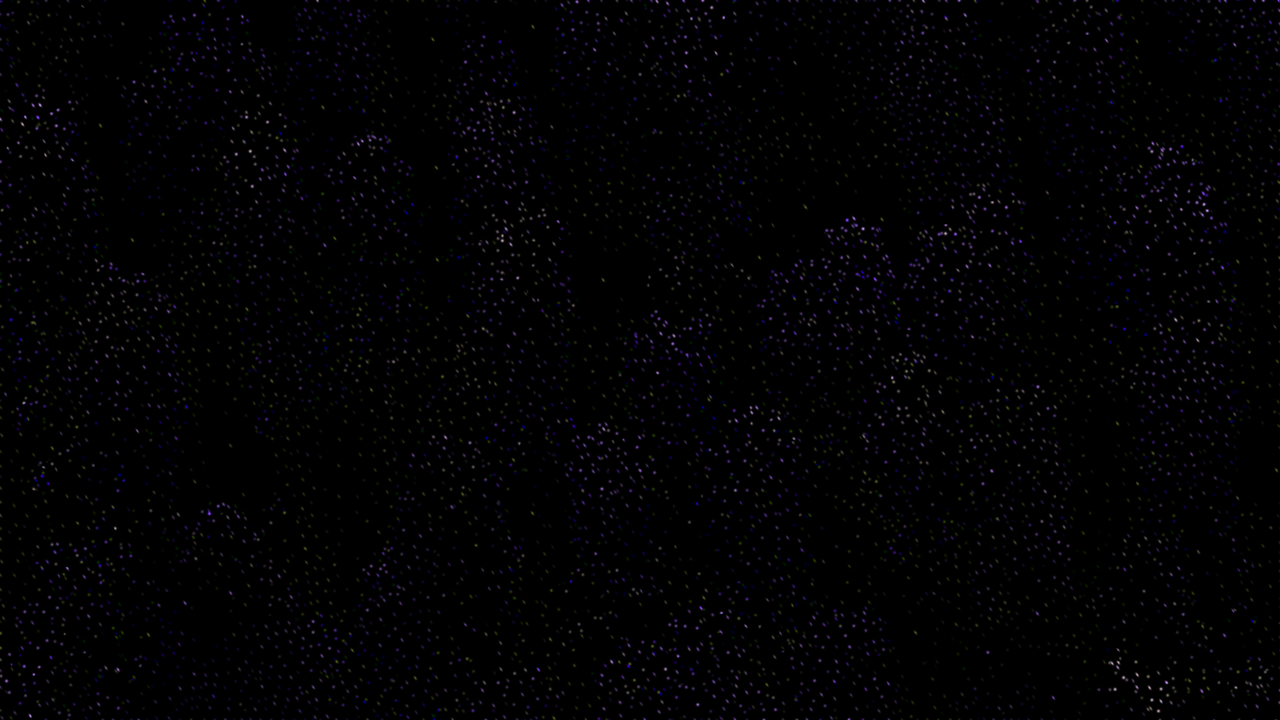}
        \label{fig:cov_init_render_wo}
    }
    \subfloat[\re{w/ quantization-stage CI}]{
        \includegraphics[width=0.23\textwidth]{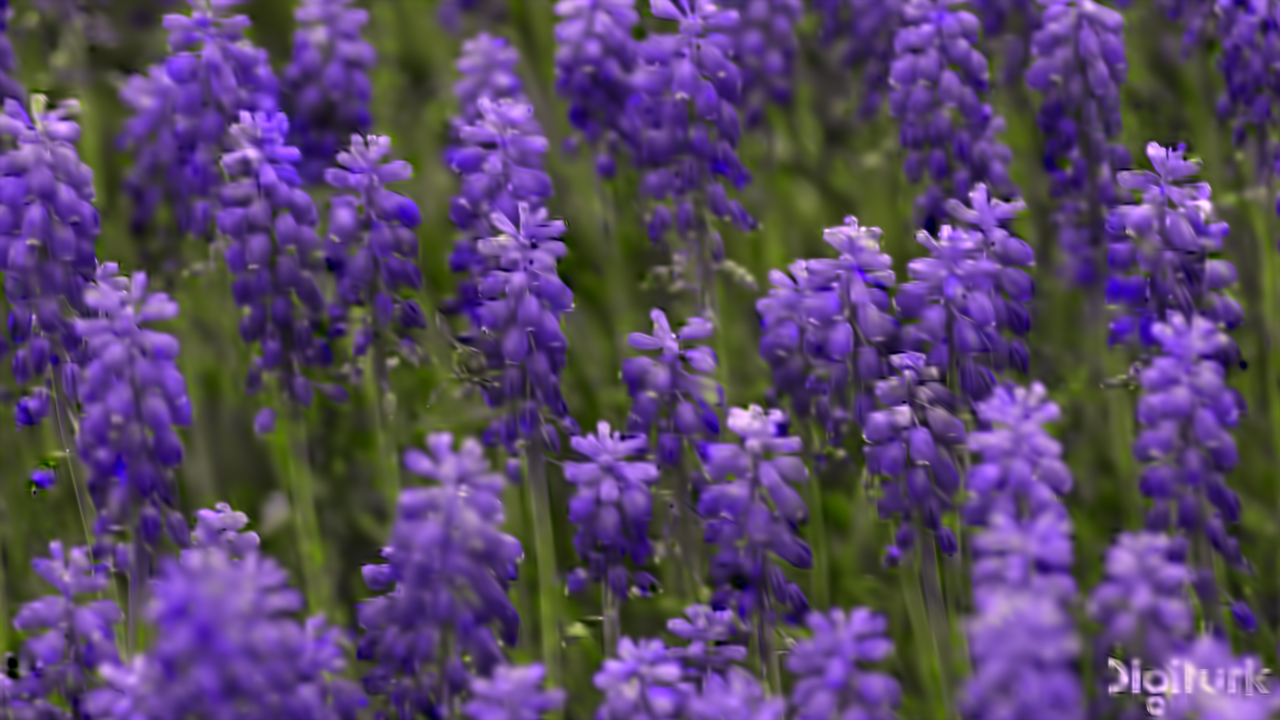}
        \label{fig:cov_init_render_w}
    }
    \caption{\re{Initial rendering results with and without quantization-stage CI, which improves the spatial coverage.}}
    \label{fig:cov_init_rendering_quantization}
\end{figure}

\textbf{Aggressive yet stable optimization.}
A direct way to reduce optimization steps is to use larger updates, but the step size cannot be increased arbitrarily without risking unstable optimization~\cite{ahn2022understanding}.
As discussed in~\Cref{Sec:Introduction}, iterative 2DGS optimization already exhibits substantial gradient fluctuations and visible flickering.
To support aggressive online updates, we introduce EMA-based Parameter Averaging (EPA), which maintains a smoothed copy of the Gaussian parameters using exponential moving average (EMA)~\cite{tarvainen2017mean}.
EPA keeps the online parameters free to update aggressively while maintaining a separate exponentially averaged parameter trajectory outside the gradient-update loop.
Specifically, we maintain two sets of Gaussian parameters: online parameters $\boldsymbol{\theta}^{(k)}$, which are optimized with aggressive updates, and EMA parameters $\bar{\boldsymbol{\theta}}^{(k)}$, which are maintained outside the gradient-update loop.
The EMA parameters are updated from the historical online parameters as
\begin{equation}
\bar{\boldsymbol{\theta}}^{(k)}
=
\beta\bar{\boldsymbol{\theta}}^{(k-1)}
+
(1-\beta)\boldsymbol{\theta}^{(k)},
\label{eq:ema}
\end{equation}
where $\beta\in[0,1)$ controls the averaging strength.
Expanding~\Cref{eq:ema} gives
\begin{equation}
\bar{\boldsymbol{\theta}}^{(k)}
=
(1-\beta)
\sum_{j=1}^{k}
\beta^{k-j}\boldsymbol{\theta}^{(j)}
+
\beta^k\bar{\boldsymbol{\theta}}^{(0)}.
\label{eq:ema_expanded}
\end{equation}
The EMA parameters therefore form an exponentially weighted average along the optimization trajectory, smoothing short-term parameter fluctuations while retaining the long-term optimization trend.
\re{We periodically evaluate the loss using EMA parameters for convergence detection and retain the better-performing state between the final online parameters and the best EMA parameters.}
Since $\bar{\boldsymbol{\theta}}$ is excluded from gradient computation, EPA does not alter the online updates of $\boldsymbol{\theta}$. 
\re{Given the substantial gradient fluctuations in 2DGS shown in~\Cref{fig:gradient_norm}, we adopt a large EMA decay factor to smooth parameter fluctuations.
EMA parameters do not require gradients; the gradients shown in~\Cref{fig:gradient_norm} are EMA-smoothed online gradients used for visualization only.}

\begin{figure}[t]
    \centering
    \subfloat[\re{w/o quantization-stage CI}]{
        \includegraphics[width=0.33\columnwidth]{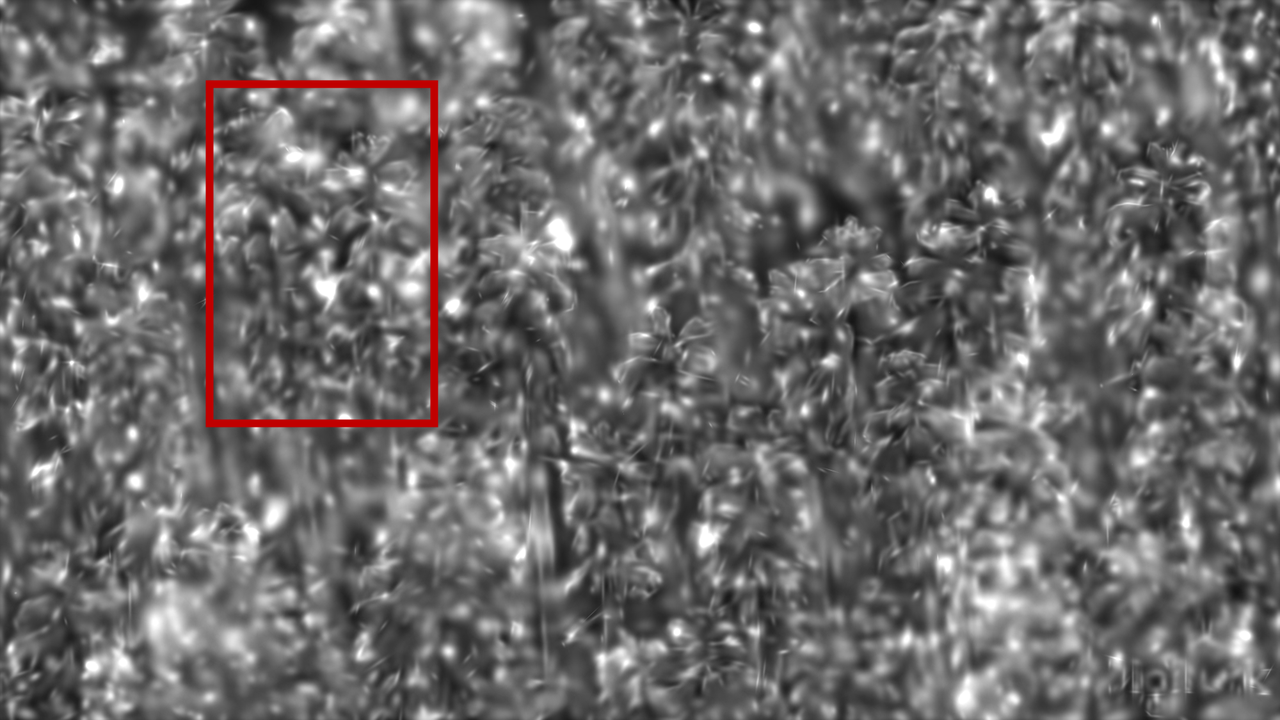}
        \includegraphics[width=0.15\columnwidth]{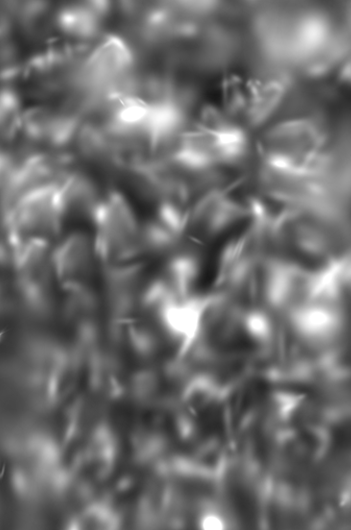}
        \label{fig:cov_init_structure_wo}
    }
    \subfloat[\re{w/ quantization-stage CI}]{
        \includegraphics[width=0.33\columnwidth]{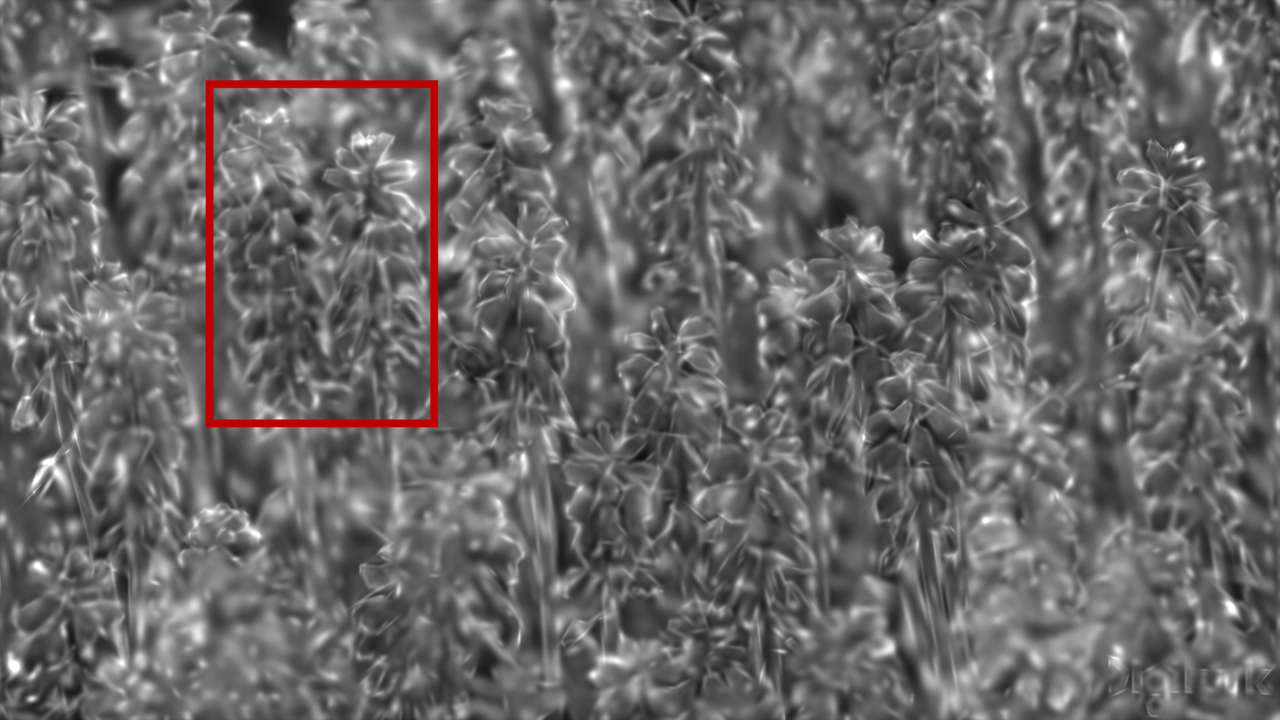}
        \includegraphics[width=0.15\columnwidth]{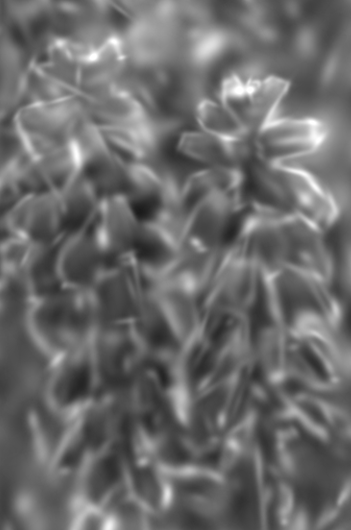}
        \label{fig:cov_init_structure_w}
    }
    \caption{Gaussian structures after quantization-aware fine-tuning with and without quantization-stage CI.}
    \label{fig:cov_init_structure}
\end{figure}

\subsection{Fast-2DGS for Video Encoding \label{sec:Fast_2DGS}}
\textbf{Gaussian video coding.}
We begin with the coding pipeline in GSVC, serving as our backbone.
At the representation level, frames are separated into anchor and propagation frames, analogous to intra- and inter-coded frames, respectively.
In single-pass mode, an anchor is trained once from scratch with newly initialized intra-Gaussians, while propagation frames inherit and refine inter-Gaussians from the previous frame through consecutive learning.
In two-pass mode, GSVC first optimizes the anchor and the following $k$ propagation frames, then uses the refined $k$-th frame to reinitialize and further optimize the anchor before propagating forward again. Only the final anchor is stored, adding no storage overhead over single-pass encoding.
To adapt to video dynamics, Gaussian Splat Augmentation (GSA) adds
$\rho_{\mathrm{dyn}}N_t$ new intra-Gaussians, while Gaussian Splat Pruning (GSP) progressively removes the same number of low-contribution Gaussians from $K_r^{\mathrm{start}}$ to $K_r^{\mathrm{end}}$ every $\Delta K_r$ iterations. Dynamic Keyframe Selection (DKS) selects a new anchor when a scene change is detected.
At the coding level, frames are encoded as either I- or P-frames.
Anchor frames are encoded as I-frames, where all Gaussians are coded independently.
Propagation frames can be encoded as P-frames: newly added Gaussians are intra-coded, while inherited inter-Gaussians are coded as residuals using fewer bits.
To limit quantization-error propagation within a GOP, selected propagation frames can instead be intra-coded as I-frames, where all Gaussians are coded independently rather than as inter-frame residuals.

\textbf{Initialization.}
\re{
CI targets different components across the two optimization stages: Gaussian parameters during representation optimization and quantizers during quantization-aware fine-tuning.
At representation stage, CI initializes newly added intra-Gaussians, while inter-Gaussians are inherited from the
previous frame.
To eliminate seed dependence, we replace random sampling, including the uniform component of gradient-guided initialization, with a fixed, unscrambled Sobol sequence~\cite{sobol1967distribution}.
At the quantization stage, Gaussians are already optimized, so CI instead initializes the quantizers.
Gaussian shape parameters, represented by Cholesky coefficients $\ell$, are quantized using LSQ+~\cite{DBLP:journals/corr/abs-2004-09576}:
$
\hat{\ell}=s\cdot Q_b\left(\frac{\ell-b}{s}\right)+b,
$
where $Q_b$ denotes $b$-bit rounding and clipping, and $s$ and $b$ are the learnable quantization scale and offset, respectively.
GSVC initializes $s$ and $b$ with small values to reduce rounding error, but this shrinks the quantized Gaussians and disrupts their learned shapes after fine-tuning, as shown in~\Cref{fig:cov_init_render_wo,fig:cov_init_structure_wo}.
We instead initialize $s=1$ and $b=0$ to preserve Gaussian shapes before fine-tuning and their spatial structure afterward, as shown in~\Cref{fig:cov_init_render_w,fig:cov_init_structure_w}.}

\textbf{Optimization.}
We apply EPA to representation optimization and quantization-aware fine-tuning. Since Gaussians can be dynamically added or removed during training, the EMA parameter set is updated accordingly to remain aligned with the online model.
EPA starts after a warm-up of $K_{\mathrm{w}}$ iterations, avoiding early unstable parameters from contaminating the EMA trajectory.
We further adapt the learning rate to each stage: anchor frames use a larger rate $\eta_{\mathrm{A}}$ to accelerate from-scratch reconstruction, while propagation frames and quantization-aware fine-tuning use lower rates $\eta_{\mathrm{P}}$ and $\eta_{\mathrm{Q}}$ to preserve inherited or already optimized structures.

\subsection{Coarse-to-Fine Reconstruction \label{sec:progressive}}

\begin{figure}[t]
    \centering
    \includegraphics[width=0.22\textwidth]{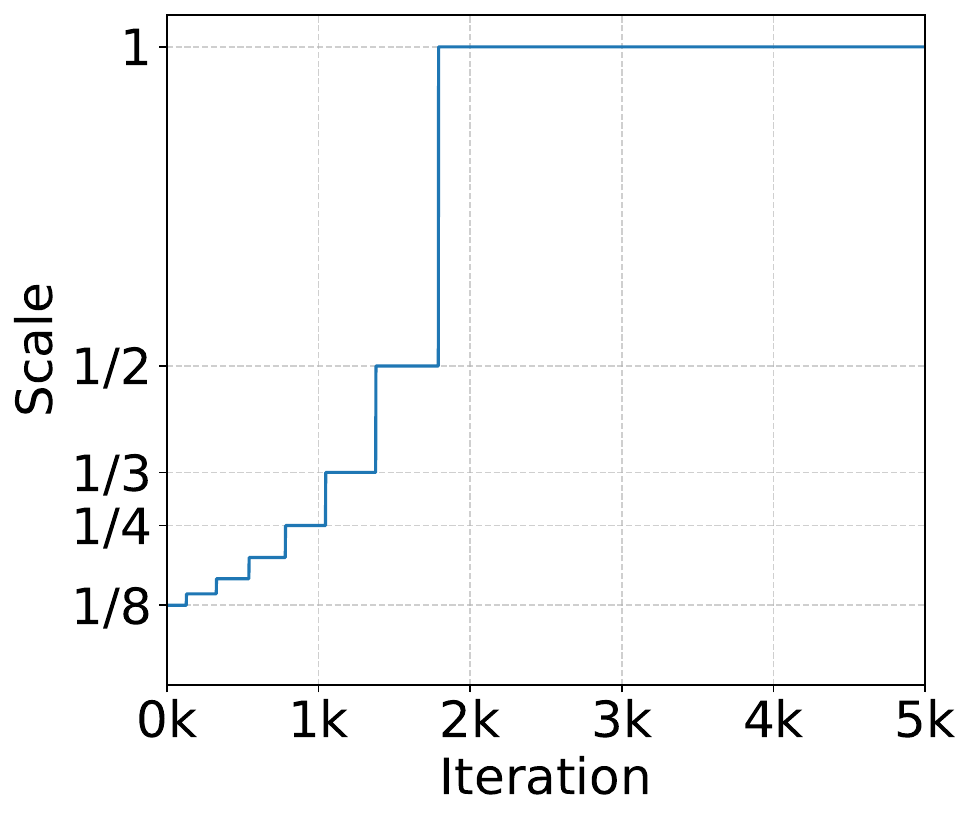}
    \hfill
    \includegraphics[width=0.22\textwidth]{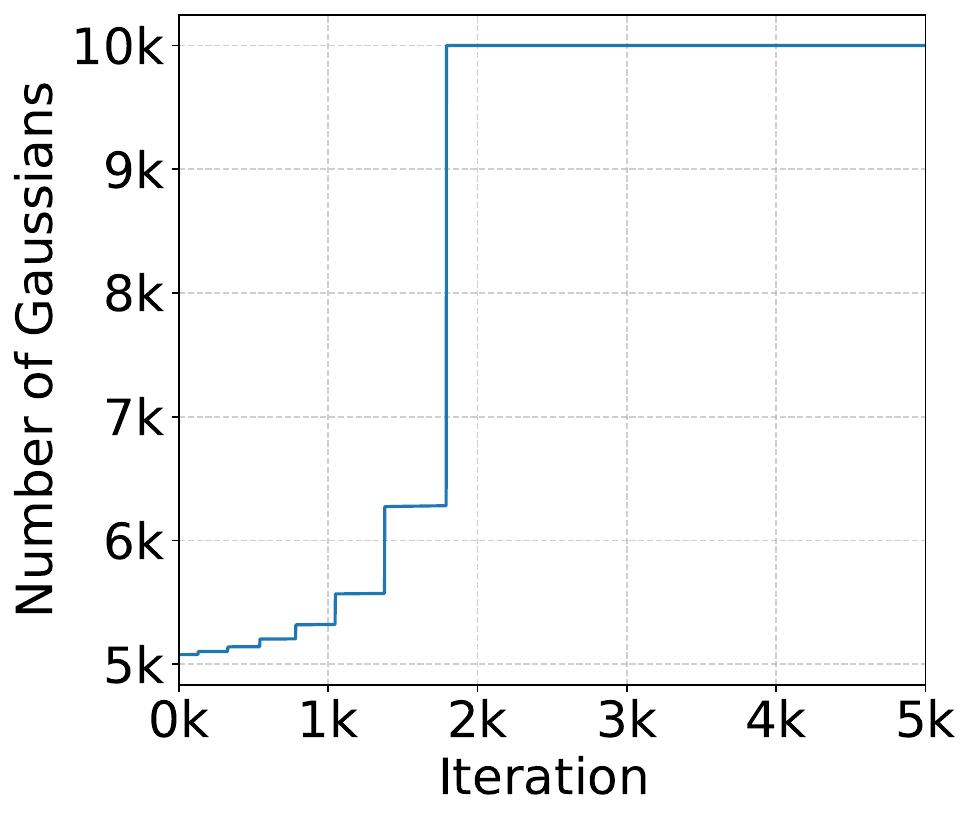}
    \caption{\re{A sample progressive schedule showing resolution scale (left) and number of Gaussians (right) over iterations.}}
    \label{fig:progressive_scheduling}
\end{figure}

To further improve training effectiveness in training-from-scratch settings for anchor frames, we complement CI with coarse-to-fine progressive training (PT).
Rather than optimizing all Gaussians toward the full-resolution target simultaneously, we first establish coarse structures and progressively refine details.
Inspired by DashGaussian~\cite{chen2025dashgaussian}, we progressively increase reconstruction complexity by jointly increasing the frame resolution and Gaussian budget.

We first construct multi-resolution reconstruction targets. Following DashGaussian~\cite{chen2025dashgaussian}, we perform anti-aliased downsampling in the Fourier domain. Given an anchor frame $\mathbf{F}_t$ with a centered Fourier spectrum $\mathbf{Z}_t$, the target at downsampling factor $r$ is obtained by retaining the central frequency region $\Omega(r)$ of size $H/r \times W/r$ and applying the inverse Fourier transform. As $r$ decreases toward $1$, higher-frequency details are progressively introduced.

Rather than using a fixed resolution schedule, we adapt the downsampling factor \(r\) according to the sum of retained Fourier magnitudes, defined as
\begin{equation}
\chi_t(r)
=
\sum_{\omega\in\Omega(r)}
\left|\mathbf{Z}_t(\omega)\right|.
\label{eq:freq_significance}
\end{equation}
We first determine the coarsest downsampling factor $r_{\max}$ by solving
$\chi_t(r_{\max})=\chi_t(1)/4$ via binary search.
This adapts the starting resolution to the spectral distribution of each anchor frame.
We then construct the downsampling schedule from $r_{\max}$ to $r=1$ (full resolution).
Specifically, we uniformly sample $m$ significance levels $\{\chi_i\}_{i=0}^{m-1}$ between
$\chi_{\min}=\chi_t(r_{\max})$ and $\chi_{\max}=\chi_t(1)$, and solve the corresponding downsampling factors $r_i$ from $\chi_t(r_i)=\chi_i$ via binary search.
Each significance level is mapped to a transition iteration as
\begin{equation}
s_i
=
K_{\mathrm{w}}
\frac{\log(\chi_i/\chi_{\min})}
{\log(\chi_{\max}/\chi_{\min})},
\label{eq:resolution_transition}
\end{equation}
where full resolution is reached by the EMA warm-up iteration $K_{\mathrm{w}}$.
Between adjacent transition points $(r_i,s_i)$ and $(r_{i+1},s_{i+1})$, we follow DashGaussian~\cite{chen2025dashgaussian} and interpolate the inverse squared downsampling factor:
\begin{equation}
r_k
=
\left[
(1-\lambda_k)\frac{1}{r_i^2}
+
\lambda_k\frac{1}{r_{i+1}^2}
\right]^{-\frac{1}{2}},
\qquad
\lambda_k
=
\frac{k-s_i}{s_{i+1}-s_i}.
\label{eq:resolution_interpolation}
\end{equation}
Together, the logarithmic transition mapping, interpolation, and rounding allocate fewer iterations to coarse resolutions and progressively more iterations as the resolution increases as shown in~\Cref{fig:progressive_scheduling}.
Finally, we couple the Gaussian budget with the resolution progression.
At iteration $k$, the Gaussian number is
\begin{equation}
N_k
=
N_0+
\frac{N_t-N_0}
{r_k^{\,2-k/K_w}},
\qquad
N_0=\frac{N_t}{2}.
\label{eq:gaussian_schedule}
\end{equation}
As resolution increases, the Gaussian budget grows toward $N_t$ to capture finer details.

\subsection{Flow-Guided Propagation \label{sec:optical_flow}}

\begin{figure}[t]
    \centering
    \subfloat[Random]{
        \includegraphics[width=0.23\textwidth]{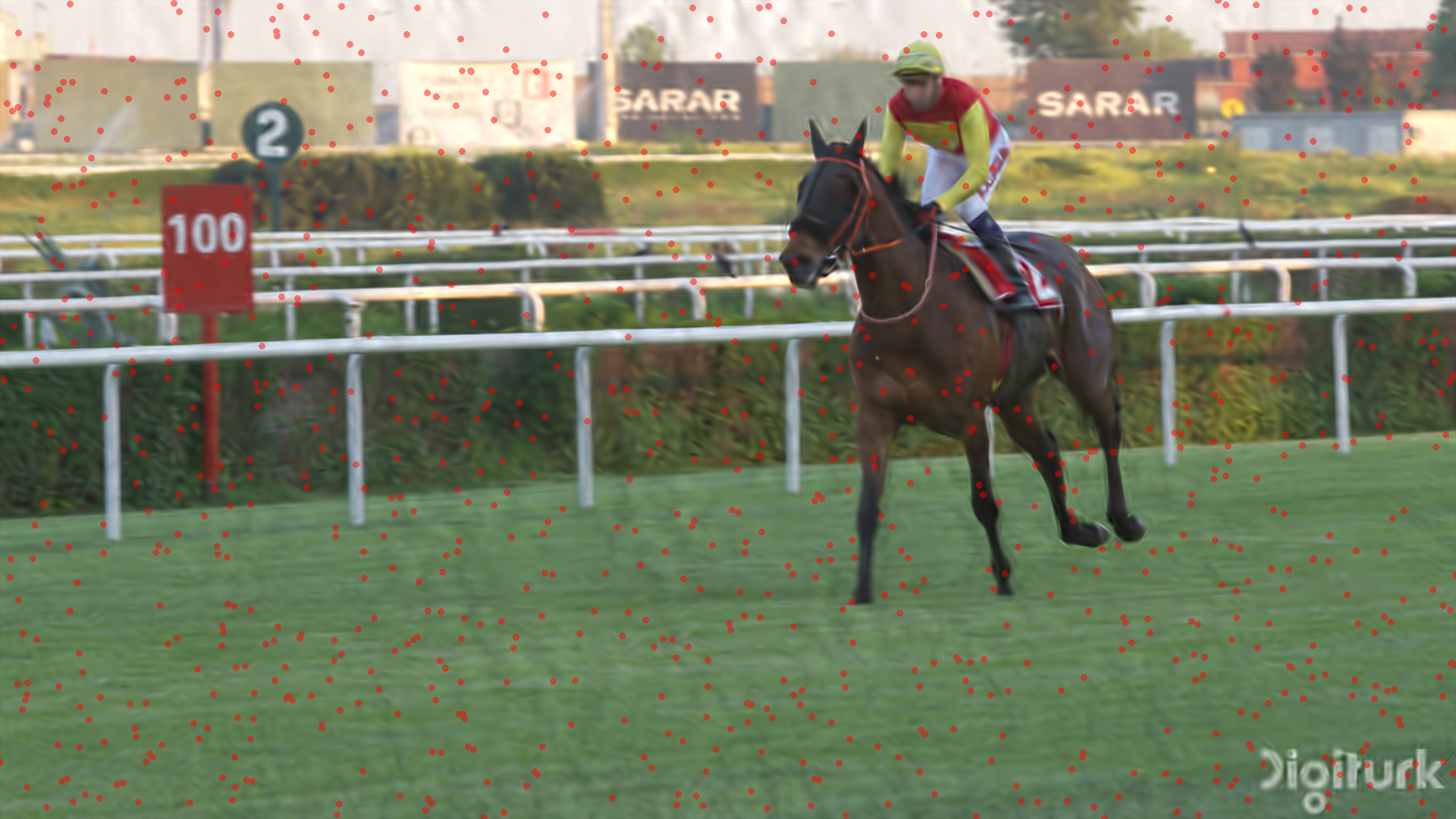}
        \label{fig:gsa_wo_of}
    }
    \subfloat[Optical Flow guided]{
        \includegraphics[width=0.23\textwidth]{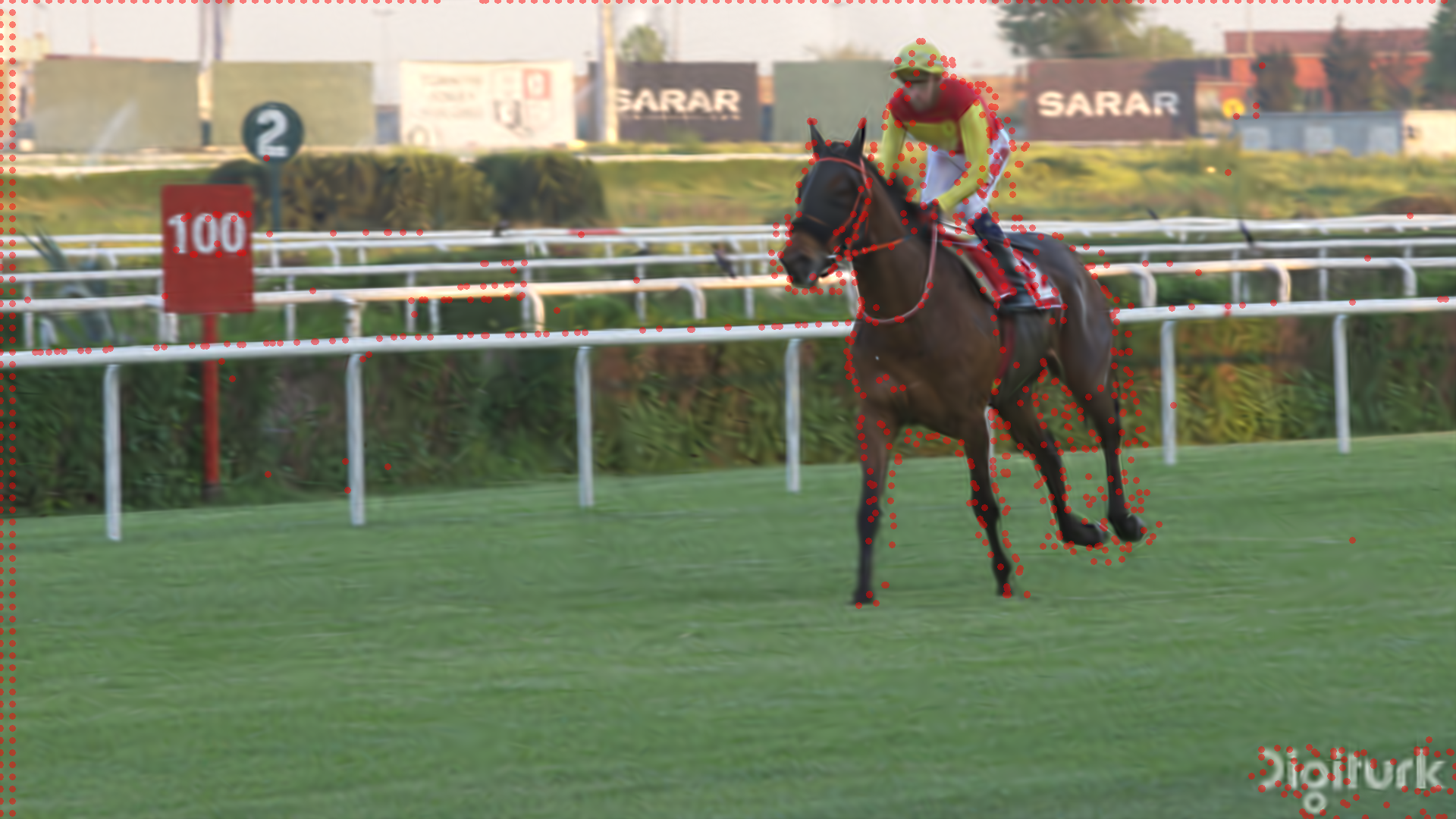}
        \label{fig:gsa_w_of}
    }
    \caption{Propagation frame initialization with different Gaussian injection strategies. Red dots denote newly added intra-Gaussians' center.}
    \label{fig:gsa_of}
\end{figure}

For consecutive learning in propagation frames, we use optical flow to better exploit temporal coherence and accelerate training.

\textbf{Flow-guided initialization.}
Before training, we add $\rho_{\mathrm{dyn}}N_t$ new intra-Gaussians.
We target regions with large warping error using optical-flow guidance, as shown in~\Cref{fig:gsa_of}.
Specifically, for a propagation frame $\mathbf{F}_t$ and its preceding frame $\mathbf{F}_{t-1}$, we compute the backward optical flow $\mathbf{U}_{t\rightarrow t-1}$.
For each pixel $\mathbf{p}$ in $\mathbf{F}_t$, the flow maps it to the corresponding source in $\mathbf{F}_{t-1}$:
\begin{equation}
\mathbf{q}
=
\mathbf{p}
+
\mathbf{U}_{t\rightarrow t-1}(\mathbf{p}).
\end{equation}
We sample $\mathbf{F}_{t-1}$ at $\mathbf{q}$ to construct the warped previous frame
$\tilde{\mathbf{F}}_{t-1\rightarrow t}$, and compute the pixel-wise warping error as
\begin{equation}
e_t(\mathbf{p})
=
\left\|
\mathbf{F}_t(\mathbf{p})
-
\tilde{\mathbf{F}}_{t-1\rightarrow t}(\mathbf{p})
\right\|_1.
\label{eq:flow_warp_error}
\end{equation}
In this way, newly appearing content, disocclusions, and inaccurate motion correspondence tend to produce larger warping errors, indicating regions that are poorly explained by temporal propagation and may benefit from new Gaussians.
To avoid excessive overlap in highly dynamic regions, we divide the error map into tile-sized cells and place one Gaussian at the maximum-error pixel of each top-ranked cell. Any remaining Gaussians are placed randomly. Since flow serves only as coarse guidance and its errors can be corrected by newly introduced Gaussians, we choose the flow estimator primarily for its lightweight design, as detailed in~\Cref{sec:experiment}.

\begin{figure}[t]
    \centering
    \subfloat[w/o spatiotemporal constraints]{
        \includegraphics[width=0.23\textwidth]{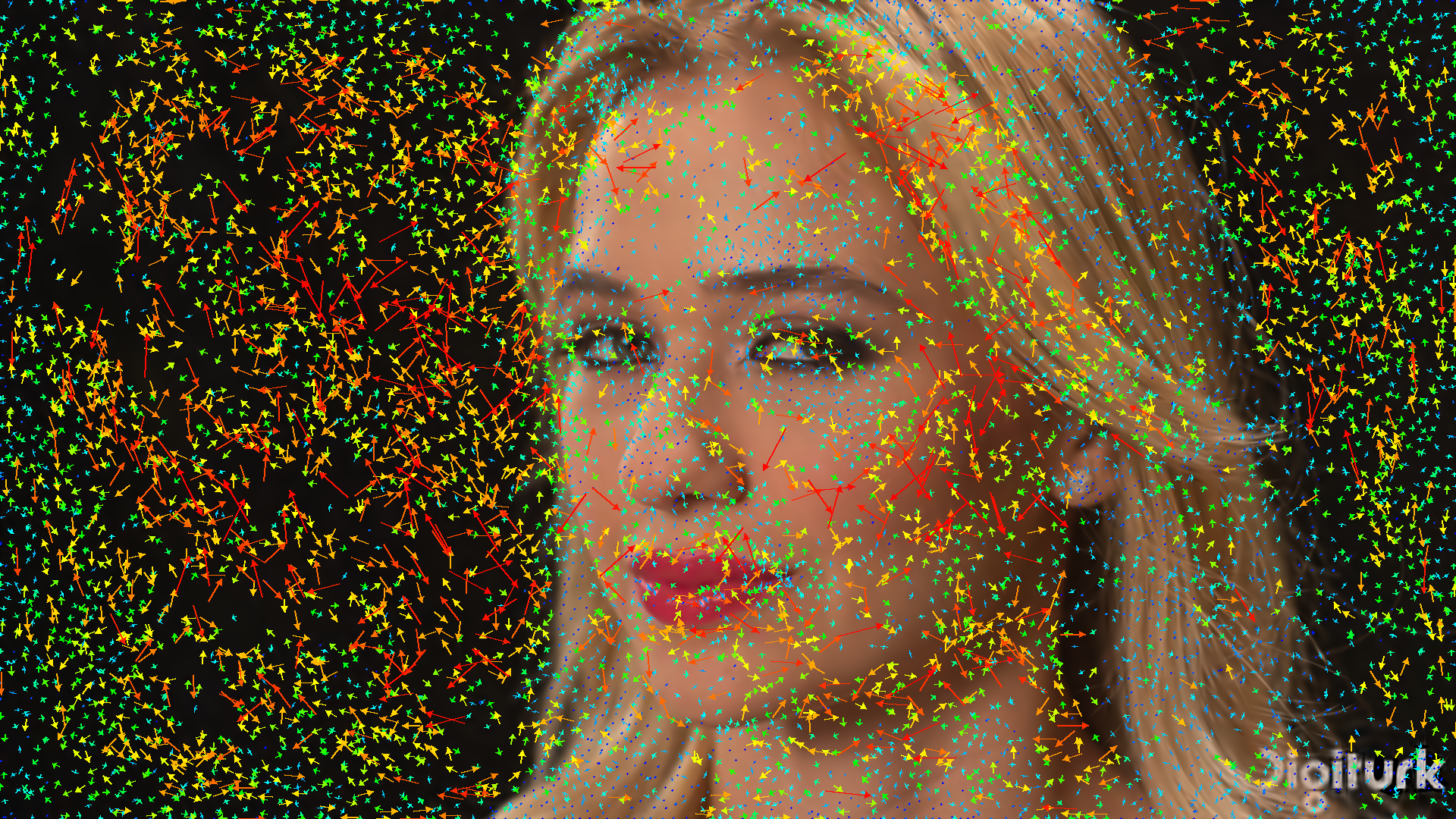}
        \label{fig:motion_reg_wo}
    }
    \subfloat[w/ spatiotemporal constraints]{
        \includegraphics[width=0.23\textwidth]{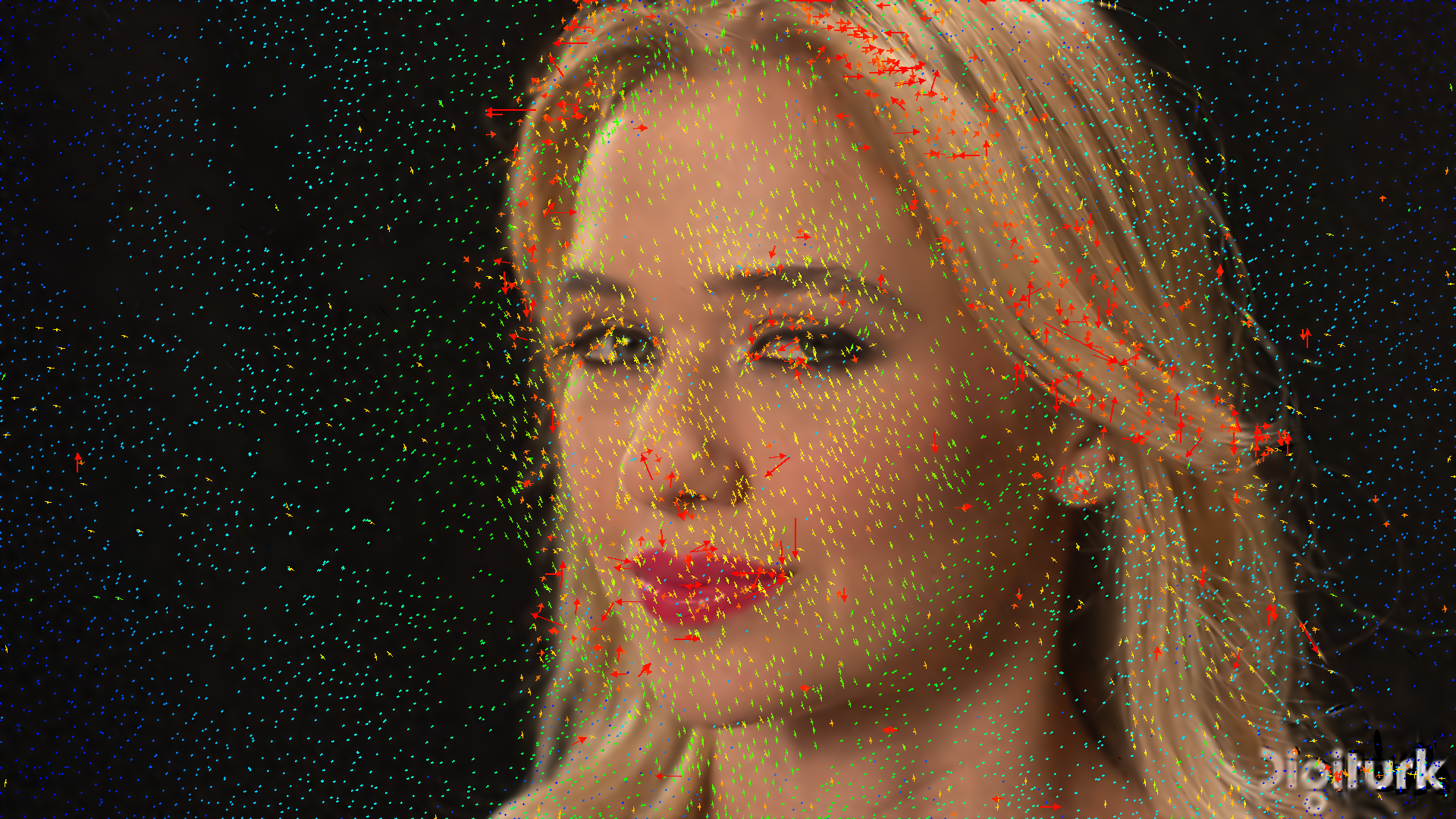}
        \label{fig:motion_reg_w}
    }
    \caption{Inter-Gaussian motion after quantization fine-tuning.}
    \label{fig:motion_regularization}
\end{figure}

\begin{figure}[t]
    \centering

    \subfloat[w/o spatiotemporal constraints]{
        \includegraphics[width=0.23\textwidth]{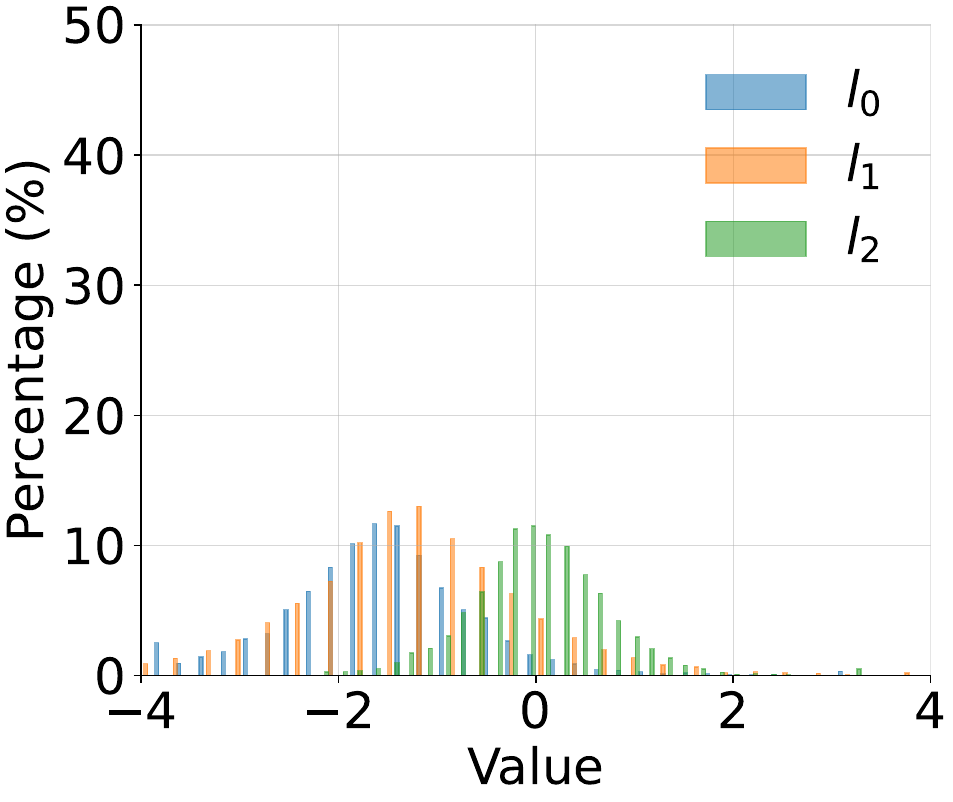}
        \includegraphics[width=0.23\textwidth]{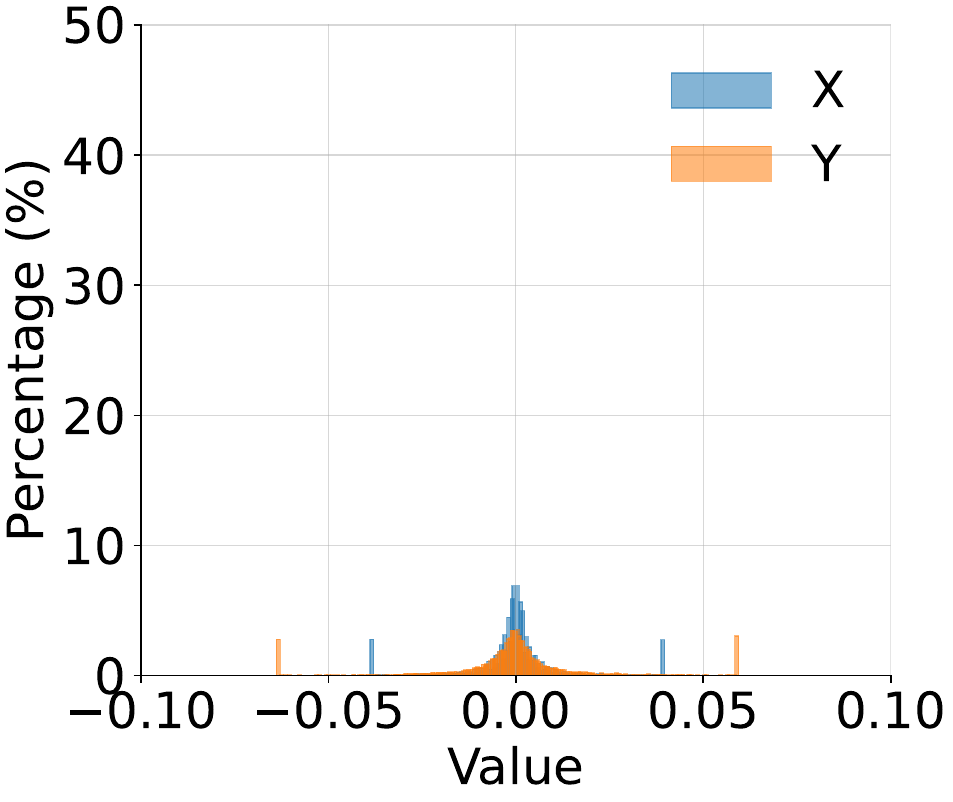}
        \label{fig:parameter_distribution_wo}
    }

    \subfloat[w/ spatiotemporal constraints]{
        \includegraphics[width=0.23\textwidth]{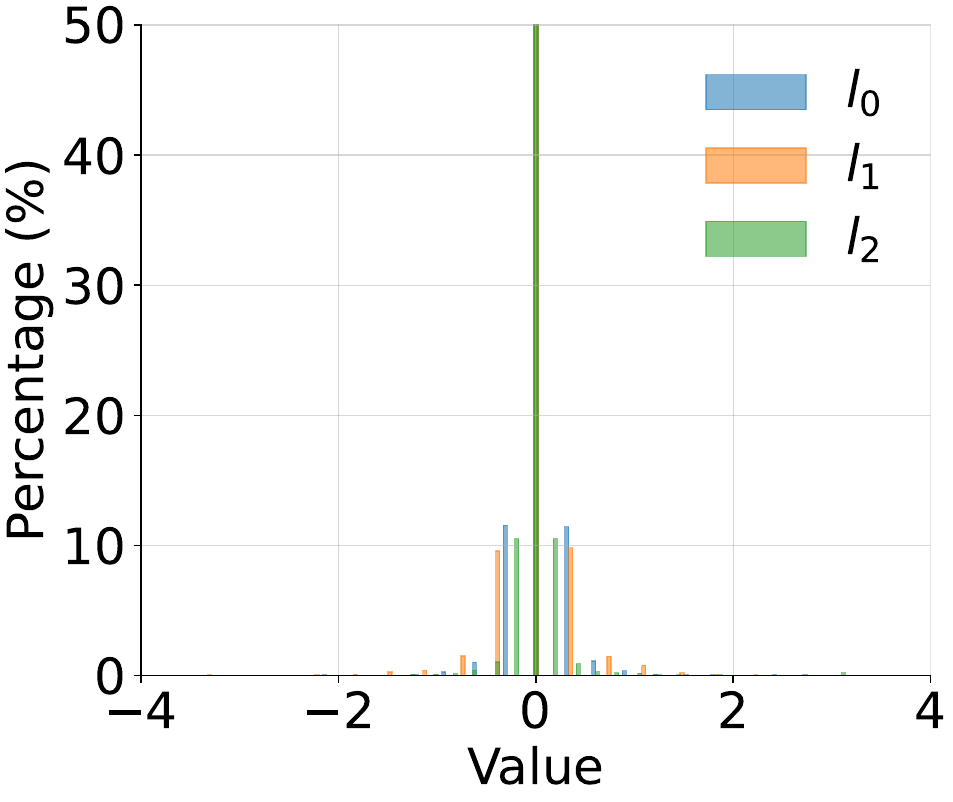}
        \includegraphics[width=0.23\textwidth]{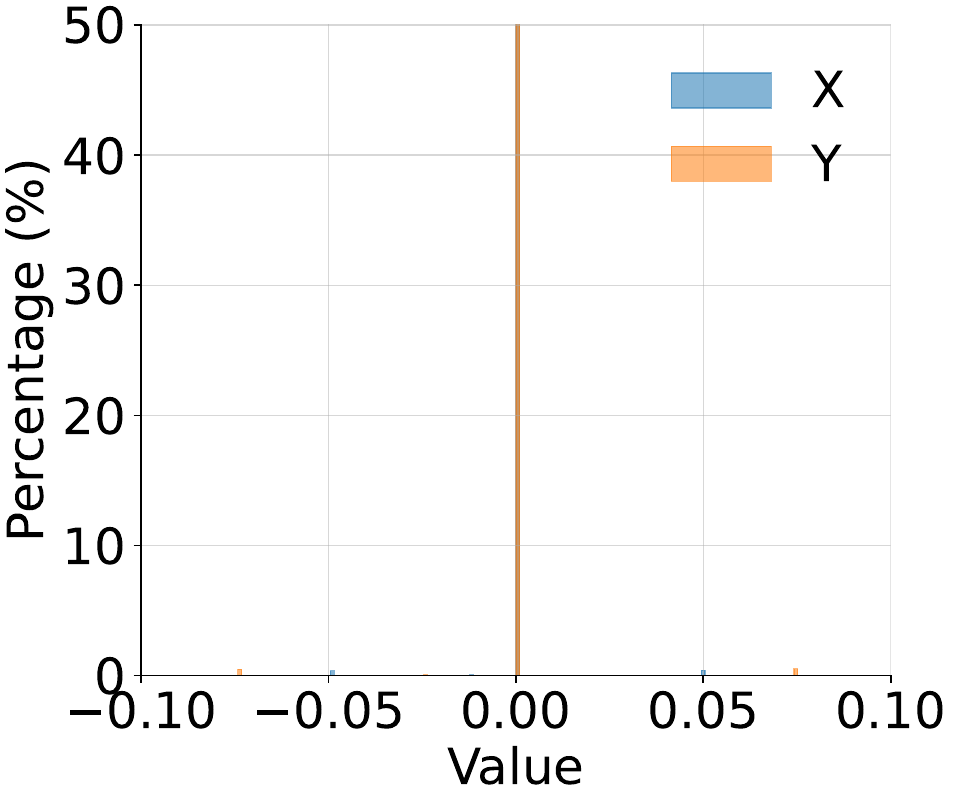}
        \label{fig:parameter_distribution_w}
    }
    \caption{Distribution of inter-Gaussian parameter residuals after quantization fine-tuning.}
    \label{fig:parameter_distribution}
\end{figure}

\begin{algorithm}[t]
\caption{Stage I: Representation Optimization}
\label{alg:overfitting}
\small
\begin{algorithmic}[1]
\REQUIRE Video frames $\mathcal{F}=\{\mathbf{F}_t\}_{t=1}^{T}$, Gaussian budgets $\{N_t\}_{t=1}^{T}$.
\ENSURE Optimized Gaussians $\{\mathcal{G}_t^{*}\}_{t=1}^{T}$.

\IF{$T>1$}
    \STATE Select anchor frames using DKS.
\ENDIF

\FOR{$t=1$ \TO $T$}
    \IF{$\mathbf{F}_t$ is an anchor frame}
        \STATE Initialize $\mathcal{G}_t$ using CI; compute progressive resolution and Gaussian schedules using~\Cref{eq:resolution_interpolation,eq:gaussian_schedule}.
    \ELSE
        \STATE Inherit inter-Gaussians from $\mathcal{G}_{t-1}$; add $\rho_{\mathrm{dyn}}N_t$ intra-Gaussians with CI in flow-based high-error regions using~\Cref{eq:flow_warp_error}.
    \ENDIF
    \WHILE{not converged, $k=0,1,2,\dots$}
        \IF{$\mathbf{F}_t$ is an anchor frame}
            \STATE Update progressive schedules and optimize $\mathcal{G}_t$ with $L_2$ loss.
        \ELSE
            \STATE Optimize $\mathcal{G}_t$ with the motion prior using~\Cref{eq:optical_flow_prior}.
        \ENDIF
        \IF{$K_r^{\mathrm{start}}\leq k\leq K_r^{\mathrm{end}}$ \AND $(k-K_r^{\mathrm{start}})\bmod\Delta K_r=0$}
            \STATE Remove low-contribution Gaussians.
        \ENDIF
        \IF{$k\geq K_w$}
            \STATE Initialize/update $\bar{\mathcal{G}}_t$ using EPA in~\Cref{eq:ema}.
        \ENDIF
        \IF{\re{convergence check interval reached}}
            \STATE \re{Evaluate loss using EMA parameters.}
            \STATE \re{Update convergence status and best EMA checkpoint.}
        \ENDIF
    \ENDWHILE
    \STATE Select the best model between final $\mathcal{G}_t^{last}$ and best $\bar{\mathcal{G}}_t$ as $\mathcal{G}_t$.
    \IF{$\mathbf{F}_t$ is an anchor frame \AND two-pass encoding}
    \STATE Propagate $\mathcal{G}_t$ through the next $k$ frames to obtain $\mathcal{G}_{t+k}$.
    \STATE Reinitialize the anchor with $\mathcal{G}_{t+k}$  and re-optimize to obtain $\bar{\mathcal{G}}_t$.
\ENDIF
\ENDFOR

\end{algorithmic}

\end{algorithm}

\textbf{Flow-Guided Optimization.}
We further use optical flow as a motion prior for inter-Gaussian optimization. For each inter-Gaussian $i$, we sample the forward flow at its projected center and four half-radius offsets along the horizontal and vertical directions, and average the valid samples to obtain its motion target:
\begin{equation}
\bar{\mathbf{u}}_i
=
\frac{1}{|\mathcal{V}_i|}
\sum_{\mathbf{q}\in\mathcal{V}_i}
\mathbf{U}_{t-1\rightarrow t}(\mathbf{q}),
\label{eq:gaussian_flow}
\end{equation}
where $\mathcal{V}_i$ denotes the valid sampling locations for inter-Gaussian $i$.
During optimization, we use the Gaussian flow as a motion prior:
\begin{equation}
\mathcal{L}_{\mathrm{flow}}
=
\frac{1}{N_{\mathrm{inter}}}
\sum_{i=1}^{N_{\mathrm{inter}}}
\left\|
(\mathbf{p}_i^{(k)}-\mathbf{p}_i^{(0)})
-
\bar{\mathbf{u}}_i
\right\|_2^2,
\label{eq:flow_prior}
\end{equation}
where $\mathbf{p}_i^{(0)}$ and $\mathbf{p}_i^{(k)}$ denote the initial and current positions of inter-Gaussian $i$, respectively.
The motion prior is jointly optimized with the reconstruction loss:
\begin{equation}
\mathcal{L}^{(k)}
=
\mathcal{L}_{\mathrm{rec}}
+
\lambda_{\mathrm{flow}}^{(k)}
\mathcal{L}_{\mathrm{flow}},
\qquad
\lambda_{\mathrm{flow}}^{(k)}
=
\begin{cases}
\lambda_{\mathrm{flow}}, & k < K_{\mathrm{flow}},\\
0, & k \geq K_{\mathrm{flow}},
\end{cases}
\label{eq:optical_flow_prior}
\end{equation}
where $K_{\mathrm{flow}}=\min\left(K_{\mathrm{flow}}^{\mathrm{conv}}, K_r^{\mathrm{start}}\right)$, with $K_{\mathrm{flow}}^{\mathrm{conv}}$ denoting the convergence iteration of the flow prior and $K_r^{\mathrm{start}}$ the start of Gaussian removal.
During early optimization, $\mathcal{L}_{\mathrm{flow}}$ guides inherited inter-Gaussians toward their estimated motion, reducing the search for their new positions. Once the flow prior converges or Gaussian removal begins, whichever occurs first, optimization continues solely with $\mathcal{L}_{\mathrm{rec}}$.
Newly added intra-Gaussians are unaffected by the motion prior, as $\mathcal{L}_{\mathrm{flow}}$ is defined only over inter-Gaussians.

\begin{algorithm}[t]
\caption{Stage II: Quantization-Aware Fine-Tuning}
\label{alg:quantization}
\small
\begin{algorithmic}[1]
\REQUIRE Video frames $\mathcal{F}=\{\mathbf{F}_t\}_{t=1}^{T}$, Optimized Gaussians $\{\mathcal{G}_t^{*}\}_{t=1}^{T}$, anchor frame list, GOP settings.
\ENSURE Quantized Gaussians $\{\tilde{\mathcal{G}}_t\}_{t=1}^{T}$.
\FOR{$t=1$ \TO $T$}
    \STATE Initialize quantizers with CI 
    \IF{$\mathbf{F}_t$ is an I-frame}
        \STATE Quantize all Gaussians as intra-Gaussians.
    \ELSE
        \STATE Quantize intra-Gaussians and inter-Gaussian residuals w.r.t.\ $\hat{\mathcal{G}}_{t-1}$.
    \ENDIF
    \WHILE{not converged, $k=0,1,2,\dots$}
        \STATE Quantize and dequantize $\mathcal{G}_t$.
        \IF{$\mathbf{F}_t$ is an I-frame}
            \STATE Optimize $\mathcal{G}_t$ with the reconstruction $L_2$ loss.
        \ELSE
            \STATE Recover $\mathcal{G}_t$ w.r.t.\ $\hat{\mathcal{G}}_{t-1}$ and optimize with~\Cref{eq:quantization_loss}.
        \ENDIF
        \IF{$k\geq K_w$}
            \STATE Initialize/update $\bar{\mathcal{G}}_t$ using EPA in~\Cref{eq:ema}.
        \ENDIF
        \IF{\re{convergence check interval reached}}
            \STATE \re{Evaluate loss using EMA parameters.}
            \STATE \re{Update convergence status and best EMA checkpoint.}
        \ENDIF
    \ENDWHILE
    \STATE Select the best model between final $\mathcal{G}_t^{last}$ and best $\bar{\mathcal{G}}_t$,and quantize it as the output $\tilde{\mathcal{G}}_t$; dequantize to $\hat{\mathcal{G}}_t$ for subsequent P-frames.
\ENDFOR
\end{algorithmic}
\end{algorithm}

\subsection{Temporal-Aware Quantization Fine-Tuning\label{sec:quantization}}

For compression, we exploit temporal redundancy through Temporal-Aware Quantization (TAQ), which uses different precisions for intra-Gaussians and temporally propagated inter-Gaussians.
Specifically, for intra-Gaussians, which are encoded independently, we use 16-bit precision for position, two-stage 8-bit RVQ~\cite{DBLP:journals/taslp/ZeghidourLOST22,gray1984vector} for color, and 6-bit LSQ+~\cite{DBLP:journals/corr/abs-2004-09576} for Cholesky parameters.
For inter-Gaussians, we encode their residuals.
Temporal coherence keeps these residuals concentrated around zero, allowing lower-precision quantization while preserving reconstruction quality: 8-bit LSQ+ for position, two-stage 7-bit RVQ for color, and 5-bit LSQ+ for Cholesky parameters.
The codebook is optimized with a commitment loss~\cite{zhang2024gaussianimage}, and we deterministically construct it using Linde--Buzo--Gray (LBG) clustering~\cite{linde1980algorithm}, replacing the random K-means initialization used by GSVC~\cite{wang2025gsvc_v2}. 
We refer readers to GSVC~\cite{wang2025gsvc_v2} for details of the quantizer formulations.

Since low-bit quantization introduces precision loss relative to full-precision representation, quantization-aware fine-tuning is needed to recover reconstruction quality. However, when compensating for quantization errors, inter-Gaussians may deviate from their references, weakening the previously established temporal coherence.
As shown in~\Cref{fig:motion_reg_wo}, the Gaussian motion becomes more irregular after fine-tuning, resulting in more dispersed residuals.
To preserve temporal coherence during quantization-aware fine-tuning, we introduce Spatiotemporal Constraints (STC), which regularize the position and shape residuals of inter-Gaussians.
For inter-Gaussian $n$ in frame $t$, we denote its position and Cholesky residuals with respect to its reference Gaussian by
$\Delta\boldsymbol{\mu}_n^t$ and $\Delta\boldsymbol{l}_n^t$, respectively.
We then optimize
\begin{equation}
\mathcal{L}_{\mathrm{q}}
=
\mathcal{L}_{\mathrm{rec}}
+
\lambda_{\mathrm{q}}
\frac{1}{|\mathcal{N}_t^{\mathrm{Inter}}|}
\sum_{n\in\mathcal{N}_t^{\mathrm{Inter}}}
\left(
\eta_t
\left\|
\Delta\boldsymbol{\mu}_n^t
\right\|_1
+
\left\|
\Delta\boldsymbol{l}_n^t
\right\|_1
\right),
\label{eq:quantization_loss}
\end{equation}
where $\mathcal{N}_t^{\mathrm{Inter}}$ denotes the set of indices of inter-Gaussians in frame $t$.
Since reconstruction is more sensitive to Gaussian positions, we leave the inter-Gaussian positions in the first P-frame unconstrained to absorb quantization errors propagated from the anchor frame. Accordingly, the position constraint is applied from the second P-frame onward ($\eta_t=0$ for the first P-frame and $1$ otherwise), while the Cholesky residual is constrained for all P-frames. As shown in~\Cref{fig:motion_reg_w,fig:parameter_distribution}, this regularization promotes more coherent Gaussian motion and produces more concentrated residuals.

\subsection{Encoding Pipeline and Execution\label{sec:pipeline}}
Together, the {\name} pipeline consists of representation optimization shown in~\Cref{alg:overfitting}, quantization-aware fine-tuning shown in~\Cref{alg:quantization}, and entropy coding. 
The first two stages both apply CI and EPA.
Representation optimization further uses progressive training for anchor frames and optical-flow guidance for propagation frames, while quantization-aware fine-tuning applies Temporal-Aware Quantization and Spatiotemporal Constraints.
To reduce runtime scheduling overhead, we employ CUDA Graphs~\cite{nvidia_cuda_graphs} to replay repeated GPU operations with fewer CPU launches.
During representation optimization, graph replay starts once the image resolution and Gaussian number become fixed.
During quantization-aware fine-tuning, where the model structure and Gaussian number remain fixed, the repeated optimization steps are captured after initialization.
Since CUDA Graphs only change execution scheduling while preserving the underlying optimization operations, they reduce kernel-launch overhead without affecting reconstruction quality.
Finally, all quantized parameters are directly entropy-coded, except for the 16-bit position parameters, whose high and low 8-bit bytes are coded separately.
The symbol ranges are hard-coded for all parameters to further reduce the entropy-coding overhead.

\begin{figure}
  \includegraphics[width=0.9\linewidth]{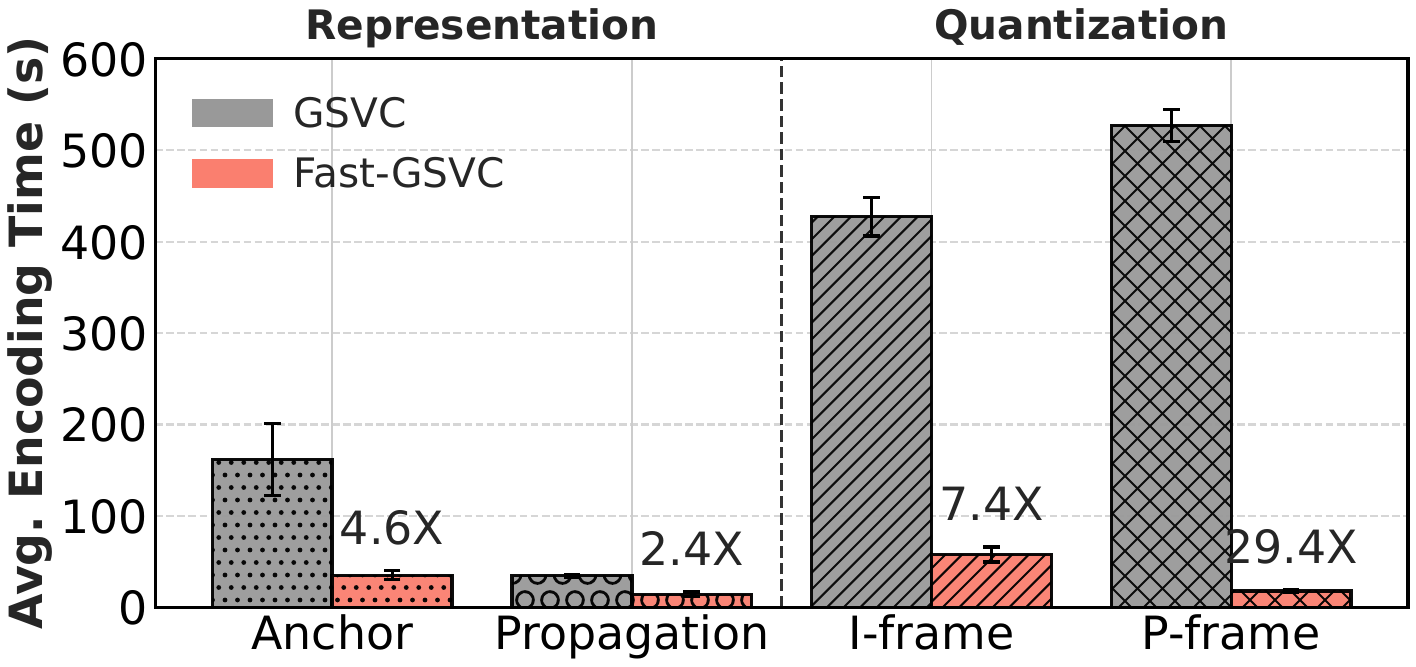}
  \caption{Average encoding time of GSVC and {\name} across anchor, propagation, I-frame, and P-frame processing. Error bars indicate $\pm$ 1 SD.
 }
 \label{fig:speed}
\end{figure}

\begin{figure*}[t]
    \centering
    \subfloat[Anchor frame]{
        \includegraphics[width=0.22\textwidth]{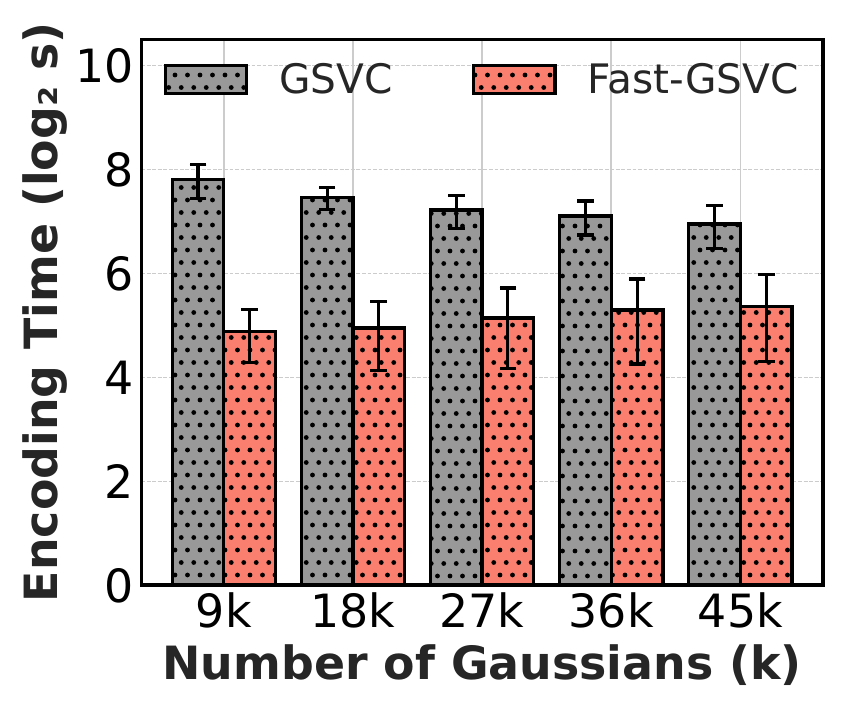}
        \label{fig:speed_anchor}
    }
    \hspace{0.01\textwidth}
    \subfloat[Propagation frame]{
        \includegraphics[width=0.22\textwidth]{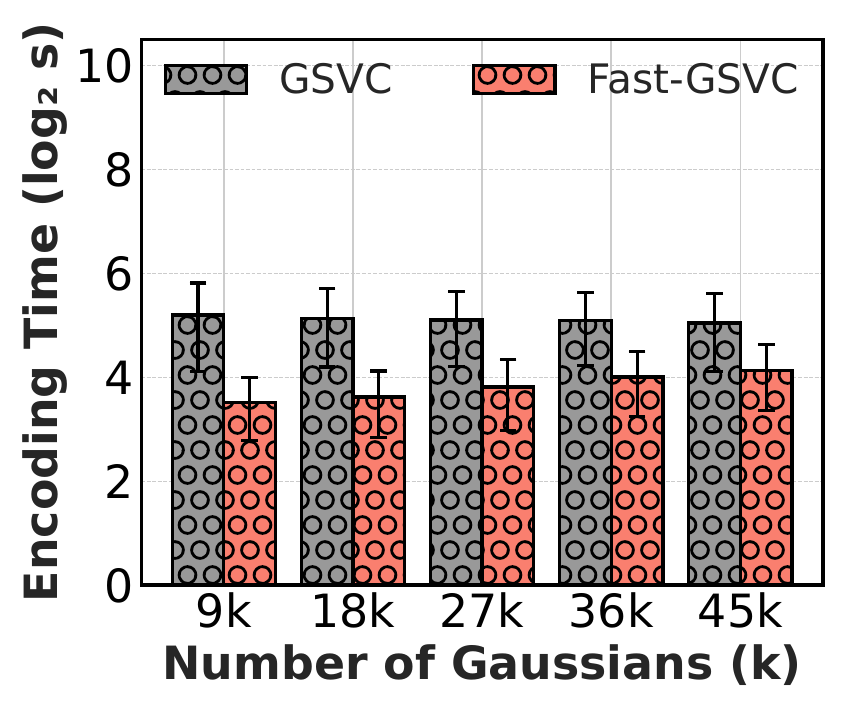}
        \label{fig:speed_propagation}
    }
    \hspace{0.01\textwidth}
    \subfloat[I-frame]{
        \includegraphics[width=0.22\textwidth]{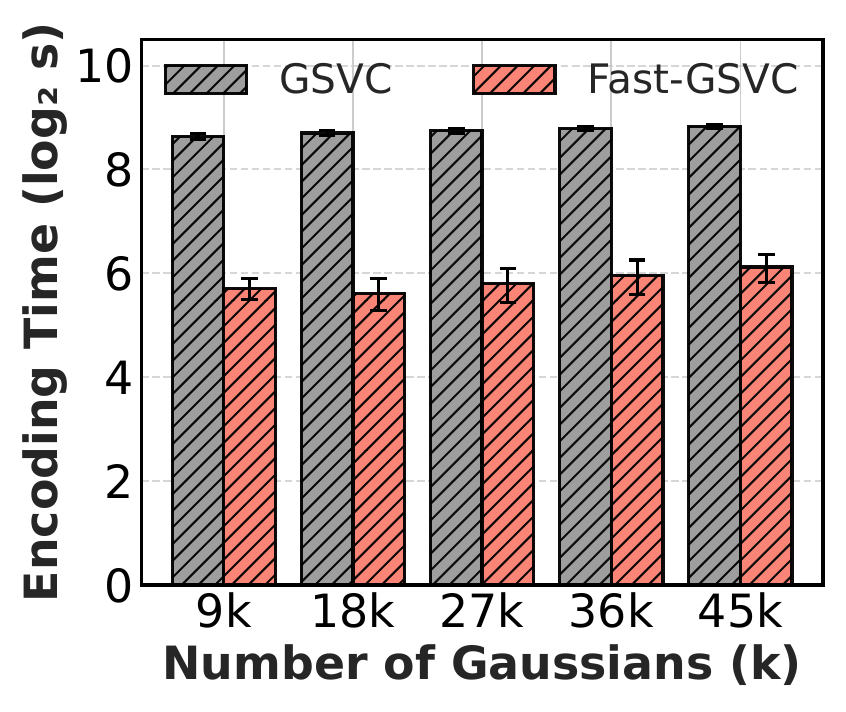}
        \label{fig:speed_iframe}
    }
    \hspace{0.01\textwidth}
    \subfloat[P-frame]{
        \includegraphics[width=0.22\textwidth]{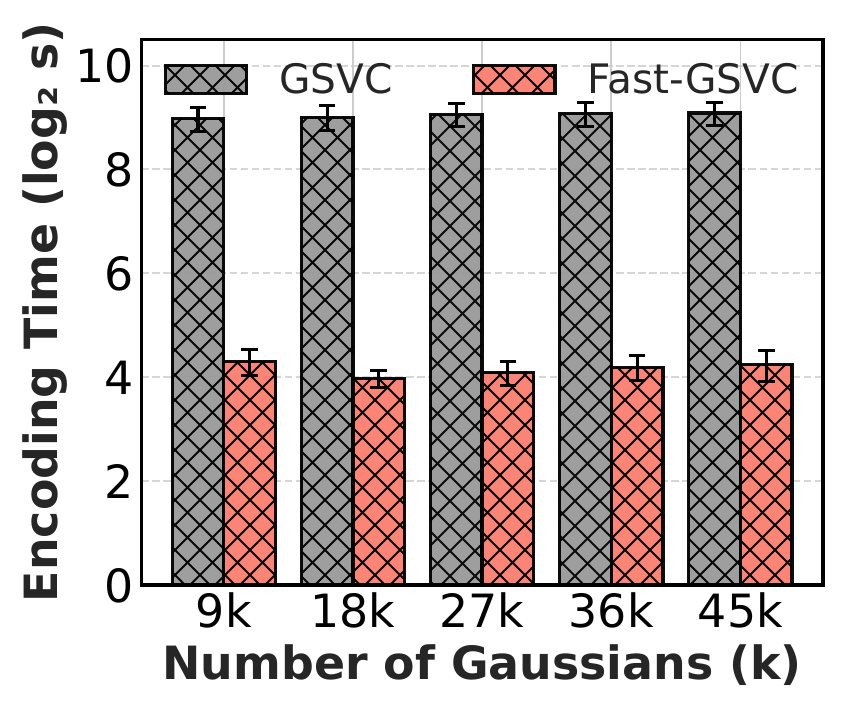}
        \label{fig:speed_pframe}
    }
    \caption{Encoding time across Gaussian budgets on a $\log_2$ scale: representation optimization (a, b) and quantization (c, d). Error bars indicate $\pm$ 1 SD.}
    \label{fig:speed_per_buget_bar}
\end{figure*}

\begin{figure*}[t]
    \centering
    \includegraphics[width=0.23\textwidth]{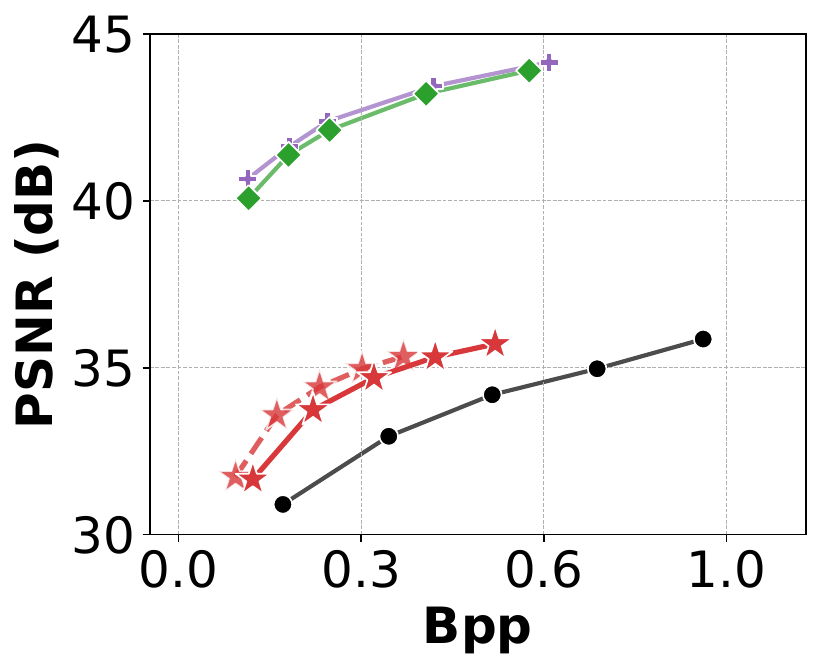} 
    \hspace{0.03\textwidth}
    \includegraphics[width=0.23\textwidth]{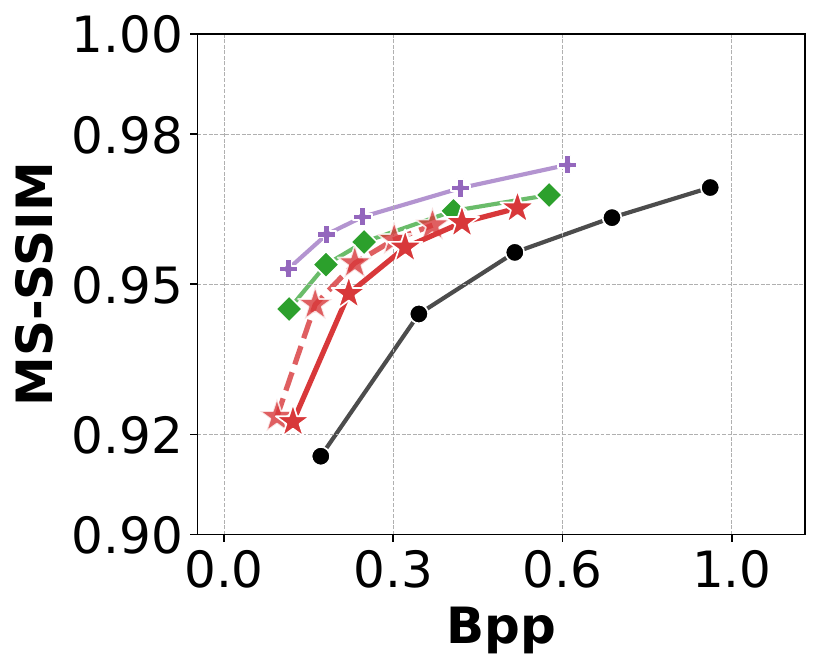} 
    \hspace{0.03\textwidth}
    \includegraphics[width=0.23\textwidth]{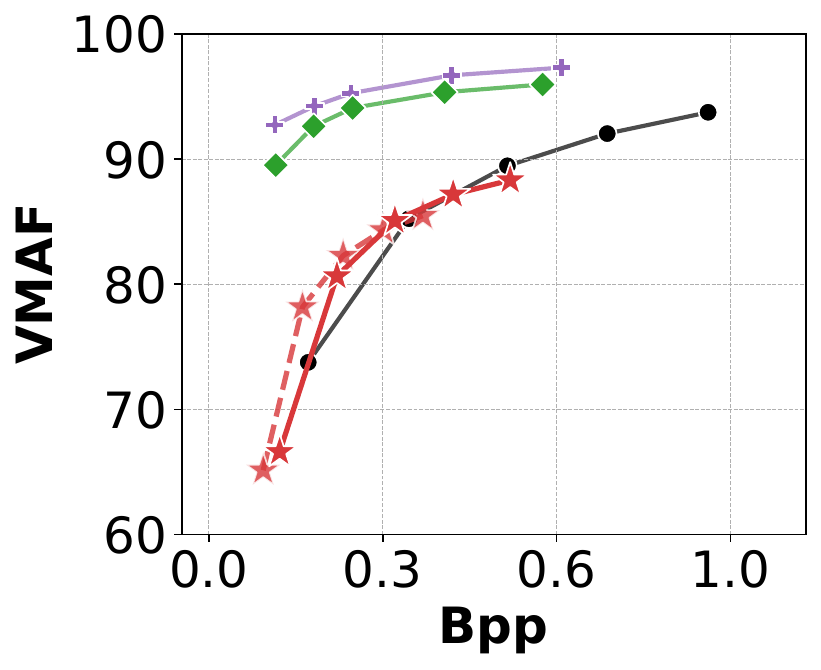}
    \hspace{0.01\textwidth}
    \raisebox{0.9cm}{%
    \includegraphics[width=0.13\textwidth]{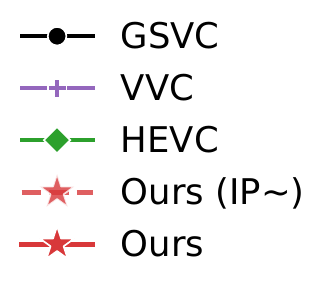}}
    \caption{\re{Quantized rate–distortion results of~\name~versus baselines on UVG dataset.}}
    \label{fig:quantization}
\end{figure*}

\begin{figure}
    \centering
    \includegraphics[width=0.23\textwidth]{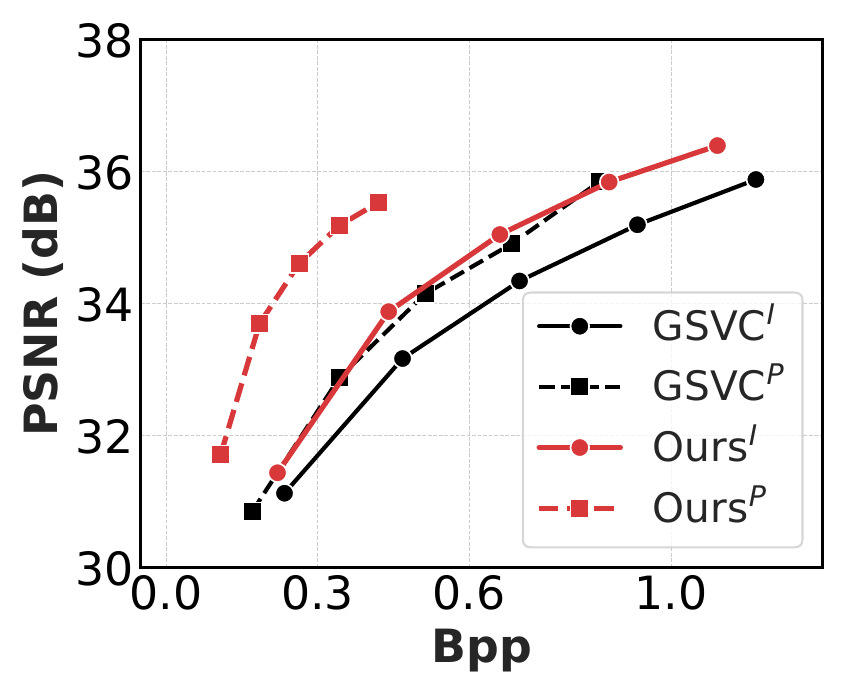}\hspace{0.01\textwidth}
    \includegraphics[width=0.23\textwidth]{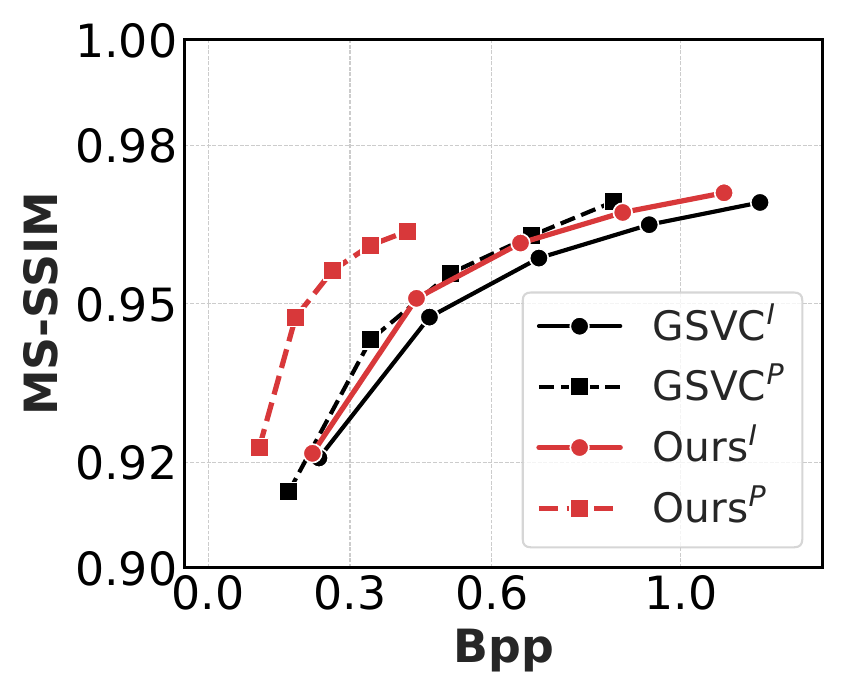}
    \caption{Rate-distortion comparison between GSVC and {\name} for I- and P-frames.}
    \label{fig:speed_per_buget_rd}
\end{figure}

\section{Experiments} \label{sec:experiment}
\subsection{Experimental Setup}

\textbf{Datasets and Metrics}.
We evaluate {\name} on all 1080p videos in the UVG dataset~\cite{DBLP:conf/mmsys/MercatVV20}, provided at $1920\times1080$ resolution and 120 fps in 8-bit YUV 4:2:0 format. Following GSVC~\cite{wang2025gsvc_v2}, we use the first 50 frames of each video.
We report PSNR, MS-SSIM~\cite{wang2003multiscale}, and VMAF~\cite{bampis2018spatiotemporal} for reconstruction quality, bpp for compression efficiency, and end-to-end wall-clock encoding time, including Gaussian initialization, optical-flow processing, representation optimization, quantization-aware fine-tuning, and entropy coding.
Higher PSNR, MS-SSIM, and VMAF indicate better reconstruction quality, while lower bpp and encoding time correspond to smaller compressed size and faster encoding, respectively.

\textbf{Implementation.}
We use an IPPPP GOP structure with Gaussian budgets of $\{9\mathrm{k}, 18\mathrm{k}, 27\mathrm{k}, 36\mathrm{k}, 45\mathrm{k}\}$.
\re{Convergence is checked using the EMA reconstruction loss every 100 iterations and declared when the loss improves by less than $1\%$ for 10 consecutive checks.}
\re{For CI, the balancing coefficient $\alpha$ is set to 32, and Gaussian centers are sampled using a mixture of $90\%$ gradient-guided sampling and $10\%$ uniform sampling.}
\re{For EPA, we use a warm-up of $K_w=2500$ iterations and an EMA decay factor of $\beta=0.99$, consistent with common EMA smoothing practices~\cite{DBLP:journals/tmlr/Morales-Brotons24}.}
The anchor-frame learning rate is initialized as $10^{-3}$ before $K_w$ for stable warm-up and increased to $10^{-2}$ afterwards, while the propagation-frame learning rate is fixed at $10^{-3}$.
The two-pass encoding is set to $k=1$.
For progressive training, we use $m=32$ downsampling levels.
We allocate an additional dynamic Gaussian budget of $\rho_{\mathrm{dyn}}=10\%$ and remove Gaussians every $\Delta K_r=100$ iterations, from $K_r^{\mathrm{start}}=0$ to $K_r^{\mathrm{end}}=4000$ for anchor frames and from $K_r^{\mathrm{start}}=500$ to $K_r^{\mathrm{end}}=1000$ for propagation frames.
The optical-flow prior is weighted by $\lambda_{\mathrm{flow}}=10^{-3}$.
During quantization-aware fine-tuning, we use a learning rate of $10^{-3}$ and set $\lambda_{\mathrm{q}}=10^{-8}$.
For I-frames, the learning rate is further halved twice after convergence for refinement.
Meanwhile, we lower-bound the quantization interval at $5\%$ of its initial value to avoid excessive code clipping.
The entropy-friendly loss weight is set to $10^{-5}$ for YachtRide and Jockey due to their more complex motion, and to $10^{-4}$ for other sequences.
Since we only require a coarse motion prior and flow errors can be further corrected by newly added Gaussians, we use the lightweight CPU-based OpenCV DIS estimator to compute bidirectional flow at half resolution, taking $\sim$102 ms.
All experiments are conducted on an NVIDIA H100 GPU.

\begin{figure*}[t]
    \centering
    \subfloat[Training from Scratch]{
        \includegraphics[width=0.23\textwidth]{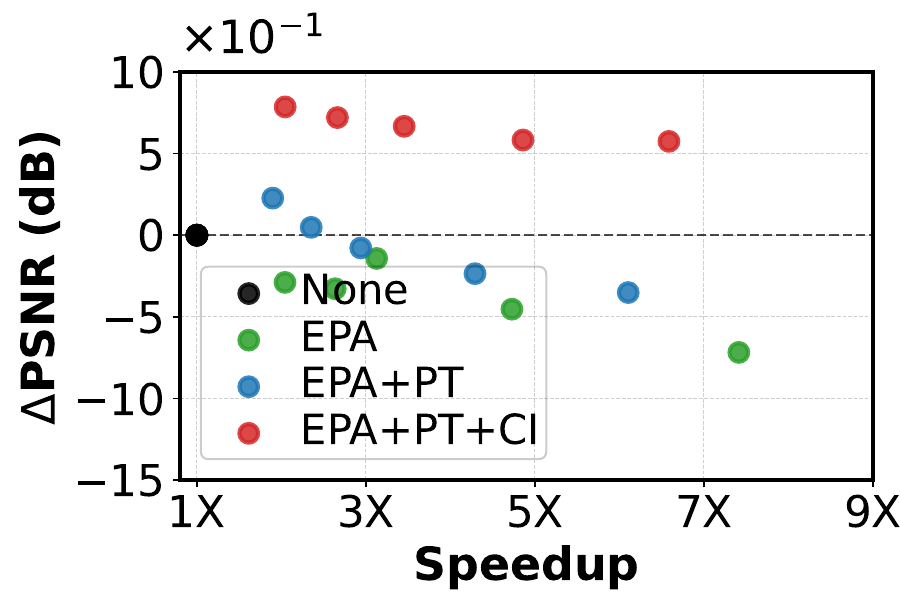}
        \label{fig:ab_sc}
    }\hspace{0.02\textwidth}
    \subfloat[Fine-tuning]{
        \includegraphics[width=0.23\textwidth]{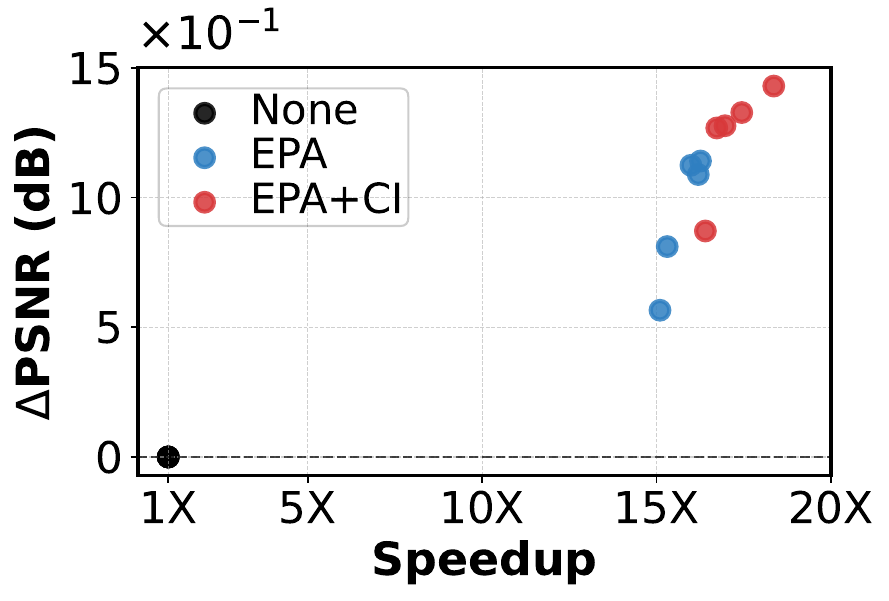}
        \label{fig:ab_ft}
    }\hspace{0.02\textwidth}
    \subfloat[Propagation]{
        \includegraphics[width=0.23\textwidth]{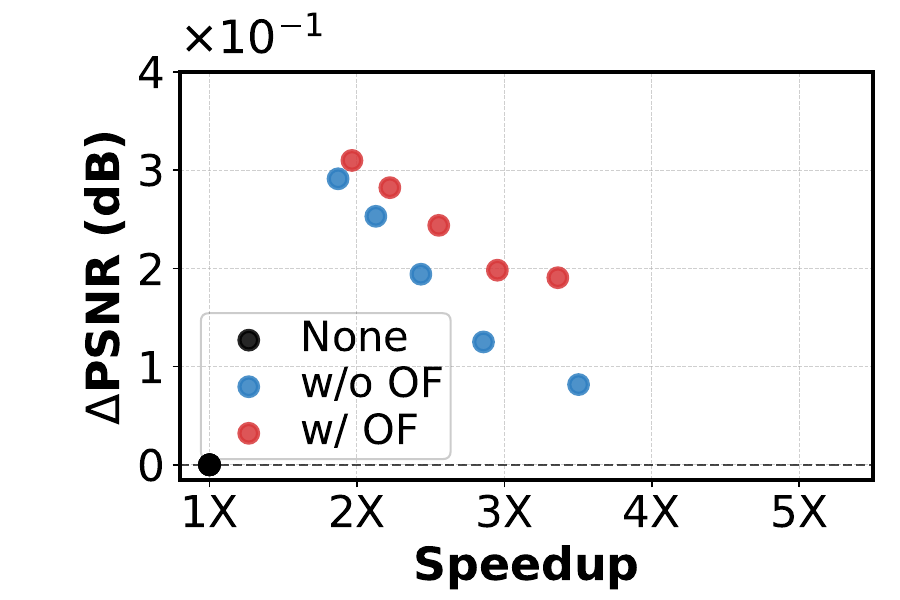}
        \label{fig:ab_of}
    }
    \caption{\re{Ablation of acceleration components on the speed--quality trade-off across three optimization settings: (a) training from scratch, (b) quantization fine-tuning, and (c) propagation-frame optimization.}}
    \label{fig:ab_speed_quality}
\end{figure*}

\textbf{Comparison Methods.}
We first compare with GSVC~\cite{wang2025gsvc_v2}, the SOTA iterative 2D Gaussian video codec, using the same Gaussian budgets and GOP structure (IPPPP).
Since the two methods employ different optimization procedures, we do not enforce the same number of iterations. Instead, we measure the encoding time required by each method to achieve its best performance under its original optimization settings.
Accordingly, we first compare encoding time and then evaluate rate-distortion performance at the corresponding encoding time, demonstrating that {\name} achieves the goal formulated in~\Cref{sec:formulation}: better rate-distortion performance with lower encoding time.
We additionally report \textit{Ours (IP\textasciitilde)}, where only the first frame is encoded as an I-frame and all remaining frames as P-frames, to evaluate long-term temporal propagation.
\re{We also compare with standardized video codecs HEVC~\cite{HEVC} and VVC~\cite{VVC}, using HM-16.2 and VTM-23.6, respectively.
Both adopt the same IPPPP GOP structure, with QP values ranging from 18 to 28.}
Moreover, we include the SOTA feed-forward 2DGS method AIR~\cite{zeng2026air} to examine feed-forward prediction as an alternative.

\subsection{Results and Analysis}

\textbf{Encoding Acceleration.}
We first report the encoding time for two-pass encoded anchor, propagation, I-, and P-frames, each averaged across all tested Gaussian-budget settings.
As shown in~\Cref{fig:speed}, {\name} substantially accelerates encoding.
During representation optimization, it achieves a 4.6$\times$ speedup for anchor frames after two-pass encoding and a 2.4$\times$ speedup for propagation frames.
During quantization-aware fine-tuning, the speedups further reach 7.4$\times$ for I-frames and 29.4$\times$ for P-frames.
Notably, the dominant encoding bottleneck shifts from P-frames in GSVC to I-frames in {\name}, indicating that the proposed P-frame design more effectively exploits temporal coherence.
For a 50-frame video with an IPPPP coding structure, {\name} reduces the total processing time from approximately 7.56 hours to 34 minutes, achieving a 13.4$\times$ overall speedup, and a 92.5\% reduction in encoding time.

For a more detailed comparison, we further report the encoding time for each Gaussian budget in~\Cref{fig:speed_per_buget_bar}.
{\name} consistently accelerates encoding across all tested Gaussian budgets.
Specifically, the speedup ranges from $3.1$--$8.0\times$ for anchor frames after two-pass encoding, $2.0$--$3.3\times$ for propagation frames, $6.5$--$7.7\times$ for I-frames, and $25.9$--$31.0\times$ for P-frames.
Interestingly, for GSVC anchor-frame optimization, the encoding time decreases as the Gaussian budget increases, whereas the encoding time of {\name} increases with the budget.
This is because, at smaller budgets, for GSVC, the randomly initialized Gaussians provide insufficient image coverage and, as discussed in~\Cref{Sec:Introduction}, require more early-stage optimization steps to establish adequate coverage.
In contrast, coverage-aware initialization in {\name} provides sufficient coverage at the beginning, eliminating this early-stage optimization inefficiency.

\textbf{Rate-Distortion Trade-off.}
We then show that {\name} achieves better rate-distortion performance with the substantially reduced encoding time reported above.
\Cref{fig:quantization} reports the overall rate-distortion performance.
When combined into the full pipeline, {\name} achieves comparable (VMAF) or better (PSNR, MS-SSIM) performance than GSVC across all three metrics. At the same PSNR, {\name} reduces bpp by approximately 46\% (38--51\%); at the same bpp, it improves PSNR by about 1.9 dB (1.5--2.2 dB).
With \textit{Ours (IP\textasciitilde)}, the gains further increase to 58\% (51--63\%) bpp reduction and 2.5 dB (2.2--2.8 dB) PSNR improvement.
Although \textit{Ours (IP\textasciitilde)} has slightly lower post-quantization quality at the same Gaussian budget, its longer P-frame chain better exploits temporal redundancy, yielding lower-entropy residuals and smaller bitstreams.
This further demonstrates the effectiveness of temporal prediction and residual coding in {\name}.
Moreover, as shown in~\Cref{fig:speed_per_buget_rd}, {\name} consistently outperforms GSVC on both I- and P-frames. 
While GSVC runs for 100k iterations and selects its best checkpoint, {\name} requires only 23.7k and 6.2k iterations on average for I- and P-frames, respectively, showing that better rate-distortion performance can be achieved with substantially fewer optimization iterations.
\re{For completeness, we also compare with standardized video codecs.
As shown in~\Cref{fig:quantization}, while {\name} remains behind HEVC and VVC in PSNR and VMAF, it achieves comparable MS-SSIM at higher bitrates.
More importantly, {\name} narrows the rate-distortion gap in PSNR and MS-SSIM to these standardized codecs while substantially reducing encoding time compared with GSVC.
Currently, we adopt an $L_2$ reconstruction term for its low computational cost.
Incorporating perceptual losses may improve visual quality but increase encoding time, as observed in dynamic Gaussian reconstruction~\cite{DBLP:journals/fcsc/XuYSW25}.
Stricter convergence criteria and additional iterations could further improve PSNR at the cost of encoding time~\cite{wang2025gsvc_v2}.
Beyond rate-distortion performance, the explicit 2D Gaussian representation enables direct control over individual primitives, offering flexibility for adaptive Gaussian allocation, refinement, and progressive coding~\cite{wang2026p}.}

\begin{figure}[t]
    \centering
    \includegraphics[width=0.23\textwidth]{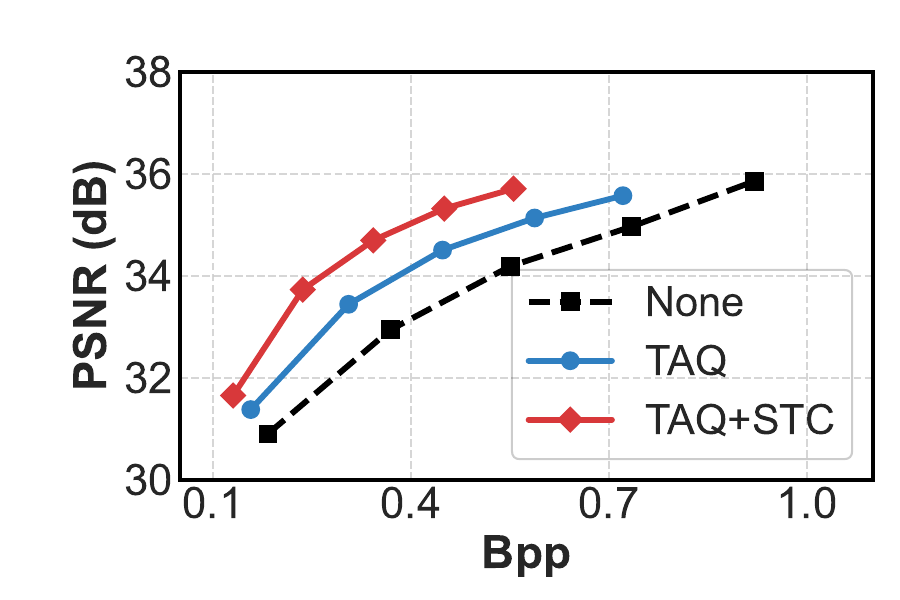}%
    \hfill
    \includegraphics[width=0.23\textwidth]{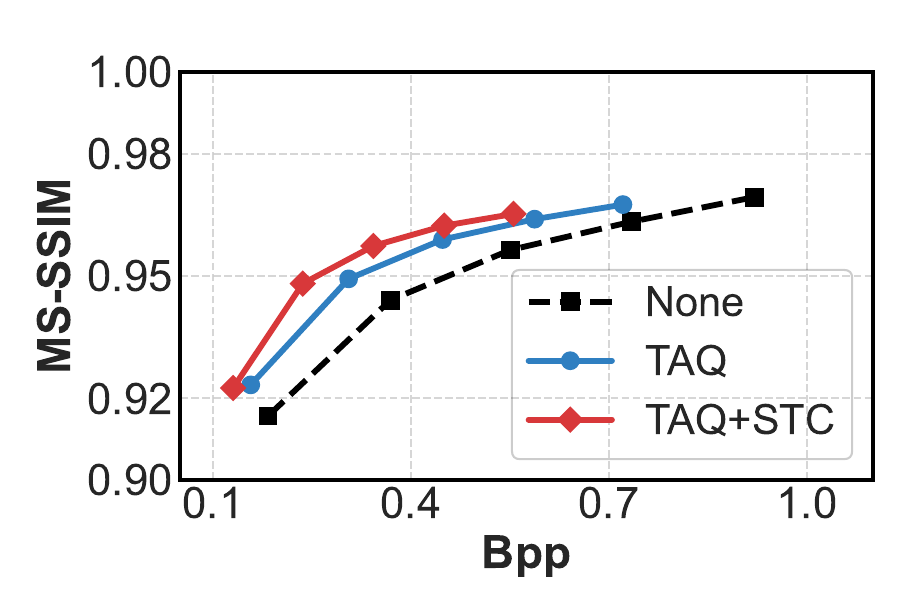}
    \caption{Rate-distortion ablation of Temporal-Aware Quantization (TAQ) and Spatiotemporal Constraints (STC).}
    \label{fig:ab_rd}
\end{figure}

\textbf{Ablation Study.}
We conduct ablation studies at two levels: acceleration and compression performance.
For acceleration, we evaluate the speed-quality trade-off under three settings: training from scratch (single-pass anchor-frame optimization), fine-tuning (quantization), and consecutive optimization (propagation-frame optimization). \Cref{fig:ab_speed_quality} reports the results, where \textit{None} denotes the baseline without acceleration components. Each point corresponds to one of the five Gaussian budgets and reports the PSNR change relative to \textit{None} at the same budget.
As shown in~\Cref{fig:ab_sc,fig:ab_ft}, all accelerated variants reduce training time, while progressively adding EPA, PT, and CI improves reconstruction quality, validating their effectiveness in achieving a better speed--quality trade-off. Since optical-flow guidance is specific to propagation-frame optimization, we ablate it separately in~\Cref{fig:ab_of}, where it provides further gains in the speed--quality trade-off.
For compression performance, we ablate Temporal-Aware Quantization (TAQ) and Spatiotemporal Constraints (STC) in~\Cref{fig:ab_rd}. TAQ improves rate-distortion performance over the baseline, while STC provides further gains, validating the effectiveness of both components.

\textbf{Feed-forward Initialization.}
Feed-forward methods enable fast reconstruction without per-instance optimization, but such approaches are less suitable for video coding, where ``knobs'' for fine-grained rate-quality control are required.
A per-video-optimized 2DGS video codec naturally supports both constant-rate encoding with arbitrary Gaussian budgets and constant-quality encoding through progressive pruning of Gaussians to a target quality~\cite{wang2025gsvc_v2}.
In contrast, the number of Gaussians generated per pixel in feed-forward methods is tightly coupled with the learned network, making budget changes require architectural modifications and retraining.
Direct pruning can also be challenging for feed-forward predictions, as observed in our experiments below.

\begin{figure}[t]
    \centering
      \includegraphics[width=0.8\linewidth]{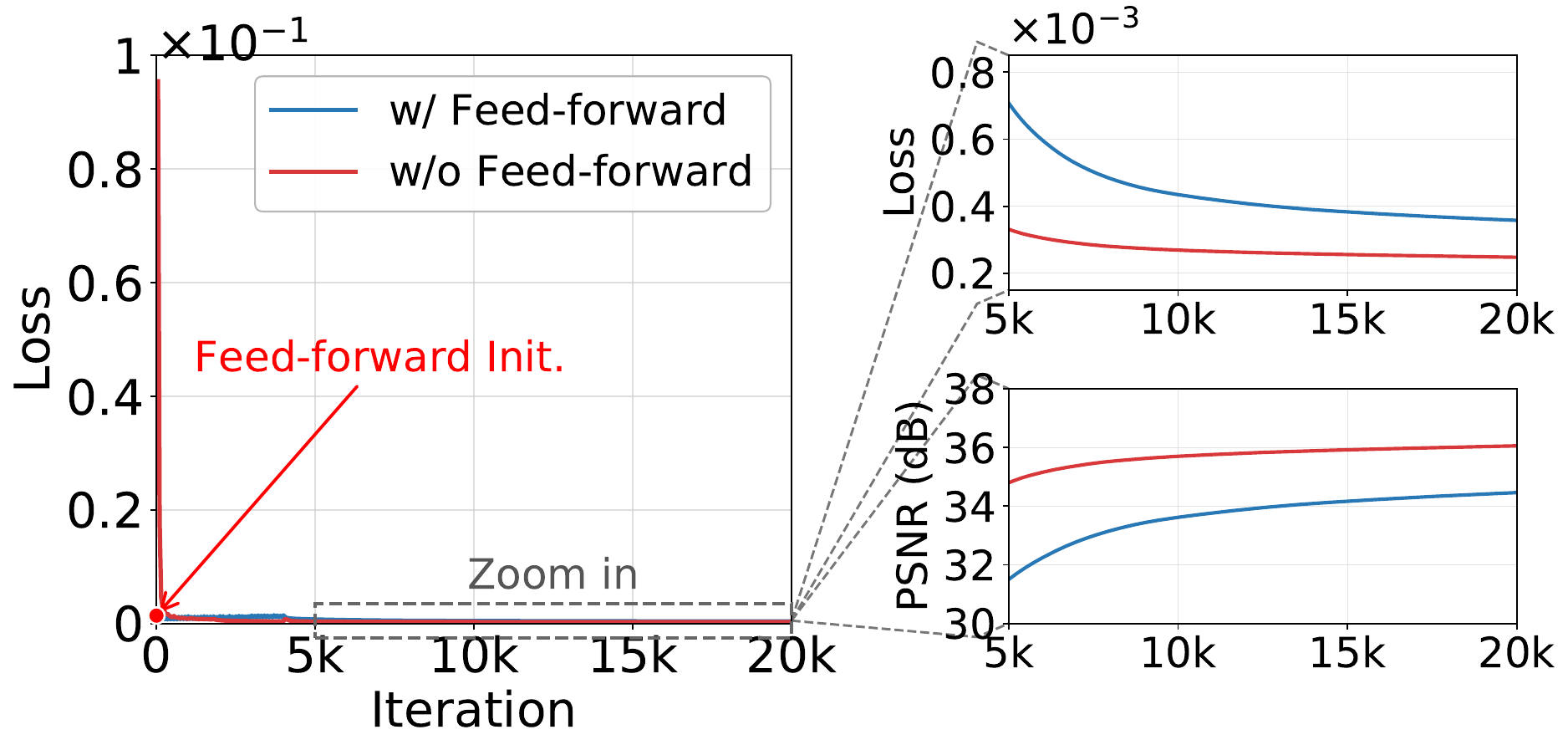}
    \caption{Optimization dynamics of {\name} w/ vs. w/o feed-forward initialization. }
\label{fig:training_feed_forward_dynamics}
\end{figure}

A possible way to combine their advantages is to use feed-forward prediction as initialization and progressively prune the predicted Gaussians during subsequent optimization.
We evaluate this strategy using the SOTA feed-forward 2DGS method AIR~\cite{zeng2026air}. Specifically, AIR first predicts 10,764 Gaussians for the anchor frame of Jockey, which are then progressively pruned to the target budget of 9K; our initialization starts from the same number of Gaussians and follows the same pruning schedule. 
As shown in~\Cref{fig:training_feed_forward_dynamics}, feed-forward initialization provides a better starting point with lower loss. However, AIR produces a much more overlapping Gaussian initialization, with each pixel covered by 8.3 Gaussians on average, compared with only 3.9 for Fast-GSVC.
After applying the same progressive pruning schedule, the AIR initialization converges to a solution with an approximately 2~dB lower PSNR. 
This suggests that, in our experiment, AIR initialization is more sensitive to pruning, while the underlying cause requires further investigation.
Although {\name} does not currently incorporate off-the-shelf feed-forward initialization, jointly designing feed-forward prediction, Gaussian pruning, and subsequent optimization remains a promising direction for future work.

\textbf{Computational Cost.}
With all acceleration components for optimization effectiveness enabled, peak GPU memory increases from 824--838 MiB to 874--892 MiB across 9k--45k Gaussians, corresponding to only about 50 MiB of overhead. This increase mainly comes from maintaining detached EMA parameters outside the optimization loop; since they require neither gradients nor optimizer states, the additional memory remains small.
Enabling CUDA Graphs further reduces training time by 36\%--46\% (1.56$\times$--1.85$\times$ speedup), while the peak GPU memory usage increases to 1,278--1,338 MiB. 
The additional memory mainly comes from persistent graph buffers required for replay.
In our experiment, we reserve buffer capacity at three times the peak memory demand observed during warm-up to accommodate runtime variation without dynamic reallocation; tighter capacity estimation could further reduce this overhead.
In comparison, the feed-forward initialization of AIR~\cite{zeng2026air} requires approximately 6.6 GB of GPU memory to generate about 10k Gaussians using its smallest patch-size-14 model. \re{For decoding computational cost between 2DGS-based and standardized video codecs, we refer readers to the extended GSVC study~\cite{wang2025gsvc_v2} for decoding latency and energy efficiency.}

\section{Conclusion}
In this work, we identify slow optimization convergence as a key bottleneck in 2DGS video coding.
We address this bottleneck by improving optimization effectiveness through complementary designs, including better initialization, stable yet aggressive parameter updates, progressive optimization, and temporal propagation guidance, while further reducing runtime overhead through GPU-level optimization.
Moreover, we improve rate-distortion performance through temporal-aware quantization while preserving temporal coherence through spatiotemporal constraints.
Together, these designs enable {\name} to achieve a 13.4$\times$ encoding speedup over GSVC, reducing the average encoding time from approximately 9 minutes to 40 seconds per frame, while reducing file size by up to 58\% at comparable reconstruction quality.
This work makes 2D Gaussian video coding faster, more compact, and higher-quality, narrowing the gap toward practical deployment.
\re{We also leave a formal subjective evaluation for future work.}
Despite the acceleration, per-frame iterative optimization remains a bottleneck for real-time encoding. Future work could further explore feed-forward Gaussian prediction to reduce optimization costs while preserving compression efficiency and scalability.


\balance
\bibliographystyle{ACM-Reference-Format}
\bibliography{ref}
\appendix
\onecolumn
\section{Visual Comparisons}
\label{sec:appendix_visual}

We compare all seven UVG sequences using full-frame views and
$3.2\times$ enlarged views of identical $600\times150$-pixel regions.

\subsection{Convergence Speed}
\label{sec:appendix_speed}

\begin{figure}[!ht]
    \centering
    \includegraphics[width=\linewidth]{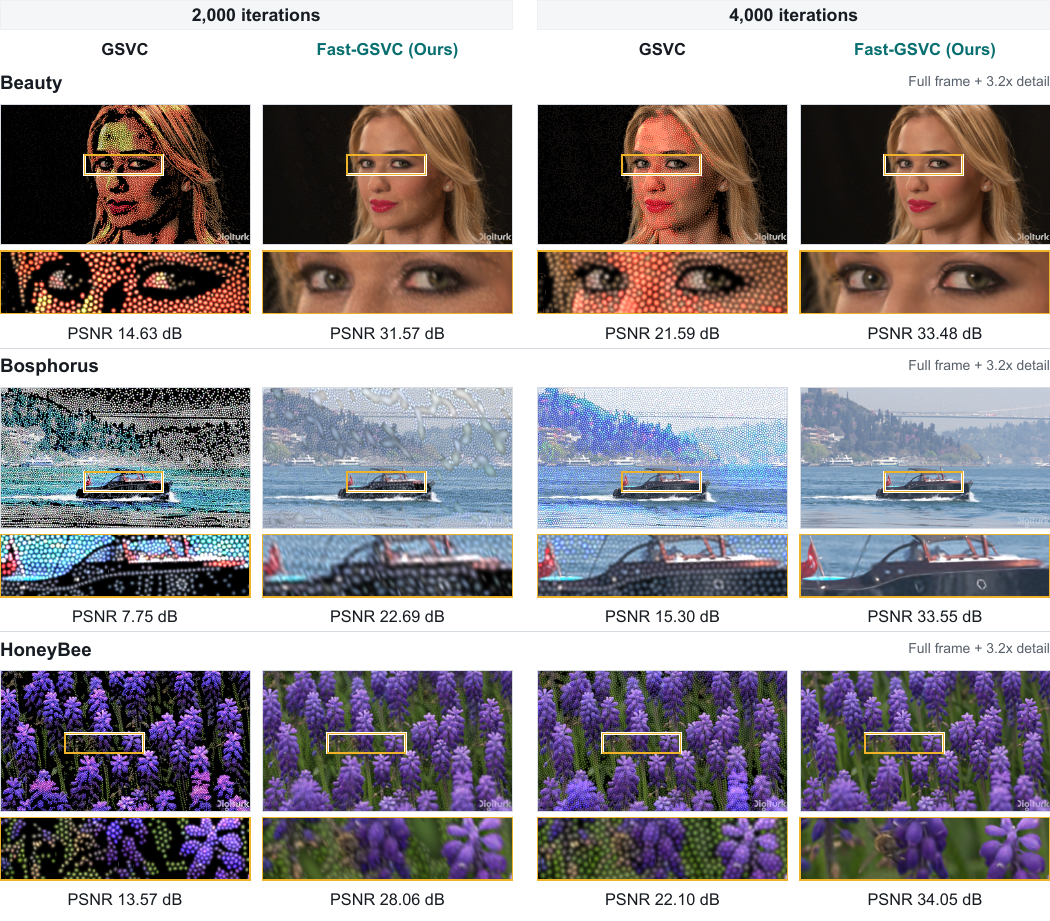}
    \caption{Convergence comparison on Beauty, Bosphorus, and HoneyBee at 2,000 (left) and 4,000 (right) iterations. Each group compares GSVC and {\name}, with PSNR (dB) and enlarged details shown below. All reconstructions use 10,000 Gaussians.}
    \label{fig:appendix_speed_1}
\end{figure}

\begin{figure}[!ht]
    \centering
    \includegraphics[width=\linewidth]{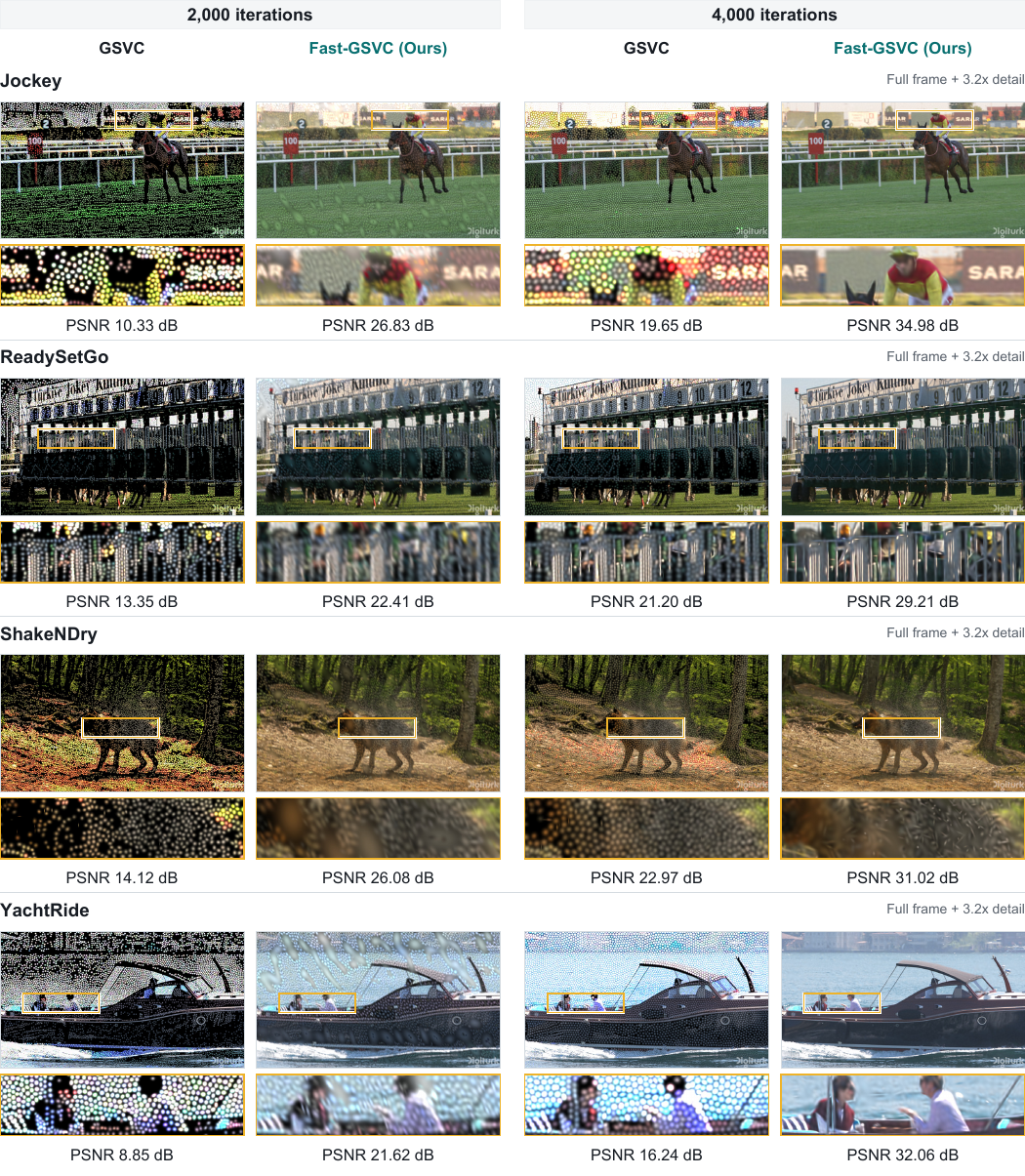}
    \caption{Convergence comparison on Jockey, ReadySetGo, ShakeNDry, and YachtRide at 2,000 (left) and 4,000 (right) iterations. Each group compares GSVC and {\name}, with PSNR (dB) and enlarged details shown below. All reconstructions use 10,000 Gaussians, except {\name} on ShakeNDry at 2,000 iterations (6,767 Gaussians) due to dynamic progressive scheduling.}
    \label{fig:appendix_speed_2}
\end{figure}

\Cref{fig:appendix_speed_1,fig:appendix_speed_2}
compare reconstructions after 2,000 and 4,000 iterations.
At both checkpoints, {\name} consistently achieves higher PSNR
than GSVC, demonstrating faster convergence.

\clearpage

\clearpage
\subsection{I-Frame Quality}
\label{sec:appendix_iframe}

\begin{figure}[!ht]
    \centering
    \includegraphics[width=\linewidth]{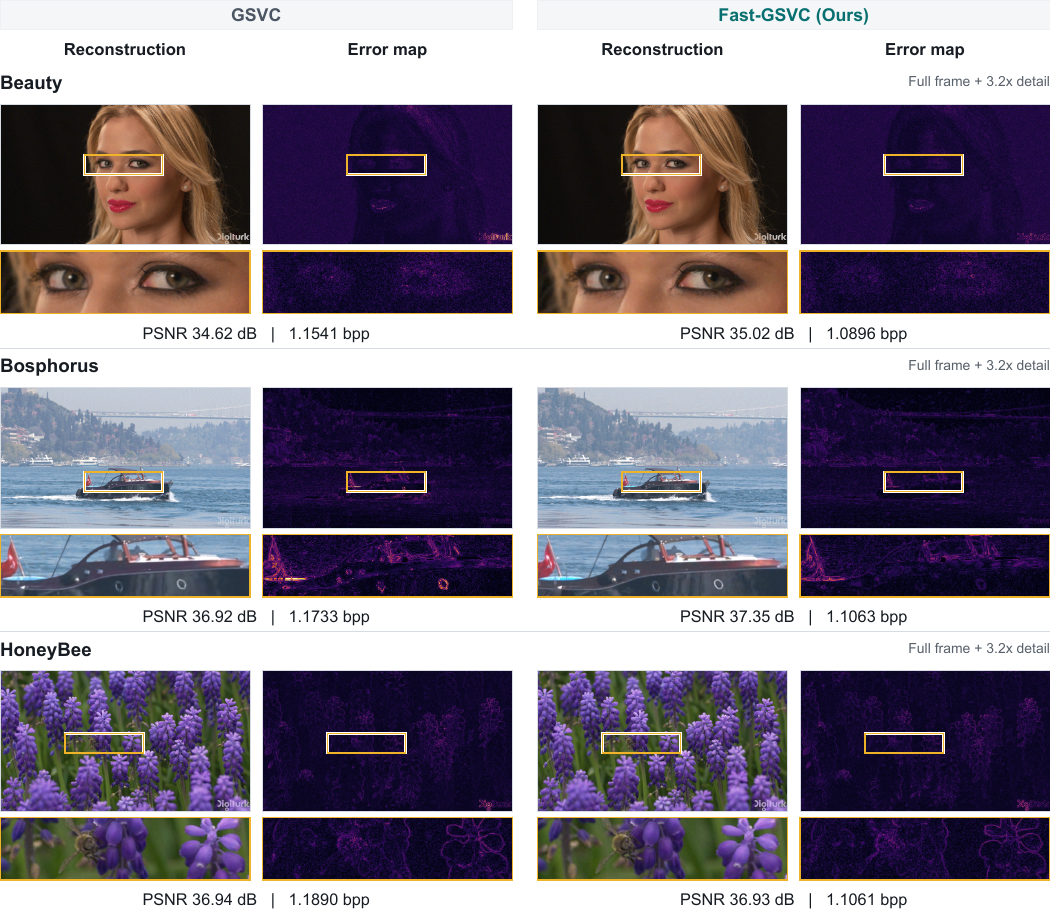}
    \caption{I-frame (frame 1) reconstruction quality on Beauty, Bosphorus, and HoneyBee. GSVC and {\name} are compared using reconstructions, error maps, and enlarged details, with PSNR (dB) and bpp reported for each method.}
    \label{fig:appendix_iframe_1}
\end{figure}

\begin{figure}[!ht]
    \centering
    \includegraphics[width=\linewidth]{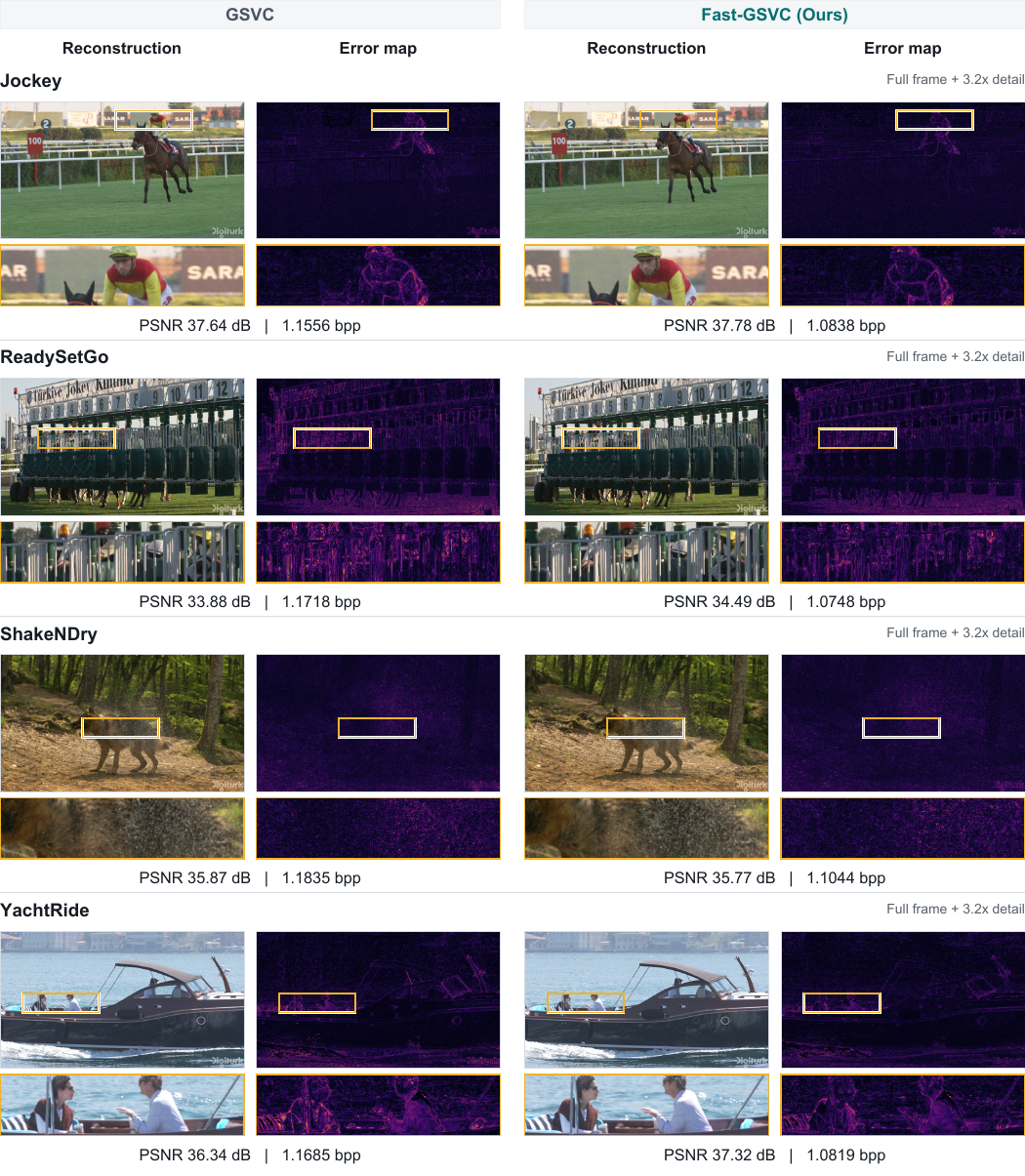}
    \caption{
    I-frame (frame 1) reconstruction quality on Jockey, ReadySetGo, ShakeNDry, and YachtRide. GSVC and {\name} are compared using reconstructions, error maps, and enlarged details, with PSNR (dB) and bpp reported for each method.}
    \label{fig:appendix_iframe_2}
\end{figure}

\Cref{fig:appendix_iframe_1,fig:appendix_iframe_2} compare I-frame quality across seven UVG sequences. Despite substantially faster encoding, {\name} achieves comparable reconstruction quality at similar or lower bitrates, as illustrated by the reconstructed frames and error maps.

\clearpage

\clearpage
\subsection{P-Frame Quality}
\label{sec:appendix_pframe}

\begin{figure}[!ht]
    \centering
    \includegraphics[width=\linewidth]{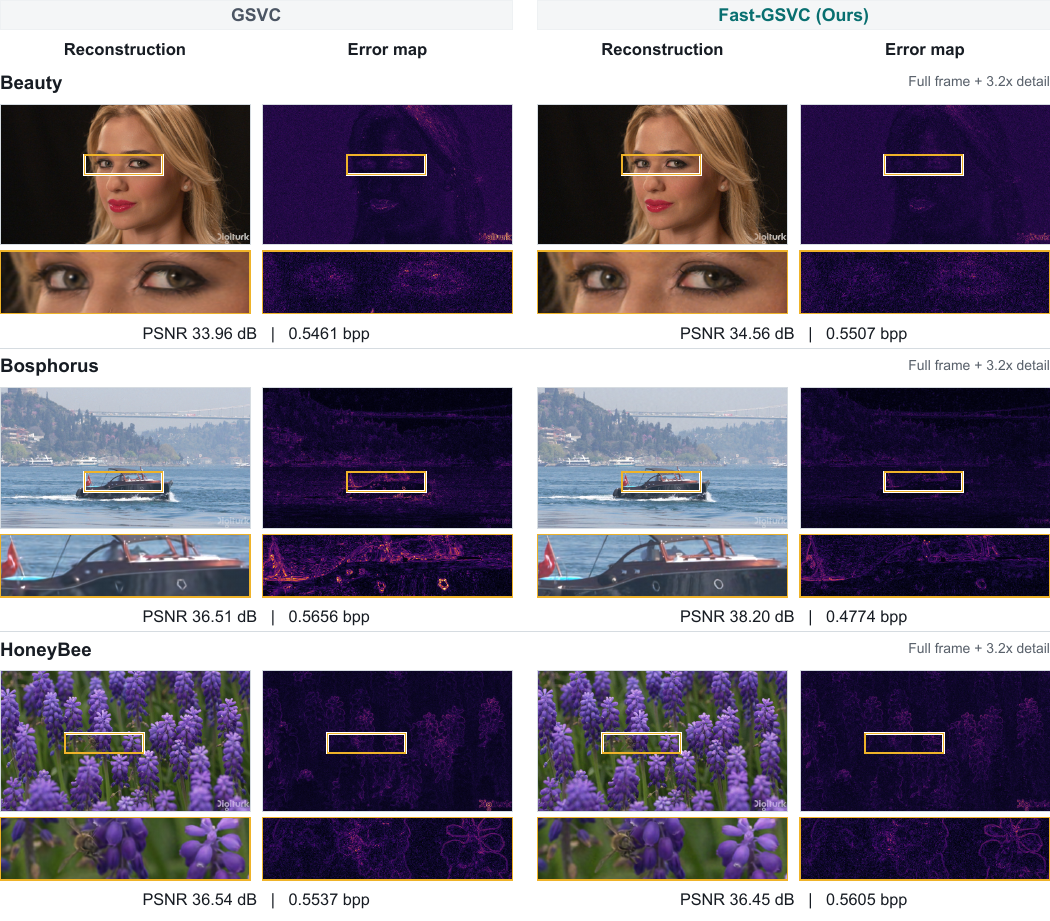}
    \caption{P-frame (frame 2) reconstruction quality on Beauty, Bosphorus, and HoneyBee. GSVC and {\name} are compared using reconstructions, error maps, and enlarged details, with PSNR (dB) and bpp reported for each method.}
    \label{fig:appendix_pframe_1}
\end{figure}

\begin{figure}[!ht]
    \centering
    \includegraphics[width=\linewidth]{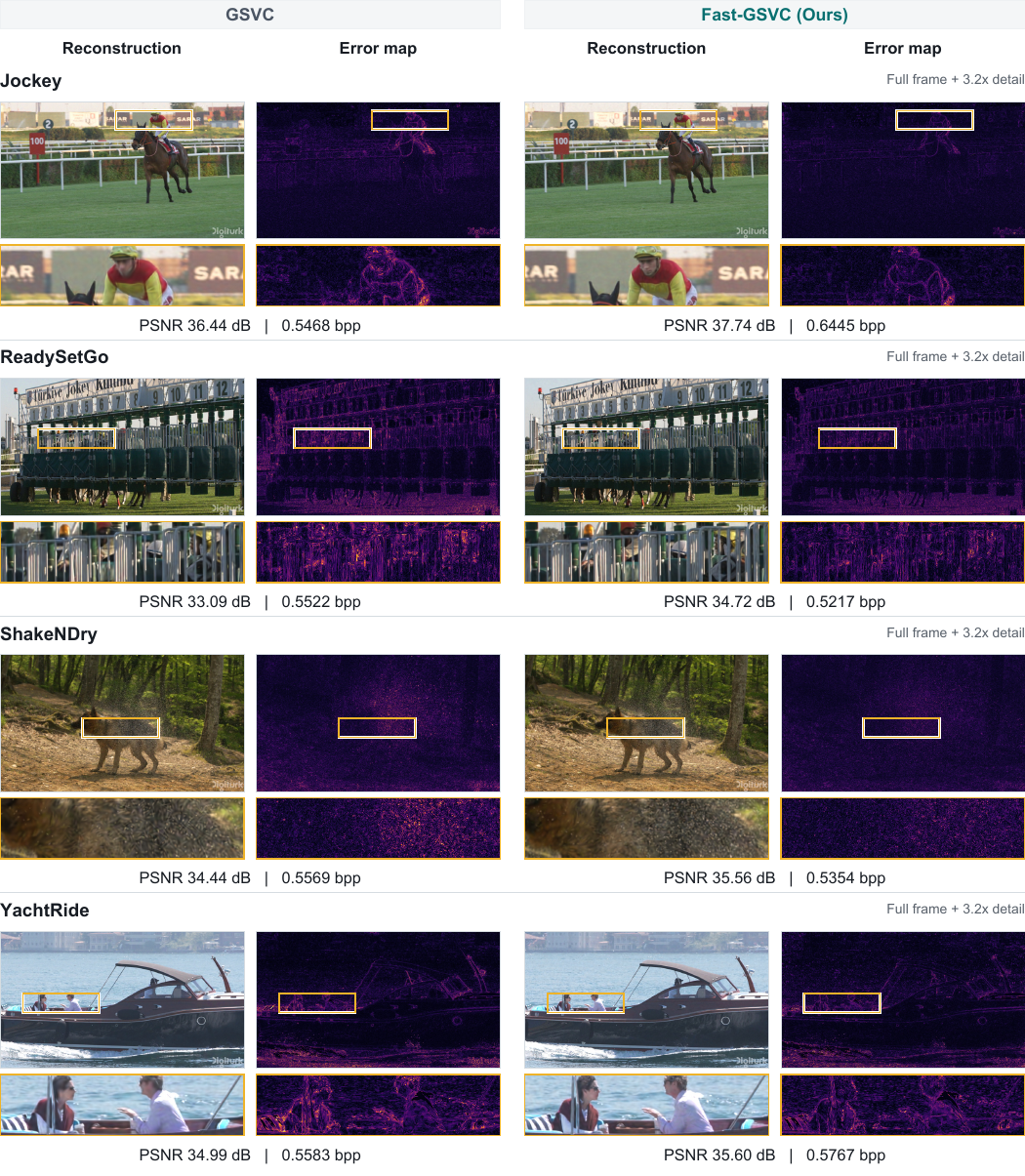}
    \caption{P-frame (frame 2) reconstruction quality on Jockey, ReadySetGo, ShakeNDry, and YachtRide. GSVC and {\name} are compared using reconstructions, error maps, and enlarged details, with PSNR (dB) and bpp reported for each method.}
    \label{fig:appendix_pframe_2}
\end{figure}

\Cref{fig:appendix_pframe_1,fig:appendix_pframe_2} compare P-frame quality across seven UVG sequences. Despite substantially faster encoding, {\name} achieves better or comparable reconstruction quality at similar or lower bitrates, as illustrated by the reconstructed frames and error maps.

\clearpage

\end{document}